%% file: main.tex
\documentclass{paper}

\lprepnumber{%
DFTT 46/09\\
EPHOU 09-003\\
July, 2009}
\ltitle{Formulation of Supersymmetry on a Lattice\\
as a Representation of a Deformed Superalgebra}
\lauthor{%
Alessandro D'Adda\inst{a}\email{dadda}{to.infn.it},
Noboru Kawamoto\inst{b}\email{kawamoto}{particle.sci.hokudai.ac.jp}
\lastauthor\
Jun Saito\inst{b}\email{saito}{particle.sci.hokudai.ac.jp}%
}
\linstitute{%
\inst{a} INFN sezione di Torino,\\ and\\
Dipartimento di Fisica Teorica,
Universita di Torino\\
I-10125 Torino, Italy\\[2ex]
\inst{b} Department of Physics, Hokkaido University\\
Sapporo, 060-0810 Japan%
}
\ldate{March 31, 2009}

\begin{document}
\lmaketitle

\input{abst}
\newpage
\input{intro}

\input{rev}

\input{hopf}

\input{concl}

\appendix
\input{app-hopf}

\input{app-wz}

\input{bib}

\end{document}

%% file: abst.tex
\begin{abstract}
The lattice superalgebra of the link approach is 
shown to satisfy a Hopf algebraic supersymmetry where the 
difference operator is introduced as a momentum operator. 
The breakdown of the Leibniz rule for the lattice 
difference operator is accommodated as a coproduct operation 
of (quasi)triangular Hopf algebra and the associated field theory is 
consistently defined as a braided quantum field theory.
Algebraic formulation of path integral is perturbatively 
defined and Ward--Takahashi identity can be derived on the lattice.
The claimed inconsistency of the link approach leading to 
the ordering ambiguity for a product of fields is solved 
by introducing an almost trivial braiding structure
corresponding to the triangular structure of the Hopf algebraic
superalgebra. This could be seen as a generalization of spin and 
statistics relation on the lattice. From the consistency of
this braiding structure of fields a grading nature for the momentum
operator is required.
\end{abstract}

%% file: intro.tex
\section{Introduction}



We consider that a constructive definition of regularized supersymmetric 
field theory is getting 
increasingly important. There are several reasons: 
Phenomenologically there is an expectation that superparticles might be 
discovered by LHC experiment in the near future. If the supersymmetry 
becomes reality, obviously we need to formulate supersymmetric field theory 
constructively. The formulation should provide a basis for the numerical 
study of nonperturbative supersymmetry phenomenology. It is natural 
to expect that a lattice formulation of supersymmetry may play a 
crucial \role just like the lattice QCD is playing an important \role as 
the only numerical mean for the strong interaction phenomenology. 
We expect that a fermionic counterpart of QCD region may exist in the 
new energy scale. 

Secondly, it is not obvious that the lattice fermion
problems~\cite{No-Go, Anomaly} are well understood from the lattice 
regularization 
point of view. It is, however, a general consensus that the chiral 
fermion problem is solved for lattice 
QCD~\cite{Gins-Wil,Domai-Wall,chiral-fermion}. One may say 
that species doublers of chiral fermion on a lattice are lattice 
artifacts so that it would have been better if they were not there. 
It was, however, claimed that these 
extra species doubler degrees of freedom is exactly the needed one 
and corresponds to the extended twisted 
supersymmetry degrees of freedom~\cite{DKKN:1, DKKN:2, DKKN:3, Catt:1}. 
It was shown that the twisted supersymmetry can be derived by the 
Dirac-\Kahler twisting procedure~\cite{Kawamoto:DK} in any dimensions: 
$\MC{N}=2$ in two dimensions,
$\MC{N}=4$ in three dimensions, and
$\MC{N}=4$ in four dimensions which 
coincides with the twisting derived by Marcus~\cite{Marcus}. 
In these formulations the fermionic internal degrees of freedom 
can be defined semilocally on a lattice to be compatible with 
differential form nature of Dirac-\Kahler 
fermion~\cite{DKKN:1,DKKN:2,DKKN:3,Catt:1,Nagata:CS,Nagata:nclattice}. 
This type of the correspondence has been anticipated by the 
old works~\cite{old-Dirac-Kahler}.
It turns out that the lattice Dirac-\Kahler fermion 
formulation~\cite{Becher-Joos} 
was, however, proved to be equivalent to the 
staggered fermion formulation~\cite{Kanamori-Kawamoto} with an 
introduction of 
mild noncommutativity between differential forms and fields to 
accommodate the modified Leibniz rule for lattice difference 
operator.
These results suggest that the regularization of 
fermions on a lattice naturally leads to a necessity of supersymmetry 
in a fundamental way. 

In the path integral formulation of field theory, fermionic fields 
are treated as Grassmann odd variables and thus have an anti-commuting 
nature in compatible with spin and statistics theorem which 
requires Lorentz invariance exactly~\cite{Pauli}. 
Since the Lorentz invariance is 
broken on the lattice it is not obvious that this anti-commuting 
nature of fermions at the lattice constant level is the mandatory 
requirement. In this paper we explore a possibility that commuting 
and anti-commuting nature of fields are modified with an introduction 
of mild noncommutativity in compatible with the lattice Leibniz rule 
of difference operator. This may be identified as a generalization of 
spin and statistics on a lattice. 

Trials for the formulation of supersymmetry on a lattice have a long 
history. 
Since lattice does not have infinitesimal translational invariance, 
there are various difficulties to formulate supersymmetry algebra which 
includes an infinitesimal translation generator. 
To overcome the difficulties, various approaches and formulations
have been proposed so far.
If we focus on the treatment of the algebraic aspects of lattice 
supersymmetry, there are essentially three possible approaches:
\begin{enumerate}
\item keeps the continuum superalgebra approximately with lattice 
corrections; 
%
\item keeps exactly only a subalgebra of the continuum superalgebra which
doesn't contain the momentum operators;

\item deforms the continuum superalgebra into a lattice version of
superalgebra, and keeps exactly
the full sector of this lattice superalgebra.

\end{enumerate}
There is a long list of many trials of the first 
approach summarized in \cite{Dondi-Nicolai, Montvay, Feo} and the 
previous references are 
therein, where superalgebra is kept up to the lattice corrections. 
There are some later 
developments~\cite{other-lattsusy1, other-lattsusy2, Japanese-recent,
Fujikawa:Leibniz}.
In this approach one has to see how the superalgebra
is restored in the continuum limit. 
In order to see the recovery of supersymmetry for the whole range 
of coupling constant, it is inevitable to 
find reliable methods for numerical analyses. 
Influenced by the developments of renormalization group analyses and 
Ginsparg-Wilson relation of chiral symmetry analyses for lattice QCD, 
there have been recent systematic applications of the methods to 
supersymmetric models~\cite{D-W-lattsusy, Igarashi-So, G-W Bruckmann}.

In the second approach \cite{deconstruction, Unsal, twist-lattsusy, Sugino,
Takimi, Giedt, Damgaard-Matsuura, Unsal2, review-lat-SUSY},
only one or two nilpotent supercharges are 
preserved exactly. It is again particularly important to examine 
how the full continuum
supersymmetry is recovered in the continuum limit. 
In this approach the importance of accidental symmetry is stressed 
so that the recovery of supersymmetry in the infrared region is 
expected due to the suppression of relevant operators by the 
partial exactness of the total supersymmetry algebra%
~\cite{accidental symmetry, review-lat-SUSY}.
In fact in some specific cases it can be shown that 
only the part of supersymmetry which is realized on the lattice
is enough to suppress non-manageable fine-tuning
in the continuum limit \cite{deconstruction, twist-lattsusy, Sugino,
Takimi, review-lat-SUSY, Kanamori}.
One can then
use such models as constructive definition of corresponding
continuum supersymmetric models.
We should, however, note that in extracting the sector of superalgebra
which can be preserved on the lattice, extended supersymmetry
and its twisting procedure~\cite{Witten:tft, Kawamoto:DK}
play an important \role.

In the third approach \cite{DKKN:1, DKKN:2, DKKN:3, Nagata:CS,
Nagata:nclattice, Damgaard-Matsuura, ADFKS}, one defines lattice version
of superalgebra where the momentum operators 
in the continuum superalgebra are replaced by finite
difference operators on the lattice.
This seems the most natural and \naive
deformation of continuum algebra.
Nevertheless, it is not straightforward due to the following
obvious reason.
Since a finite difference operator, not being an infinitesimal
operator, is not rigorously an element of algebra,
that deformed ``superalgebra'' is not strictly an algebra in the
usual sense.
This is actually not simply a terminological issue,
but is crucial in the formulation.
Namely, finite difference operator does not obey the Leibniz
rule, which is nothing but equivalent to say that the operator
is not an element of algebra.
On the other hand, because of the nilpotency of Grassmann parameters,
``normal'' supercharge would always obey exactly the Leibniz
rule. This mismatch of natures of finite difference operators
and supercharges makes the \naive realization of
lattice deformed superalgebra in the above sense difficult.
What we are going to follow in this paper is in fact the formulation
struggling to give an answer to this difficult situation,
which was originally proposed in \cite{DKKN:1, DKKN:2, DKKN:3}.
We call this formulation as link approach as a whole or DKKN formalism 
when we have more stress on the algebraic aspect of the formulation. 
In this approach, we introduce the notion of \emph{modified Leibniz rule}
to overcome this difficulty. 

Despite of the invention of the modified Leibniz rule,
the link approach still faces its incompleteness. Firstly,
it is rather unclear whether this additionally introduced modified
Leibniz rule is totally consistent and acceptable to express
the ``symmetry'' of a quantum field theory,
because it is in any case different
from the standard Lie algebraic symmetry.
As an answer to this issue we will present that this modification is indeed
consistently introduced, utilizing the fact that
the deformed algebra forms a \emph{Hopf algebra} which generalizes
the Lie algebraic symmetry, and that a quantum field theory
which has a Hopf algebraic symmetry can be constructed at least
perturbatively thanks to the previously formulated framework
known as braided quantum field theory (BQFT) \cite{Oeckl}.
For this argument,
it is important
to identify the DKKN superalgebra as a rigorous Hopf algebra
which obeys a set of axioms to prescribe the Hopf algebra.
It is also crucial
to correctly determine the braiding structure on the representation
space of the Hopf algebra.
With the use of the BQFT formulation, we can derive
a series of Ward--Takahashi identities corresponding to the Hopf
algebraic symmetry, which would give a clear physical interpretation
of the deformed symmetry.

The second and most crucial aspect to be clarified in the link 
approach is on the ``inconsistency'' raised in \cite{Dutch},
which claims that the modified Leibniz rule inevitably leads
to a problem due to the ordering ambiguity caused essentially
by the asymmetric nature of the deformation.
In order to clarify the ordering problem, a matrix formulation
for one dimensional model was explicitly analysed. It was shown
that there is no ambiguity at the superfield level
but the problem remains at the component level \cite{ADFKS}.
Fortunately,
our Hopf algebraic
description also resolves this problem: given the
Hopf algebraic symmetry together with the
appropriate treatment of the braiding structure of BQFT,
we will show that this difficulty no longer exists.
The braiding structure, which could be interpreted as
a kind of generalized statistics or a ``mild noncommutativity'',
is again the key ingredient for this argument. The treatment for 
the gauge theory is outside of the scope of this paper. 

In the recent investigations it is stressed that the second approach 
formulated by orbifold construction and the third approach of the 
link construction are equivalent~\cite{Damgaard-Matsuura, Unsal2}.
It was already noticed 
that a particular choice of a constant vector parameter makes 
the scalar supercharge shiftless and $\MC{N}=D=2$ super Yang-Mills action 
of the link construction coincides with that of orbifold 
construction~\cite{DKKN:2}. It was, however, stressed
that supercharges carrying shifts are not supersymmetry invariant
\cite{Damgaard-Matsuura, Unsal2}.
Therefore in the symmetric choice of the 
parameters in the link construction, all the supercharges carry 
shifts and thus no supersymmetry exists, although the corresponding
action has larger discrete chiral and 
spacetime symmetries than that of orbifold construction~\cite{Unsal2}.
This criticism is due to the non-standard definition of the 
shifted (anti-)commutators. 
In this paper we intend to stress exactly on this point that all 
supercharges of the link construction preserves deformed lattice 
supersymmetry exactly, where shifting nature plays a 
crucial \role. 

Recently it was shown as a no-go theorem that a proper definition 
of product of lattice fields naturally leads to a breakdown of 
Leibniz rule of lattice differential operator 
under the conditions of translational invariance and 
locality on the lattice~\cite{No-go:Kato-Sakamoto-So}.
It was also shown that an 
extension of blocked symmetry transformation realizing 
Ginsparg-Wilson relation to supersymmetric case leads 
to only consistent solution of SLAC-type derivative~\cite{SLAC} 
which is in fact consistent with the above no-go
theorem~\cite{G-W Bruckmann}. 
It is known, however, that SLAC-derivative is a highly non-local 
differential operator. It should be noted that the modified nature 
of the lattice Leibniz rule or equivalently the deformation of 
supersymmetry algebra is compatible with the results of those 
analyses.

This paper is organized as follows.
In Section~\ref{rev} we will make a brief review of the fundamental
structures of the link formulation in somewhat generalized
form and its above mentioned difficulties. 
In Section~\ref{hopf} we will give a description
how to treat the superalgebra on the lattice
as a deformed/modified algebra in the scheme of the Hopf algebra theory.
We will list all the necessary and sufficient formulae which
form the whole structures of a Hopf algebra. We also
derive the explicit form of the braiding which is necessary
for the total consistency of the representation.
Twisting of our Hopf algebra will be discussed, too,
which naturally explains our lattice theory should
have the braiding or equivalent noncommutativity.
We will illustrate how a quantum field theory with this Hopf
algebraic symmetry can be perturbatively defined entirely
based on the general formulation of BQFT. 
As a concrete example we show a two dimensional $\MC{N}=2$ 
Wess-Zumino model.
We then give the conclusion of the paper and
discuss some remaining issues in the last section.
In the appendix we give a brief summary of Hopf algebra 
to fix the notation and terminology appeared in the text.

%% file: rev.tex

\section{General Framework of the Link Formulation
of the Dirac--\Kahler Twisted Supersymmetry on a Lattice}
\label{rev}

\subsection{Generality of the Formulation}

The principle of the link approach~\cite{DKKN:1, DKKN:2, DKKN:3}
to a realization of a supersymmetric theory on a lattice
is based on a simple assumption that the superalgebra
in the continuum%
\footnote{The notation here is schematic; indices
$A$ and $B$ could contain both spinor and internal
d.o.f.,\ and their conjugate as well.}
\begin{equation}
	\begin{gathered}
		\{Q_A, Q_B\}= 2\tau_{AB}^\mu P_\mu,\\
		[Q_A, P_\mu] = [P_\mu, P_\nu] = 0
	\end{gathered}
\end{equation}
has some natural counterpart on the lattice 
\begin{equation}
	\begin{gathered}
		\{\Ql_A, \Ql_B\}
		= 2\tau_{AB}^\mu \Pl_\mu,\\
		[\Ql_A, \Pl_\mu]
		= [\Pl_\mu, \Pl_\nu] = 0.
	\end{gathered}
\label{alg-lat-gen}
\end{equation}
Here $\tau_{AB}^\mu$
is just a constant coefficient, and
$\Ql_A$ and $\Pl_\mu$
are understood both as \emph{deformed} operators on the lattice
which come back to $Q_A$ and $P_\mu$, respectively, in
the \naive continuum limit;
\begin{equation}
	\lim_{a\RA 0}\Ql_A = Q_A,\qquad
	\lim_{a\RA 0}\Pl_\mu = P_\mu.
\end{equation}
We require that
\begin{equation}
	\sum_x \Pl_\mu\Vp(x) = 0
\end{equation}
for the ``momentum'' operator $\Pl_\mu$ and any field $\Vp(x)$
on the lattice. This is because, in the continuum,
the general superinvariance of Lagrangian
is up to total divergence which vanishes under the integral,
and the same structure should be true for the ``exact'' supersymmetry
on the lattice, for which the property above is necessary.
We would also require the translational invariance
and (semi-)locality for the operator
$\Pl_\mu$ so that the whole theory would have these properties.
Another possible requirement might be the Hermiticity
(or the reflection (Osterwalder--Schrader) positivity
\cite{Osterwalder-Schrader}
of transfer matrices on the lattice \cite{reflection}),
but we don't force it here because it is related to the subtlety
of the doubling phenomenon \cite{No-Go, Anomaly}
for which we defer the discussion
to later sections.

The simplest candidates for the ``momentum''
operator $\Pl_\mu$ would be the finite
difference operators on the lattice,
\begin{equation}
	\Pl_\mu
	= i\del_{\pm\mu},\quad i\del_\mu^\MR{s},\quad \text{etc.},
\label{diffops}
\end{equation}
where
\begin{alignat}{2}
	&\del_{+\mu}\Vp(x)
		:= \frac{1}{a}\bigl( \Vp(x+a\muhat) - \Vp(x) \bigr)
		&\qquad
		&(\text{forward difference operator}),\\
	&\del_{-\mu}\Vp(x)
		:= \frac{1}{a}\bigl( \Vp(x) - \Vp(x-a\muhat) \bigr)
		&\qquad
		&(\text{backward difference operator}),\\
	&\begin{aligned}
		\del_\mu^\MR{s}\Vp(x)
		&:= \frac{1}{2}\bigl( \del_{+\mu} + \del_{-\mu} \bigr)\Vp(x) \\
		&=  \frac{1}{2a}\bigl( \Vp(x+a\muhat) - \Vp(x-a\muhat) \bigr)
	\end{aligned}
		&\qquad
		&(\text{symmetric difference operator}),
\end{alignat}
where $\muhat$ is the unit vector to the direction of $x^\mu$
and $a$ is the lattice constant%
\footnote{We always keep the lattice constant $a$ explicitly in this paper
unless otherwise specified.}. The symmetric difference is self anti-Hermitian:
$(\del_\mu^\MR{s})^\dagger = -\del_\mu^\MR{s}$, while the others
are anti-Hermitian conjugate to each other;
$(\del_{\pm\mu})^\dagger = -\del_{\mp\mu}$.
An immediate consequence of using
these finite difference operators is that they break
the Leibniz rule, or put it milder, obey
the \emph{modified} Leibniz rule as in
\begin{equation}
	\begin{aligned}
		\del_{\pm\mu}(\Vp_1\Vp_2)(x)
		&= \del_{\pm\mu}\Vp_1(x) \Vp_2(x)
			+ \Vp_1(x \pm a\muhat) \del_{\pm\mu}\Vp_2(x) \\
		&= \del_{\pm\mu}\Vp_1(x) \Vp_2(x \pm a\muhat)
			+ \Vp_1(x) \del_{\pm\mu}\Vp_2(x) \\
		&= \del_{\pm\mu}\Vp_1(x) \Vp_2(x)
			+ \Vp_1(x) \del_{\pm\mu}\Vp_2(x)
			\pm a \del_{\pm\mu}\Vp_1(x) \del_{\pm\mu}\Vp_2(x),
	\end{aligned}
\end{equation}
and
\begin{equation}
	\begin{aligned}
		\del_\mu^\MR{s}(\Vp_1\Vp_2)(x)
		&= \del_\mu^\MR{s}\Vp_1(x) \Vp_2(x-a\muhat)
			+ \Vp_1(x+a\muhat) \del_\mu^\MR{s}\Vp_2(x) \\
		&= \del_\mu^\MR{s}\Vp_1(x) \Vp_2(x+a\muhat)
			+ \Vp_1(x-a\muhat) \del_\mu^\MR{s}\Vp_2(x) \\
		&= \del_\mu^\MR{s}\Vp_1(x) \Vp_2(x)
			+ \Vp_1(x) \del_\mu^\MR{s}\Vp_2(x)
			+ \frac{a}{2}\Bigl(
				\del_{+\mu}\Vp_1(x) \del_{+\mu}\Vp_2(x)
				-
				\del_{-\mu}\Vp_1(x) \del_{-\mu}\Vp_2(x)
			\Bigr).
	\end{aligned}
\label{sym-diff-Leibniz}
\end{equation}
Although superficially the breaking term in each case of the Leibniz rule
is proportional to the lattice constant $a$,
it is not in general of higher order in the continuum limit,
due to the contributions from the cut off scale region of the momentum
$\del_{\pm\mu}\Vp(x)\sim\Order{1/a}$ \cite{Fujikawa:Leibniz}.
Note also that the last term of \Ref{sym-diff-Leibniz} is proportional
to a total difference
$\del_{-\mu}\bigl(\del_{+\mu}\Vp_1(x)\del_{+\mu}\Vp_2(x)\bigr)$,
so that one may consider that this breaking of the Leibniz rule
is irrelevant under a summation over the whole lattice sites.
But this is true only for the product of two fields,
so that might be a good property only in a free theory,
not in an interacting case. (Even in the free case
there is an associated doubler problem for the anti-Hermitian
symmetric difference. We will see this later in more detail.)
One might also try to impose a constraint on the fields to
make the breaking terms vanish, but this would only result
in a nonlocal formulation \cite{Dondi-Nicolai}.
Thus, as long as we use the simple difference operators \Ref{diffops},
we can't \naively neglect the breaking of the Leibniz rule.
In fact, it is more generally shown
\cite{G-W Bruckmann, No-go:Kato-Sakamoto-So}
that we have to admit
the breaking of the Leibniz rule of any ``momentum'' operators
on a lattice, unless we allow nonlocal operators like so-called
SLAC derivative \cite{SLAC} or, say, many multiflavors.
These facts are already enough for the lattice counterpart of the
superalgebra \Ref{alg-lat-gen} to lose the nature of
strict Lie superalgebra, which is the most evident and crucial
obstacle to formulate supersymmetry entirely based on the superalgebra
on a lattice.

One possibility to overcome the situation is to interpret
the superalgebra on the lattice \Ref{alg-lat-gen} as a
``deformed'' Lie superalgebra with the deformation parameter
that vanishes in the continuum limit.
This is in fact the basic strategy in the link approach as we
can see in what follows.

Since the r.h.s.\ of \Ref{alg-lat-gen} obeys the modified
Leibniz rule, it is natural to deform the algebra
so that the generators in the l.h.s.\ also obeys
a modified Leibniz rule. In the link approach, the central
ansatz is that the supercharge $\Ql_A$ obeys
the Leibniz rule of the form%
\footnote{Here $\stat{\Vp}$ is $0$ or $1$, depending
on whether $\Vp$ is bosonic or fermionic, respectively.}
\begin{equation}
	\Ql_A(\Vp_1\Vp_2)(x)
	= \Ql_A\Vp_1(x)\Vp_2(x)
	+ (-1)^\stat{\Vp_1}\Vp_1(x+a_A) \Ql_A\Vp_2(x),
\label{Leibniz-Q}
\end{equation}
where $x+a_A$ is to be interpreted as denoting an extended
lattice site which goes to $x$ in the \naive continuum limit.
Introducing a translation or shift operator $T_{a_A}$ in a 
``fundamental'' representation such that
\begin{equation}
	T_{a_A}\Vp(x)=\Vp(x+a_A),
\label{fund-T}
\end{equation}
the Leibniz rule can be written as
\begin{equation}
	\begin{aligned}
		\Ql_A(\Vp_1\Vp_2)(x) 
		&= \Ql_A\Vp_1(x)\Vp_2(x)
			+ (-1)^\stat{\Vp_1}
			T_{a_A}\Bigl(\Vp_1 T_{a_A}^{-1}\Ql_A\Vp_2\Bigr)(x), \\
		\IE
		T_{a_A}^{-1}\Ql_A(\Vp_1\Vp_2)(x) 
		&= \bigl(T_{a_A}^{-1}\Ql_A\Vp_1\bigr)(x)\Vp_2(x-a_A)
			+ (-1)^\stat{\Vp_1}
			\Vp_1(x) \bigl(T_{a_A}^{-1} \Ql_A\Vp_2\bigr)(x),
	\end{aligned}
\label{Leibniz-Q-right}
\end{equation}
showing that the operator $T_{a_A}^{-1}\Ql_A$ obeys
a slightly different modified Leibniz rule. We may also write
it in a symmetric form as
\begin{equation}
		T_{a_A}^{-1/2}\Ql_A(\Vp_1\Vp_2)(x) 
		= \bigl(T_{a_A}^{-1/2}\Ql_A\Vp_1\bigr)(x)\Vp_2(x-a_A/2)
			+ (-1)^\stat{\Vp_1}
			\Vp_1(x+a_A/2) \bigl(T_{a_A}^{-1/2} \Ql_A\Vp_2\bigr)(x),
\label{Leibniz-Q-sym}
\end{equation}
which is still a modified version of Leibniz rule.
We could have begun with a little more generalized modification
such as
\begin{equation}
	\Ql_A(\Vp_1\Vp_2)(x)
	= \Ql_A\Vp_1(x)\Vp_2(x+\ar_A)
	+ (-1)^\stat{\Vp_1}\Vp_1(x+\al_A) \Ql_A\Vp_2(x),
\label{Leibniz-Q-gen}
\end{equation}
but with a redefinition of $\Ql_A$ to $T_{\ar_A}^{-1}\Ql_A$
it is always equivalent to the original one \Ref{Leibniz-Q},
which can be seen in a similar fashion as in the above.
Such a redefinition only causes a total difference in the algebra
\Ref{alg-lat-gen}, hence the original form \Ref{Leibniz-Q} suffices
in general.

The field in the fundamental representation of the
translation/shift operator \Ref{fund-T} could be
interpreted as a normal function on the lattice.
If, by contrast, we introduce
the ``adjoint'' representation of the translation/shift operator
as in
\begin{equation}
	T_{a_A}\Vp(x)T_{a_A}^{-1} = \Vp(x+a_A),
\label{adjoint}
\end{equation}
the Leibniz rule \Ref{Leibniz-Q} can be written as
\begin{equation}
		\begin{aligned}
		\Ql_A(\Vp_1\Vp_2)(x) 
		&= \Ql_A\Vp_1(x)\Vp_2(x)
			+ (-1)^\stat{\Vp_1} T_{a_A}
			\Vp_1(x) T_{a_A}^{-1} \Ql_A\Vp_2(x), \\
		\IE
		T_{a_A}^{-1}\Ql_A(\Vp_1\Vp_2)(x) 
		&= T_{a_A}^{-1}\Ql_A\Vp_1(x)\Vp_2(x)
			+ (-1)^\stat{\Vp_1}
			\Vp_1(x) T_{a_A}^{-1} \Ql_A\Vp_2(x).
	\end{aligned}
\label{Leibniz-Q-usual}
\end{equation}
Now we can see that the operator $T_{a_A}^{-1}\Ql_A$ obeys the usual
exact Leibniz rule.
We could write this further as
\begin{equation}
	T_{a_A}^{-1}\Ql_A(\Vp_1\Vp_2)(x)T_{a'_A}
	= T_{a_A}^{-1}\Ql_A\Vp_1(x)T_{a'_A} \Vp_2(x-a'_A)
	+ (-1)^\stat{\Vp_1}
	\Vp_1(x) T_{a_A}^{-1} \Ql_A\Vp_2(x)T_{a'_A},
\end{equation}
which shows the operator $T_{a_A}^{-1}\Ql_A \overleftarrow{T}_{a'_A}$,
where the arrow denotes the multiplication from the right, follows
a different Leibniz rule. We would thus again find that the original
modified Leibniz rule \Ref{Leibniz-Q} itself is equivalent
to the more general form \Ref{Leibniz-Q-gen}, and even
to the usual Leibniz rule \Ref{Leibniz-Q-usual} with a suitable
redefinition of the operator $\Ql_A$. Notice
that in this adjoint representation such a redefinition
would change the algebra \Ref{alg-lat-gen} in a nontrivial way,
except for the case $\ar_A = \al_A$ for which
the algebra, or more precisely the ``momentum'' operator,
would remain unchanged up to a total difference
so that the ``momentum'' operator still obeys the modified
Leibniz rule.
In other words, unless $\ar_A = \al_A$, we have a possibility
to redefine both the operators $\Ql_A$ and $\Pl_\mu$ so as to follow
the usual Leibniz rule for which the usual representation would exist.
This fact may play an important \role for an explicit representation of 
the lattice superalgebra.
Another point to observe is that,
in the adjoint representation, field
itself should be identified as an operator or a matrix
which formally belongs, together with the shift operator, to an
algebra (which would be a universal enveloping
algebra of a Lie superalgebra, just as in a canonical quantization
scheme.). The operator $T_{a_A}$, when multiplying from the left/right
on a field, changes the property under a commutation of the field.
For instance, 
suppose $\Vp_1(x)$ and $\Vp_2(x)$ commute with each other;
$\Vp_1(x)\Vp_2(x)=\Vp_2(x)\Vp_1(x)$. Then
$T_{a_A}\Vp_1(x)$ and $\Vp_2(x)$ no longer commute strictly, but
commute with a shift in the sense that
$T_{a_A}\Vp_1(x)\Vp_2(x) = T_{a_A}\Vp_2(x)T_{a_A}^{-1}T_{a_A}\Vp_1(x)
 = \Vp_2(x+a_A)T_{a_A}\Vp_1(x)$. 
This type of mild noncommutative nature doesn't simply occur
in the fundamental case since
$(T_{a_A}\Vp_1)(x)\Vp_2(x)=\Vp_2(x)(T_{a_A}\Vp_1)(x)$. We will see in 
the next chapter how the modified Leibniz rules in the fundamental
representation \Ref{Leibniz-Q}, \Ref{Leibniz-Q-right} and
\Ref{Leibniz-Q-sym} can be more systematically
treated in the framework of Hopf
algebraic symmetry. Here in what follows
we continue the general description mainly with the adjoint
representation case.

Suppose now $\Ql_A$ also belongs to this same algebra
that $T_{a_A}$ and $\Vp(x)$ form.
The fact that the combination $T_{a_A}^{-1}\Ql_A$ in \Ref{Leibniz-Q-usual}
follows the usual Leibniz rule thus motivates us to write formally
\begin{equation}
	T_{a_A}^{-1}\Ql_A =: \Qlh_A \equiv i\ad(\Qlch_A),
\end{equation}
which acts on a field as%
\footnote{Here in the equation below $[A,B]_\pm := AB\pm BA$.}
\begin{equation}
	\begin{gathered}
		T_{a_A}^{-1}\Ql_A\Vp(x)
		= i\ad(\Qlch_A)\Vp(x)
		:= i[\Qlch_A, \Vp(x)]_{(-1)^{\stat{\Vp}+1}},\\
		\OR
		\Ql_A\Vp(x) = i T_{a_A} [\Qlch_A, \Vp(x)]_{(-1)^{\stat{\Vp}+1}}.
	\end{gathered}
\label{def-adjoint}
\end{equation}
It can also be written as
\begin{equation}
	\begin{aligned}
		\Ql_A\Vp(x)
		&= i T_{a_A} \Qlch_A \Vp(x)
		-(-1)^\stat{\Vp} i T_{a_A}\Vp(x)T_{a_A}^{-1} T_{a_A} \Qlch_A \\
		&= i\Qlc_A\Vp(x)
		-(-1)^\stat{\Vp} \Vp(x+a_A) i\Qlc_A
		=: i[\Qlc_A, \Vp(x)]_{(-1)^{\stat{\Vp}+1}}^\MR{lat}
		=: i\ad^\MR{lat}(\Qlc_A)\Vp(x),
	\end{aligned}
\end{equation}
where $\Qlc_A:= T_{a_A}\Qlch_A$. In this last equation we have
defined a kind of deformed adjoint operation $\ad^\MR{lat}$
which was referred to as the \emph{shifted} (anti-)commutator
in the DKKN formalism.
It illustrates the general fact that an operator which obeys
a modified Leibniz
rule could be expressed with a shifted (anti-)commutator.
We have, however, introduced objects like $\Qlch_A,\ \Qlc_A$
and respectively their (anti-)commutator and shifted (anti-)commutator
$\ad(\Qlch_A)$ and
$\ad^\MR{lat}(\Qlc_A)$ only in a formal way,
neither specified the explicit forms nor even justified
the existence of them. So far we have only found that
$T^{-1}_A\Ql_A = \Qlh_A$ obeys the usual Leibniz rule,
which would be regarded as a normal operator,
and that $\Ql_A$ would be expressed as $\Ql_A = T_{a_A}\Qlh_A$.
The point here is the following: As we mentioned above,
we assume that the shift parameter $a_A$ reduces to zero
in the \naive continuum limit. Correspondingly
the translation/shift operator $T_{a_A}$ would go to unity
in the limit: $T_{a_A}\RA\vid$, and thus the formal expression
$\Ql_A = i T_{a_A}\ad(\Qlch_A) = i\ad^\MR{lat}(\Qlc_A)$ reduces to
the normal (anti-)commutator $Q_A = i\ad(\MC{Q}_A)$.
This implies that normal (anti-)commutators in the continuum,
if used in any algebraic expressions, should be simply replaced with
the shifted (anti-)commutators on the lattice to accommodate
the modified Leibniz rule \Ref{Leibniz-Q}.
This reminds us of the correspondence principle between
the Poisson bracket in the classical theory and
the commutator in the quantum theory.
We are motivated by this analogy to think the lattice
version of the superalgebra of a ``quantization'' of
the continuum superalgebra. This viewpoint of the formulation
is discussed in the next chapter.

Let us move on to the algebra \Ref{alg-lat-gen}.
Here, for generality, we consider
the modified Leibniz rule \Ref{Leibniz-Q-gen}.
The l.h.s.\ of the algebra applies on a product $\Vp_1\Vp_2$ as in
\begin{equation}
	\begin{aligned}
		\{\Ql_A, \Ql_B\}(\Vp_1\Vp_2)(x)
		&=\{\Ql_A, \Ql_B\}\Vp_1(x) \Vp_2(x+\ar_A+\ar_B)
		+\Vp_1(x+\al_A+\al_B) \{\Ql_A, \Ql_B\}\Vp_2(x) \\
		&=\sum_\mu 2\tau_{AB}^\mu\biggl(
			\Pl_\mu\Vp_1(x) \Vp_2(x+\ar_A+\ar_B)
			+
			\Vp_1(x+\al_A+\al_B) \Pl_\mu\Vp_2(x)
		\biggr),
	\end{aligned}
\label{lhs-on-product}
\end{equation}
while the r.h.s.\ as in
\begin{equation}
	2\tau_{AB}^\mu \Pl_\mu(\Vp_1\Vp_2)(x)
	= \sum_\mu 2\tau_{AB}^\mu\biggl(
	\Pl_\mu\Vp_1(x) \Vp_2(x + \ar\muhat)
	+
	\Vp_1(x + \al\muhat) \Pl_\mu\Vp_2(x)
	\biggr),
\label{rhs-on-product}
\end{equation}
where $\al$ and $\ar$ are, depending on the choice of $\Pl_\mu$,
\begin{equation}
	(\al, \ar)
	= \begin{cases}
		(\pm a, 0) \OR (0, \pm a) & \text{for}\ \Pl_\mu = i\del_{\pm\mu}, \\
		(+a,-a) \OR (-a, +a) & \text{for}\ \Pl_\mu = i\del_\mu^\MR{s}.
	\end{cases}
\label{shifts-diff}
\end{equation}
The algebra \Ref{alg-lat-gen} requires that these two
equations to be equal. As we can easily find,
the first necessary condition is that
the coefficient $\tau_{AB}^\mu$ should
have the form
\begin{equation}
	\tau_{AB}^\mu = \tau_{AB}\delta^\mu{}_{\mu(A,B)}
\label{tau}
\end{equation}
for a certain vector index $\mu(A,B)$ uniquely determined
by the combination of spinor indices $A$ and $B$.
Namely, only one, at most, of $D$ ``momenta''%
\footnote{We denote the spacetime dimension as $D$.}
$\Pl_1,\ \cdots,\ \Pl_D$
could appear in the r.h.s.\ of the algebra for each
combination of $A$ and $B$.
Then the corresponding algebra in the continuum would be such that
\(
	\{Q_A, Q_B\}
	= 2\tau_{AB}P_{\mu(A,B)},
\)
which violates the Lorentz covariance of the algebra except
that $A$ or $B$ also has a ``vector'', or more precisely
not just a spinor, index.
We know such a basis of indices in which a supercharge has
a ``vector'' index, namely as the basis of \emph{twisted} supersymmetry
\cite{Witten:tft} \cite{Kawamoto:DK}.
In fact in the link formalism and also in the other approaches
the twisted basis for the spinor indices is adopted,
and it is the twisted version of extended supersymmetry
that the lattice formulations are constructed upon
in those approaches. Here we see that the twist
is necessary for the algebraic consistency in the link
formalism, but the reason that twist, or extended
supersymmetry itself from more general point of view,
comes naturally into the lattice formulations
is deeply connected to the doubling phenomenon
on the lattice. Namely, the doubler's d.o.f.,\ 
sometimes called ``taste'', is used as the $R$-symmetry ``flavor''
of the extended supersymmetry, and is put
on the lattice in such a way that the theory becomes
free from the mismatch of d.o.f.\ between fermions
and bosons
so that comes to meet the nonperturbative criterion
for supersymmetry which reads that the partition
function becomes unity. We will see this point again later.

At any rate suppose the coefficient $\tau_{AB}^\mu$
satisfies the condition \Ref{tau}. The condition
that \Ref{lhs-on-product} coincides with \Ref{rhs-on-product}
leads in this case to
\begin{equation}
	\al_A + \al_B = \al\muhat(A,B),\qquad
	\ar_A + \ar_B = \ar\muhat(A,B).
\label{shift-relation-gen}
\end{equation}
That these have consistent solutions for $a_A^\MR{l,r}$
is the second necessary condition for the link formalism
to work. Recalling that for an operator $\Ql_A$ which satisfies
the modified Leibniz rule \Ref{Leibniz-Q-gen} the combination
$T_{\al_A}^{-1}\Ql_A \overleftarrow{T}_{\ar_A}$
follows the usual Leibniz rule in the adjoint representation,
we may consider the corresponding algebra
\begin{equation}
	\{
		T_{\al_A}^{-1}\Ql_A \overleftarrow{T}_{\ar_A},
		T_{\al_B}^{-1}\Ql_B \overleftarrow{T}_{\ar_B}
	\}
	= 2\tau_{AB}
		T_{\al_A}^{-1} T_{\al_B}^{-1} \Pl_{\muhat(A,B)}
		\overleftarrow{T}_{\ar_A} \overleftarrow{T}_{\ar_B},
\label{alg-lat-exact}
\end{equation}
where we have assumed that the condition
\Ref{shift-relation-gen} is met. The relation
\Ref{shift-relation-gen} assures that the operator
in the r.h.s.\ also follows the usual Leibniz rule.
In fact, we find
\begin{equation}
	T_{\al_A} T_{\al_B} = T_{\al\muhat(A,B)},\qquad
	T_{\ar_A} T_{\ar_B} = T_{\ar\muhat(A,B)},
\end{equation}
and we may write $\Pl_{\muhat(A,B)}$ as, up to the lattice
constant and other constant factors,
\begin{equation}
	\begin{aligned}
		\Pl_{\muhat(A,B)}
		&=\Ad(T_{\al\muhat(A,B)}) - \Ad(T_{\ar\muhat(A,B)}), \\
		\IE
		\Pl_{\muhat(A,B)}\Vp(x)
		&= T_{\al\muhat(A,B)}\Vp(x) T_{\al\muhat(A,B)}^{-1}
		- T_{\ar\muhat(A,B)}\Vp(x) T_{\ar\muhat(A,B)}^{-1},
	\end{aligned}
\end{equation}
where we define 
$\Ad(T_{\al\muhat(A,B)})\Vp(x)=
T_{\al\muhat(A,B)}\Vp(x) T_{\al\muhat(A,B)}^{-1}$ which could be 
compared with the definition of $\ad(\Qlch_A)$ in (\ref{def-adjoint}). 
Then 
\begin{equation}
	\begin{aligned}
		T_{\al_A}^{-1} T_{\al_B}^{-1} \Pl_{\muhat(A,B)}
		\overleftarrow{T}_{\ar_A} \overleftarrow{T}_{\ar_B}
		&= - \ad(T_{\al\muhat(A,B)}^{-1} T_{\ar\muhat(A,B)}), \\
		\IE
		T_{\al_A}^{-1} T_{\al_B}^{-1} \Pl_{\muhat(A,B)} \Vp(x)
		T_{\ar_A} T_{\ar_B}
		&= -\Bigl(
		T_{\al\muhat(A,B)}^{-1} T_{\ar\muhat(A,B)}\Vp(x)
		-
		\Vp(x)T_{\al\muhat(A,B)}^{-1} T_{\ar\muhat(A,B)}
		\Bigr),
	\end{aligned}
\end{equation}
which is a normal commutator.
Notice that, as mentioned before, this redefinition of
the ``momentum'' operator is possible only for $\ar_A\neq\al_A$,
which is assured here by the requirement
that \Ref{shift-relation-gen} holds.
We have seen that all ``generators'' in the algebra
\Ref{alg-lat-exact} follow the usual Leibniz rule,
so that would give a basis for the construction of
supersymmetry on the lattice in a manner quite parallel to
that of the continuum.

\subsection{Twisted Basis and the Doubling of Chiral Fermion}


When one regularizes chiral fermions on the lattice species doublers 
of chiral fermions inevitably appear~\cite{No-Go,Anomaly}. It was shown 
that the \naive fermion 
formulation where the continuum differential operators in the Dirac 
action is \naively replaced by 
the lattice difference operator can be spin diagonalized and leads to the 
staggered fermion formulation~\cite{Kawamot-Smit} which is shown to be 
essentially equivalent~\cite{G,KMN} to 
Kogut-Susskind fermion formulation~\cite{Kogut-Susskind}. The equivalence 
of the staggered 
fermion formulation and the Dirac--\Kahler fermion has been proved 
exactly with an introduction of mild noncommutativity between 
differential forms and fields~\cite{Kanamori-Kawamoto}. 
This means that all these lattice fermion formulations are equivalent 
where the mild noncommutativity seems to play an important \role.  
Among these fermion formulations the Dirac--\Kahler fermion formulation 
has clear geometrical correspondence with respect to the fields since 
the differential form and simplex of lattice have one to one 
correspondence. 

The claim of the \emph{Dirac--\Kahler}\!\! twisting procedure is that 
these species 
doublers are not just lattice artifacts but fundamental d.o.f. 
for the regularization of fermions~\cite{Kawamoto:DK}. It is exactly these 
d.o.f. which constitute the twisted extended supersymmetry: 
$\MC{N}=2$ in two dimensions,
$\MC{N}=4$ in three dimensions, and
$\MC{N}=4$ in four dimensions.
The four dimensional Dirac--\Kahler twisting procedure coincides with 
the twisting derived by Marcus~\cite{Marcus}. 
These arguments apply in higher dimensions, too, requiring
that in $D$ dimensions, which has $2^{D/2}$ (on-shell)
doubler's degeneracy , should be treated with $\MC{N}=2^{D/2}$ 
extended supersymmetry. 
(In two dimensions $\MC{N}=2$, for example, does not correspond to the 
number of total charges and thus it is sometimes denoted as 
$\MC{N}=(2,2)$ instead.)

In the Dirac--\Kahler twisting procedure spinor suffix and flavor 
suffix constitute the scalar, vector, tensor.. nature of the super 
charges. In other words the flavor d.o.f. which are 
originally the species doublers d.o.f. is now identified 
as the extended supersymmetry d.o.f.. The corresponding 
suffix can be rotated by the internal R-symmetry generator of extended 
supersymmetry. In this way the internal d.o.f. plays the 
\role of changing spin of the fields. The mechanism how the spin and 
the internal rotation are related should be understood from lattice 
point of view. This issue is fundamentally related to the spin and 
statistics problem on the lattice. Since the Lorentz invariance is 
broken on the lattice it is natural to expect that the (anti-)commuting 
nature of fields will be modified.

Let us begin with the two dimensional case.
Here we only consider the simplest cases.
Superalgebra in the Dirac--\Kahler twisted basis on the lattice
is given as
\begin{equation}
	\{\Ql, \Ql_\mu\} = \Pl_\mu, \qquad
	\{\Qtl, \Ql_\mu\} = -\epsilon_{\mu\nu} {\Pl}'_\nu, \qquad
	\{\text{others}\} = 0,
\label{alg-twist-2d}
\end{equation}
which is the twisted version of $\MC{N}=(2,2)$ superalgebra
in two dimensions. We have put a prime on the second ``momentum'' operator
to distinguish from the first one, since there is
an ambiguity for the lattice ``momentum'' operator as explained above.
Note in each commutator
the r.h.s.\ contains only one ``momentum'' operator
for each given combination of indices, which is necessary
for the algebraic consistency as claimed in the preceding
section. The reason we specified $\MC{N}=(2,2)$
is that then the corresponding supermultiplet contains
four fermions, which has the same (on-shell) d.o.f.\ as
that of the Dirac--\Kahler\!\!/staggered fermions which
originate the doubler's d.o.f.\ on the lattice in two dimensions.

The shift variable condition \Ref{shift-relation-gen}
reads in this case
\begin{equation}
	a^\MR{l,r} + a^\MR{l,r}_\mu = a^\MR{l,r}\muhat,\qquad
	\tilde{a}^\MR{l,r} + a^\MR{l,r}_\mu 
	=	|\epsilon_{\mu\nu}|{a'}^\MR{l,r}\muhat.
\end{equation}
With the same argument given in the original link formalism
these lead to
\(
a^\MR{l,r}+a^\MR{l,r}_1+a^\MR{l,r}_2+\tilde{a}^\MR{l,r}
=(a^\MR{l,r}+{a'}^\MR{l,r})\hat{1}
=(a^\MR{l,r}+{a'}^\MR{l,r})\hat{2},
\)
which is only possible if $a^\MR{l,r} = -{a'}^\MR{l,r}$.
In our simple choices of the ``momentum'' operators,
it implies that, due to \Ref{shifts-diff},
\begin{equation}
	\begin{cases}
		\Pl_\mu = i\del_{\pm\mu}, \\
		{\Pl}'_\mu = i\del_{\mp\mu},
	\end{cases}
	\qquad\text{or}\qquad
		\Pl_\mu = {\Pl}'_\mu = i\del_\mu^\MR{s}.
\end{equation}
The former possibility was considered in the original link formulation,
whereas the latter one, although a solution for the consistency,
might not be so good from the viewpoint of the doubling issue:
it would create, if \naively used, the doubling degeneracy again.
In any of these cases, the shift conditions become
\begin{equation}
	a^\MR{l,r} + a^\MR{l,r}_\mu = a^\MR{l,r}\muhat,\qquad
	\tilde{a}^\MR{l,r} + a^\MR{l,r}_\mu 
	=	-|\epsilon_{\mu\nu}|a^\MR{l,r}\muhat,
\end{equation}
which are four conditions with one constraint, so three
remaining conditions in total, for four shift
variables. It thus seems that one shift variable could be free.
In view of the lattice structure, however, this free
parameter should not be irrational, otherwise
it would lead to uncountable number of ``dual'' lattice points,
which spoils the lattice regularization! Though still any rational
numbers are allowed, it is easy to see
we will then have unnecessary d.o.f.\ again or unnatural
lattice structure, except for
the case when this free parameter is fixed to zero
or half the lattice constant.
These choices of the free parameter were referred to
as the asymmetric and symmetric choices, respectively, in the
link formalism.

Similarly in four dimensions, we take the superalgebra
\begin{equation}
	\begin{gathered}
	\{\Ql, \Ql_\mu\} = \Pl_{+\mu}, \qquad
	\{\Ql_{\mu\nu}, \Ql_\rho\} = \delta_{\mu\nu,\rho\sigma}\Pl_{-\sigma}, \\
	\{\Qtl, \Qtl_\mu\} = \Pl_{-\mu}, \qquad
	\{\Ql_{\mu\nu}, \Qtl_\rho\} = \epsilon_{\mu\nu\rho\sigma}\Pl_{+\sigma},
	\end{gathered}
\label{alg-twist-4d}
\end{equation}
and the other commutators all vanish. This is the
Dirac--\Kahler twisted superalgebra of $\MC{N}=4$ which is required,
as explained above, from the general argument on the fermionic d.o.f.
We can show that these combinations of $\Pl_{\pm\mu}$ indeed lead
to the Leibniz rule conditions for the shift variables which have
nontrivial set of solutions \cite{DKKN:2}.

\subsection{The Claimed Inconsistency}


What is intriguing in the link formalism is the algebraic
structure based on the modified Leibniz rule for the symmetry operators.
If a suitable representation of this algebra is unambiguously obtained,
it seems at first sight that it gives a formulation of supersymmetry
on the lattice.
It turns out, however, such a representation would conflict
with the conventional
component field path integral formulation on the lattice.
This problem can be seen as the fact that,
although supertransformations of single component fields
are well-defined, supertransformations of products of
fields becomes sensitive to the order of the fields
in the products. If such an order is uniquely determined,
it is nothing harmful. However, we have no
criteria to introduce such an order on the conventional
lattice, thus we have a serious difficulty that supertransformations
are not totally defined in a unique and consistent manner
as transformations of path integral variables.
In fact, this difficulty is claimed as an inconsistency of the link
formalism in \cite{Dutch}. The criticism is two-folded: one is for
the non-gauge theories \cite{DKKN:1}, the other
is, also investigated in a similar attitude in
\cite{Damgaard-Matsuura},
for the case of gauge theories \cite{DKKN:2, DKKN:3},
and both are summarized as that the supercharges in the link
formalism add nontrivial link structure on component fields
changing the original link nature of the fields
in an ordering sensitive way.

Let us see these arguments more explicitly.
In the link formalism, scalar fields $\phi(x)$ defined on
sites of the lattice are naturally assumed to be commutative
\begin{equation}
	\phi_1(x)\phi_2(x) = \phi_2(x)\phi_1(x).
\end{equation}
Applying the supertransformation on the both sides of this equation,
we have, from the left hand side, that
\begin{equation}
	\Ql_A\bigl(\phi_1(x)\phi_2(x)\bigr)
	= \psi_{1A}(x)\phi_2(x+\ar_A) + \phi_1(x+\al_A)\psi_{2A}(x),
\label{Q-phi12}
\end{equation}
where $\psi_{1,2\,A}(x):=\Ql_A\phi_{1,2}(x)$, and from the right,
\begin{equation}
	\Ql_A\bigl(\phi_2(x)\phi_1(x)\bigr)
	= \psi_{2A}(x)\phi_1(x+\ar_A) + \phi_2(x+\al_A)\psi_{1A}(x).
\label{Q-phi21}
\end{equation}
These two equations must be the same as they are the transformations
of one and the same quantity; otherwise the supertransformations
on products of fields aren't uniquely defined.
But actually these two conflicts with each other
if fermions $\psi_{1,2\,A}$ are also assumed to be
simple (anti-)commuting objects: the term containing $\psi_{1\,A}(x)$
in the first equation has the factor $\phi_{2}(x+\ar_A)$,
whereas in the second has $\phi_{2}(x+\al_A)$, and they are
different unless $\al_A = \ar_A$,
which, however, wouldn't lead to the consistent solution
for the shift variable conditions as already explained in the previous
sections. The discrepancy between these two equations cannot be
expressed as a total difference, so that it gives an essential
obstacle for the invariance of any possible action.
It causes similar difficulties also in the gauge theory actions.

In the following chapters, we will propose
a possible solution to the above mentioned
first criticism for the non-gauge theories by introducing 
the following mild noncommutativity \cite{DKKN:3}:
\begin{equation}
\Vp_A(x)\Vp_B(y) = (-1)^{|\Vp_A||\Vp_B|}\Vp_B(y+a_A) \Vp_A(x-a_B),
\label{NC-relation}
\end{equation}
where $\Vp_A(x)$ and $\Vp_B(y)$ carry a shift $a_A$ and $a_B$, 
respectively. In fact we can easily confirm that the expressions of 
(\ref{Q-phi12}) and (\ref{Q-phi21}) coincide if we identify that 
$\psi_{1,2A}$ carry a shift $\al_A-\ar_A$ while $\phi_{1,2}$ carry 
no shift and they satisfy the noncommutative 
relation (\ref{NC-relation}). 
The key point is to treat each field as a noncommutative object, 
or an object with nontrivial statistics, to uniquely define the
ordering which is necessary to avoid the conflict.

If we introduce the noncommutative nature for the fields as in 
(\ref{NC-relation}), the formulation of field theory should be 
modified from the conventional definition in such a way that 
any algebraic manipulation of fields and operators should be 
compatible with the new deformed supersymmetry.  
In the following we show that it is possible to define 
a new lattice field theory which has the exact deformed 
supersymmetry with Hopf algebraic nature. 
Addressing the similar questions to the gauge theories is out 
of the scope of this paper.


%% file: hopf.tex
\section{Hopf Algebraic Structure of the Lattice Superalgebra}
\label{hopf}

In this section, we investigate
the ``lattice superalgebra'' from a yet different
algebraic viewpoint, namely in terms of Hopf algebra.
As has been developed in recent years,
extending the notion of the symmetry in a field theory
to the Hopf algebraic one brings us still 
useful frameworks in some specific cases
especially in noncommutative theories
\cite{Oeckl, twisted-sym, Asakawa, Sasai-Sasakura, Riccardi-Szabo}.
A slightly different applications are found in
\cite{Kanamori-Kawamoto}.
In the current interest, the superalgebra on the lattice is
understood as a deformed algebra on the lattice and
forming a Hopf algebra. This identification assures us
of the mathematical consistency of the deformed algebra.
Using the general scheme
called braided quantum field theory \cite{Oeckl, Sasai-Sasakura},
we will show that the field theory whose symmetry is prescribed
by the deformed algebra can be constructed at least perturbatively.
The deformed symmetry leads to the corresponding Ward--Takahashi
identities on the lattice, which may serve as a good physical
interpretation of the deformed symmetry itself.

Appendix~\ref{app-hopf}
is devoted to a brief mathematical basis on Hopf algebra
and summarizing our notation and terminology.

\subsection{Lattice Superalgebra as a Hopf Algebra}

We begin with the lattice superalgebra \Ref{alg-lat-gen},
or those in the twisted basis \Ref{alg-twist-2d},
\Ref{alg-twist-4d}. Here we treat these as abstract
Lie superalgebra and denote as \MC{A}, so that
$\Pl_\mu,\ \Ql_A\in \MC{A}$. We then introduce
the space of fields on the lattice as $X=\Xe\oplus\Xo$,
where $\Xe$ consists of all bosonic fields and
$\Xo$ of all fermionic fields. We need
a multiplication/product of fields to construct
a field theory, which is in general noncommutative.
We assume here this multiplication is associative
for our current application.
Thus the space $X$ is supposed to be an associative
graded algebra. However, as a quantum field theory,
products of fields, i.e.\ composite fields,
could be clearly distinguished from the single
fields, i.e.\ elementary fields, because
the elementary fields are the variables
of path integral (if any defined),
or the ones obeying the canonical (anti-)commutation
relations, whose behaviour is clearly different
from that of the composite fields. We thus denote by $X$
the elementary fields and extend it to the formal space of
all tensor products of the elementary fields to include any
composite fields:
\begin{equation}
	\hat{X}:= \bigoplus_{n=0}^\infty X^n,\qquad
	X^0 := \Xe^0\oplus\Xo^0,\quad
	X^n := \underbrace{X\ot\cdots\ot X}_n,
\label{space-fields}
\end{equation}
where $\Xe^0$ and $\Xo^0$ are the space of
bosonic and fermionic constant functions, respectively.
Multiplications/products of fields are naturally
defined in $\hat{X}$ as
$m(\Vp\ot\Vp') = \Vp\cdot\Vp'\in\Xh
\quad(\Vp,\ \Vp'\in\Xh)$.

We now consider general \emph{action} (see Appendix A) 
of $\MC{A}$
on the space of fields $\hat{X}$.
We denote the action of an operator
$a\in\MC{A}$ as $a\act$.
With the successive actions, we are naturally led
to the notion of an (associative) universal enveloping algebra
$\MC{U}(\MC{A})$ of $\MC{A}$, as in
$(a\cdot b)\act := a\act\circ b\act := a\act(b\act)$,
with
$a,\ b\in\MC{A}$ and $a\cdot b\in\MC{U}(\MC{A})$.
We also introduce the identity operator $\vid$ as a unit
element of the universal enveloping algebra. We may
define the unit map by $\eta(c):=c\vid,\ c\in\MBB{C}$.

Even on the lattice,
actions or representations of the operators $\Ql_A$
and $\Pl_\mu$ on elementary fields would be well-defined
with no difficulty. We denote these formally as
\begin{equation}
	\Ql_A\act\Vp(x) = (\Ql_A\Vp)(x),\qquad
	\Pl_\mu\act\Vp(x) = (\Pl_\mu\Vp)(x),\qquad
	\Vp\in X.
\end{equation}
Explicit form of $\Ql_A\Vp$ depends on the model we take.
An example is listed in the appendix~\ref{app-WZ}.
As for the expression $\Pl_\mu\Vp$, we could essentially take
some of the difference operators as in \Ref{diffops},
but it turns out that lattice momentum operator $\Pl_\mu$ should 
carry a nontrivial grading structure, which is required from 
the Hopf algebraic consistency. We will see this point in 
the following subsection.

Actions on the trivial/constant fields
are also easily defined as in
\begin{equation}
	\Ql_A\act f = 0,\qquad
	\Pl_\mu\act f = 0,\qquad
	f\in \Xe^0.
\end{equation}
As a matter of convention, we write these equations in terms of
a map $\counit$ called \emph{counit} as in
\begin{equation}
	\begin{gathered}
		\Ql_A\act f = \counit(\Ql_A) f = 0,\IE \counit(\Ql_A) = 0, \\
		\Pl_\mu\act f = \counit(\Pl_\mu) f = 0,\IE \counit(\Pl_\mu) = 0.
	\end{gathered}
\label{counit-Q-P}
\end{equation}

The essential nontriviality comes in
the actions of the operators on composite fields,
i.e.\ product of the elementary fields, due to the
failure of the usual Leibniz rule.
The link formalism manages this difficulty
with the introduction of appropriate deformation or modification of 
Leibniz rules
when the operators act on the composite fields.
Mathematically, this is understood as equipping the universal enveloping
algebra
$\U(\A)$ with an additional structure,
the \emph{coproduct}/\emph{comultiplication}, denoted
by $\coproduct$. 
To be specific, consider the actions of $\Ql_A$ and $\Pl_\mu$
on a product of two elementary fields
$\Vp_1(x),\ \Vp_2(x)\in X$. Introducing the modified Leibniz rule
\Ref{Leibniz-Q-gen} and \Ref{rhs-on-product} is equivalent to
defining these actions to be
\begin{equation}
	\begin{aligned}
		\Ql_A\act\bigl(\Vp_1(x)\cdot\Vp_2(x)\bigr)
		&:= m\Bigl(\coproduct(\Ql_A)\act\bigl(\Vp_1(x)\ot\Vp_2(x)\bigr)\Bigr), \\
		\Pl_\mu\act\bigl(\Vp_1(x)\cdot\Vp_2(x)\bigr)
		&:= m\Bigl(\coproduct(\Pl_\mu)\act\bigl(\Vp_1(x)\ot\Vp_2(x)\bigr)\Bigr),
	\end{aligned}
\end{equation}
together with the coproducts
\begin{equation}
	\begin{aligned}
		\coproduct(\Ql_A) 
		&= \Ql_A\ot T_{\ar_A} + \statfn\cdot T_{\al_A}\ot\Ql_A, \\
		\coproduct(\Pl_\mu) 
		&= \Pl_\mu\ot T_{\ar\muhat} + T_{\al\muhat}\ot\Pl_\mu,
	\end{aligned}
\label{coproduct-Q-P}
\end{equation}
where $\fn$ is the fermion number operator
with which $\statfn$ takes care of the statistics factors,
and the shift operator $T_{b}$,
which is also assumed to belong to $\U(\A)$, acts as
\begin{equation}
	T_b\act\Vp(x) := \Vp(x+b).
\end{equation}
For these operators we set
\begin{equation}
	\counit(T_b) = 1,\qquad
	\coproduct(T_b) = T_b\ot T_b,
\label{counit-coproduct-T}
\end{equation}
and
\begin{equation}
	\counit\bigl(\statfn\bigr) = 1,\qquad
	\coproduct\bigl(\statfn\bigr) = \statfn\ot\statfn.
\label{counit-coproduct-F}
\end{equation}
Note that these definitions are natural, since
the counit essentially prescribes the action on a constant,
whereas the coproduct defines the action on a product.
We also list, though obvious, the action of the identity operator
$\vid$ on $\hat{X}$.
It must be, by definition, such that $\vid\act\Vp = \Vp,\ \Vp\in X$.
On a constant field, $f = \vid\act f = \counit(\vid) f,\ f\in X^0$, so that
\begin{equation}
	\counit(\vid) = 1.
\label{hom-counit-2}
\end{equation}
On a product of elementary fields,
\(
	\Vp_1\cdot\Vp_2 = \vid\act(\Vp_1\cdot \Vp_2)
	= m\Bigl(\coproduct(\vid)\act(\Vp_1\ot\Vp_2)\Bigr),
\)
so that
\begin{equation}
	\coproduct(\vid) = \vid\ot\vid.
\label{hom-coproduct-2}
\end{equation}

Counit $\counit$ and coproduct $\coproduct$ has to satisfy some consistency
conditions. First, we note that any single elementary field
$\Vp$ might be expressed as a product of unity and itself;
$\Vp =	m(1\ot\Vp)=m(\Vp\ot 1)$. The action should be uniquely determined
regardless of this reinterpretation of the degrees of product.
More specifically, this requires that
\begin{equation}
	\begin{aligned}
		(\Ql_A\Vp)(x)=\Ql_A\act\Vp(x)
		&= \Ql_A\act m\bigl(1\ot\Vp(x)\bigr)
		= m\Bigl(\coproduct(\Ql_A)\act \bigl(1\ot\Vp(x)\bigr)\Bigr) \\
		&= m\Bigl( \bigl(\Ql_A\act 1\bigr)\ot\bigl(T_{\ar_A}\act\Vp(x)\bigr)
		+ \bigl(T_{\al_A}\act 1\bigr)\ot(\Ql_A\act\Vp(x)\bigr)
		\Bigr) \\
		&= m\Bigl( \bigl(\counit(\Ql_A)1\bigr)\ot\bigl(T_{\ar_A}\act\Vp(x)\bigr)
		+ \bigl(\counit(T_{\al_A})1\bigr)\ot(\Ql_A\act\Vp(x)\bigr)
		\Bigr) \\
		&= m\bigl( 1\ot (\Ql_A\Vp)(x) \bigr) = (\Ql_A\Vp)(x),
	\end{aligned}
\end{equation}
which is consistently realized. The other consistency also holds:
\begin{equation}
	\begin{aligned}
		(\Ql_A\Vp)(x)=\Ql_A\act\Vp(x)
		&= \Ql_A\act m\bigl(\Vp(x)\ot 1\bigr)
		= m\Bigl(\coproduct(\Ql_A)\act \bigl(\Vp(x)\ot 1\bigr)\Bigr) \\
		&= m\Bigl( \bigl(\Ql_A\act\Vp(x)\bigr)\ot\bigl(T_{\ar_A}\act 1\bigr)
		+ \bigl(\statfn\cdot T_{\al_A}\act\Vp(x)\bigr)\ot(\Ql_A\act 1\bigr)
		\Bigr) \\
		&= m\Bigl(
			\bigl(\Ql_A\act\Vp(x)\bigr)\ot\bigl(\counit(T_{\ar_A})1\bigr)
			+
			\bigl((-1)^\stat{\Vp}T_{\al_A}\act\Vp(x)\bigr)
			\ot\bigl(\counit(\Ql_A)1\bigr)
		\Bigr) \\
		&= m\bigl( (\Ql_A\Vp)(x)\ot 1 \bigr) = (\Ql_A\Vp)(x).
	\end{aligned}
\end{equation}
Similar results hold for $\Pl_\mu$.
As for $T_b$,
\begin{equation}
	\begin{aligned}
		\Vp(x+b) = T_b\act\Vp(x)
		&= T_b\act m\bigl(1\ot\Vp(x)\bigr)
		= m\Bigl( \coproduct(T_b)\act\bigl(1\ot\Vp(x)\bigr)\Bigr) \\
		&= m\Bigl( \bigl(T_b\act 1\bigr) \ot \bigl( T_b\act\Vp(x) \bigr)\Bigr) \\
		&= m\Bigl( \bigl(\counit(T_b) 1 \bigr)\ot \Vp(x+b) \Bigr)
		= m\bigl(1\ot\Vp(x+b)\bigr) = \Vp(x+b),
	\end{aligned}
\end{equation}
which is again consistent. These result show that the definitions
of counit and coproduct in \Ref{counit-Q-P}, \Ref{coproduct-Q-P},
\Ref{counit-coproduct-T} are compatible to the trivial unital structure
of the algebra $\hat{X}$.
Second consistency condition is so-called the \emph{coassociativity}.
Since the multiplication
on $\hat{X}$ is associative, actions on products
of three elementary fields should respect this associativity.
This requires the coassociativity for the coproduct.
It also means the action on products of three elementary
fields is defined in a natural way as in
\begin{equation}
	\begin{aligned}
		& m\circ(m\ot\id)\circ(\coproduct\ot\id)\circ\coproduct(\Ql_A)\act
		\Bigl(\bigl(\Vp_1(x)\ot\Vp_2(x)\bigr)\ot\Vp_3(x)\Bigr) \\
		&=
		\Ql_A\act\Bigl(\bigl(\Vp_1(x)\cdot\Vp_2(x)\bigr)\cdot\Vp_3(x)\Bigr) \\
		&=
		\Ql_A\act\Bigl(\Vp_1(x)\cdot\Vp_2(x)\cdot\Vp_3(x)\Bigr) \\
		&=
		\Ql_A\act\Bigl(\Vp_1(x)\cdot\bigl(\Vp_2(x)\cdot\Vp_3(x)\bigr)\Bigr) \\
		&=
		m\circ(\id\ot m)\circ(\id\ot\coproduct)\circ\coproduct(\Ql_A)\act
		\Bigl(\Vp_1(x)\ot\bigl(\Vp_2(x)\ot\Vp_3(x)\bigr)\Bigr).
	\end{aligned}
\end{equation}
Since the product $m$ is associative,
\begin{equation}
	m\circ(m\ot\id) = m\circ(\id\ot m),
\end{equation}
it requires that
\begin{equation}
	(\coproduct\ot\id)\circ\coproduct(\Ql_A)	=
	(\id\ot\coproduct)\circ\coproduct(\Ql_A).
\label{coassociativity}
\end{equation}
The same condition should follow for $\Pl_\mu$ and $T_b$.
These conditions are indeed satisfied for the coproducts
in the present case.
Using \Ref{coproduct-Q-P}, we compute%
\footnote{Here we use the relation
$\coproduct\bigl(\statfn\cdot T_b\bigr)
=\coproduct\bigl(\statfn\bigr)\cdot\coproduct(T_b)$,
which will be explained shortly.}
\begin{equation}
	\begin{aligned}
		&(\coproduct\ot\id)\circ\coproduct(\Ql_A)
		= (\coproduct\ot\id) 
			\bigl(\Ql_A\ot T_{\ar_A} + \statfn\cdot T_{\al_A}\ot\Ql_A\bigr) \\
		&=	
			\bigl(\Ql_A\ot T_{\ar_A} + \statfn\cdot T_{\al_A}\ot\Ql_A\bigr)
			\ot T_{\ar_A}
			+
			\bigl(\statfn\cdot T_{\al_A}\ot\statfn\cdot T_{\al_A}\bigr)
			\ot \Ql_A \\
		&=
			\Ql_A\ot T_{\ar_A}\ot T_{\ar_A} 
			+
			\statfn\cdot T_{\al_A}\ot\Ql_A \ot T_{\ar_A}
			+
			\statfn\cdot T_{\al_A}\ot\statfn\cdot T_{\al_A}\ot \Ql_A,
	\end{aligned}
\label{coassociativity-1}
\end{equation}
and
\begin{equation}
	\begin{aligned}
		&(\id\ot\coproduct)\circ\coproduct(\Ql_A)
		= (\id\ot\coproduct) 
			\bigl(\Ql_A\ot T_{\ar_A} + \statfn\cdot T_{\al_A}\ot\Ql_A\bigr) \\
		&=
			\Ql_A\ot\bigl(T_{\ar_A}\ot T_{\ar_A}\bigr)
			+
			\statfn\cdot T_{\al_A}\ot
			\bigl(\Ql_A\ot T_{\ar_A} + 
			\statfn\cdot T_{\al_A}\ot\Ql_A\bigr) \\
		&=
			\Ql_A\ot T_{\ar_A}\ot T_{\ar_A} 
			+
			\statfn\cdot T_{\al_A}\ot\Ql_A \ot T_{\ar_A}
			+
			\statfn\cdot T_{\al_A}\ot\statfn\cdot T_{\al_A}\ot \Ql_A,
	\end{aligned}
\end{equation}
which shows that \Ref{coassociativity} holds for $\Ql_A$.
We have thus found unambiguously that
\begin{equation}
	\begin{aligned}
		&\Ql_A\act\bigl(\Vp_1(x)\cdot\Vp_2(x)\cdot\Vp_3(x)\bigr) \\
		&= (\Ql_A\Vp_1)(x) \cdot \Vp_2(x+\ar_A) \cdot \Vp_3(x+\ar_A) \\
		&\phantom{=}+ (-1)^\stat{\Vp_1}
		\Vp_1(x+\al_A) \cdot (\Ql_A\Vp_2)(x) \cdot \Vp_3(x+\ar_A) \\
		&\phantom{=}+ (-1)^{\stat{\Vp_1}+\stat{\Vp_2}}
		\Vp_1(x+\al_A) \cdot \Vp_2(x+\al_A) \cdot (\Ql_A\Vp_3)(x).
	\end{aligned}
\end{equation}
The same is true for $\Pl_\mu$:
\begin{equation}
	\begin{aligned}
		&\Pl_\mu\act\bigl(\Vp_1(x)\cdot\Vp_2(x)\cdot\Vp_3(x)\bigr) \\
		&= (\Pl_\mu\Vp_1)(x) \cdot \Vp_2(x+\ar\muhat) \cdot \Vp_3(x+\ar\muhat) \\
		&\phantom{=}+
			\Vp_1(x+\al\muhat) \cdot (\Pl_\mu\Vp_2)(x) \cdot \Vp_3(x+\ar\muhat) \\
		&\phantom{=}+
			\Vp_1(x+\al\muhat) \cdot \Vp_2(x+\al\muhat) \cdot (\Pl_\mu\Vp_3)(x).
	\end{aligned}
\end{equation}
Similarly, $T_b$ satisfies the
coassociativity, for
\begin{equation}
	\begin{aligned}
		(\coproduct\ot\id)\circ\coproduct(T_b)
		&=(\coproduct\ot\id)(T_b\ot T_b)
		=(T_b\ot T_b)\ot T_b \\
		&=T_b\ot(T_b\ot T_b)
		=(\id\ot\coproduct)(T_b\ot T_b)
		=(\id\ot\coproduct)\coproduct(T_b),
	\end{aligned}
\end{equation}
so that
\begin{equation}
	T_b\act\bigl(\Vp_1(x)\cdot\Vp_2(x)\cdot\Vp_3(x)\bigr)
	=\Vp_1(x+b)\cdot\Vp_2(x+b)\cdot\Vp_3(x+b).
\end{equation}
The result for $\statfn$ would be obvious.

Now that we have shown that our operators of the prime interest,
$\Ql_A,\ \Pl_\mu,\ T_b$ are well defined concerning to the actions
on the elementary fields, constants, and products of two or three
elementary fields, we find that any other actions are also
consistently defined (needless to say actions as well as
the maps introduced above are all linear). In particular the actions on
any numbers of elementary fields can be computed inductively
using the coassociativity.
We need further the actions of products of operators.
As we started above, the product of operators
is defined as an operator of the successive applications
of each operator in the product. On the elementary fields,
it is easily understood, because it is nothing but the
definition. On the trivial (i.e.\ constant)
fields, this implies a consistency on the counit map, as
\begin{equation}
	\counit(a\cdot b) f = (a\cdot b)\act f
	= a\act\circ b\act f
	= \counit(a)\counit(b)f,
\label{hom-coproduct}
\end{equation}
i.e.\ 
\begin{equation}
	\counit(a\cdot b) = \counit(a)\counit(b).
\label{hom-counit-1}
\end{equation}
Similarly, the product of operators should act on a product
of elementary fields with the successive operations with
\begin{equation}
	\begin{aligned}
		m\Bigl(\coproduct(a\cdot b) \act (\Vp_1\ot\Vp_2)\Bigr)
		&= (a\cdot b)\act (\Vp_1\cdot\Vp_2)
		= a\act\circ b\act (\Vp_1\cdot\Vp_2)
		= a\act m\Bigl( \coproduct(b)\act(\Vp_1\ot\Vp_2) \Bigr) \\
		&= m\Bigl( \coproduct(a)\act\coproduct(b)\act(\Vp_1\ot\Vp_2) \Bigr)
			= m\Bigl( \bigl(\coproduct(a)\cdot\coproduct(b))\act(\Vp_1\ot\Vp_2) \Bigr),
	\end{aligned}
\end{equation}
so implies
\begin{equation}
	\coproduct(a\cdot b)
	=
	\coproduct(a)\cdot\coproduct(b).
\label{hom-coproduct-1}
\end{equation}
As an example, we compute
\begin{equation}
	\begin{aligned}
		\coproduct(\Ql_A\cdot\Ql_B)
		&= \coproduct(\Ql_A)\cdot\coproduct(\Ql_B)\\
		&= (\Ql_A\ot T_{\ar_A} + \statfn\cdot T_{\al_A}\ot\Ql_A)
		\cdot
				(\Ql_B\ot T_{\ar_B} + \statfn\cdot T_{\al_B}\ot\Ql_B) \\
		&= \Ql_A\cdot\Ql_B\ot T_{\ar_A}\cdot T_{\ar_B}
			+ \Ql_A\cdot\statfn\cdot T_{\al_B}\ot T_{\ar_A}\cdot\Ql_B \\
		&\ + \statfn\cdot T_{\al_A}\cdot\Ql_B\ot \Ql_A\cdot T_{\ar_B}
			+ \statfn\cdot T_{\al_A}\cdot\statfn\cdot T_{\al_B}
				\ot \Ql_A\cdot\Ql_B.
	\end{aligned}
\label{coprod-QQ}
\end{equation}
Other simple examples are
\begin{equation}
	\begin{aligned}
		\coproduct(\Ql_A\cdot\Pl_\mu)
		&= \coproduct(\Ql_A)\cdot\coproduct(\Pl_\mu)\\
		&= (\Ql_A\ot T_{\ar_A} + \statfn\cdot T_{\al_A}\ot\Ql_A)
		\cdot
				(\Pl_\mu\ot T_{\ar\muhat} + T_{\al\muhat}\ot\Pl_\mu) \\
		&= \Ql_A\cdot\Pl_\mu\ot T_{\ar_A}\cdot T_{\ar\muhat}
			+ \Ql_A\cdot T_{\al\muhat}\ot T_{\ar_A}\cdot\Pl_\mu \\
		&\ + \statfn\cdot T_{\al_A}\cdot\Pl_\mu\ot \Ql_A\cdot T_{\ar\muhat}
			+ \statfn\cdot T_{\al_A}\cdot T_{\al\muhat}\ot \Ql_A\cdot\Pl_\mu,
	\end{aligned}
\end{equation}
and
\begin{gather}
	\begin{aligned}
		\coproduct(T_b\cdot\Ql_A)
		&= \coproduct(T_b)\cdot\coproduct(\Ql_A)
		= (T_b\ot T_b)
		\cdot(\Ql_A\ot T_{\ar_A}+\statfn\cdot T_{\al_A}\ot\Ql_A) \\
		&= T_b\cdot\Ql_A\ot T_b\cdot T_{\ar_A}
		+ T_b\cdot\statfn\cdot T_{\al_A}\ot T_b\cdot\Ql_A,
	\end{aligned}
\label{coprod-TQ}
	\\
	\coproduct(T_b\cdot T_c)
	= \coproduct(T_b)\cdot\coproduct(T_c)
	= (T_b\ot T_b)\cdot(T_c\ot T_c)
	= T_b\cdot T_c \ot T_b\cdot T_c.
\label{coprod-TT}
\end{gather}

Let us recall now the superalgebra \Ref{alg-lat-gen}, and introduce
a natural algebra with respect to $T_b$ as in
\begin{equation}
	\begin{gathered}
		\{\Ql_A, \Ql_B\}
		= 2\tau_{AB}^\mu \Pl_\mu,\\
		[\Ql_A, \Pl_\mu]
		= [\Pl_\mu, \Pl_\nu] = 0, \\
		[\Ql_A, T_b] = [\Pl_\mu, T_b] = [T_b, T_c] = 0.
	\end{gathered}
\label{algebra-Q-P-T}
\end{equation}
The last relations are in a way obvious, and states that
\begin{equation}
	\Ql_A\Vp(x+b) = T_b (\Ql_A\Vp) (x),\qquad
	T_b\Vp(x+c) =  T_c\Vp(x+b) = \Vp(x+b+c) = T_{b+c}\Vp(x),
\end{equation}
and similar for $\Pl_\mu$. We also list here the obvious algebra
for $\statfn$:
\begin{equation}
	\{\Ql_A,\statfn\}
	= [\Pl_A,\statfn]
	= [T_b,\statfn]
	= 0,\qquad
	\statfn\cdot\statfn
	= \vid.
\label{algebra-F}
\end{equation}
From these relations and using \Ref{coprod-QQ}, we find that
\begin{equation}
	\coproduct\bigl(\{\Ql_A, \Ql_B\}\bigr)
	=  \{\Ql_A, \Ql_B\}\ot T_{\ar_A}\cdot T_{\ar_B}
	+ T_{\al_A}\cdot T_{\al_B}\ot \{\Ql_A, \Ql_B\},
\end{equation}
reproducing the general result we found in \Ref{lhs-on-product}.
Just as an additional explicit check of the consistency, we compute
the action of the product $\Ql_A\cdot\Ql_B$ on the product of three
fields $\Vp_1\cdot\Vp_2\cdot\Vp_3$, which is given by the object
\begin{equation}
	\begin{aligned}
		(\coproduct\ot\id)\circ\coproduct(\Ql_A\cdot\Ql_B)
		&=(\coproduct\ot\id)
			\circ\bigl(\coproduct(\Ql_A)\cdot\coproduct(\Ql_B)\bigr) \\
		&= \coproduct(\Ql_A\cdot\Ql_B)\ot T_{\ar_A}\cdot T_{\ar_B}
			+ \coproduct(\Ql_A\cdot\statfn\cdot T_{\al_B})
				\ot T_{\ar_A}\cdot\Ql_B \\
		&\	- \coproduct(\Ql_B\cdot\statfn\cdot T_{\al_A})
			\ot\Ql_A\cdot T_{\ar_B}
			+ \coproduct(T_{\al_A}\cdot T_{\al_B})\ot\Ql_A\cdot\Ql_B
	\end{aligned}
\end{equation}
and then this can be computed using \Ref{coprod-QQ}, \Ref{coprod-TQ} and \Ref{coprod-TT}.
This of course leads to
\begin{equation}
	\begin{aligned}
		(\coproduct\ot\id)\circ\coproduct\bigl(\{\Ql_A, \Ql_B\}\bigr)
		&=(\coproduct\ot\id)
			\circ\bigl(\{\coproduct(\Ql_A), \coproduct(\Ql_B)\}\bigr) \\
		&= \coproduct\bigl(\{\Ql_A, \Ql_B\}\bigr)
			\ot T_{\ar_A}\cdot T_{\ar_B}
			+ \coproduct(T_{\al_A}\cdot T_{\al_B})\ot\{\Ql_A, \Ql_B\} \\
		&= \{\Ql_A, \Ql_B\}\ot T_{\ar_A}\cdot T_{\ar_B}\ot T_{\ar_A}\cdot T_{\ar_B} \\
		&\ + T_{\al_A}\cdot T_{\al_B}\ot \{\Ql_A, \Ql_B\}\ot T_{\ar_A}\cdot T_{\ar_B} \\
		&\ + T_{\al_A}\cdot T_{\al_B}\ot T_{\al_A}\cdot T_{\al_B}\ot\{\Ql_A, \Ql_B\}.
	\end{aligned}
\end{equation}

Equations \Ref{hom-counit-2}, \Ref{hom-counit-1} and
\Ref{hom-coproduct-2}, \Ref{hom-coproduct-1}
naturally require that the counit and coproduct, respectively, are
both consistent to the structure of the algebra $\U(\A)$,
i.e.\ both algebra maps (algebra homomorphisms).%
\footnote{These conditions are the same as imposing
the product $m$ and unit $\eta$ should be
coalgebra maps.}
With these properties, we can compute the actions
of any operators on any fields in a consistent manner.
Mathematically, all these features assures that our lattice superalgebra
actually forms a \emph{bialgebra}.

Notice that our bialgebra is a mixture of both algebra-like elements,
like $\Ql_A$ or $\Pl_\mu$, and group-like elements,
like $T_b$. The latter have their inverse, $T_b^{-1}$.
The former would also have a sort of inverse,
$-\Ql_A$ and $-\Pl_\mu$, implying the \naive
connection between group and algebra. In fact,
we need one more ingredient, namely an \emph{antipode},
to claim that the DKKN lattice superalgebra is a Hopf algebra,
and it is essentially a map to give the ``inverse''
element for each operator.
It would be introduced as a linear map
such that satisfies the identity
\begin{equation}
	\opprod\circ(S\ot\id)\circ\coproduct 
	= \opprod\circ(\id\ot S)\circ\coproduct
	= \eta\circ\counit,
\label{antipode}
\end{equation}
where we have used the notation $\opprod(a\ot b) = a\opprod b$
for the product of operators.
We define it explicitly, on the single
operators, as
\begin{equation}
	\begin{gathered}
		S(\Ql_A) = 
			-T_{\al_A}^{-1}\cdot\statfn\cdot\Ql_A\cdot T_{\ar_A}^{-1},\qquad
		S(\Pl_\mu) = -T_{\al\muhat}^{-1}\cdot\Pl_\mu\cdot T_{\ar\muhat}^{-1},\\
		S(T_b) = T_b^{-1},\qquad
		S\bigl(\statfn\bigr) = (-1)^{-\fn} = \statfn,
	\end{gathered}
\label{antipode-Q-P-T}
\end{equation}
and extend it so that it becomes linear and \emph{anti-algebraic}
in the sense
$S(a\cdot b) = S(b)\cdot S(a),\ S(\vid) = \vid,\ (a,\ b\in\U(\A))$.
In fact it is shown that the anti-algebraic nature automatically
follows if the identity \Ref{antipode} holds for the antipode.
Here we just see what this identity implies in our superalgebra,
without digging into the detail.
Applying the first two terms of \Ref{antipode} on $\Ql_A$, we find%
\footnote{We use here
$S\bigl(\statfn\cdot T_b\bigr) = S(T_b)\cdot S\bigl(\statfn\bigr)$
as explicitly shown as \Ref{antipode-gg}.}
\begin{equation}
	\begin{aligned}
		\opprod\circ(S\ot\id)\circ\coproduct(\Ql_A)
		&= \opprod\circ(S\ot\id)\bigl(
			\Ql_A\ot T_{\ar_A} + \statfn\cdot T_{\al_A}\ot\Ql_A
		\bigr) \\
		&= \opprod\bigl(
			-T_{\al_A}^{-1}\cdot\statfn\cdot\Ql_A\cdot T_{\ar_A}^{-1}
				\ot T_{\ar_A}
		+ T_{\al_A}^{-1}\cdot\statfn\ot\Ql_A \bigr) \\
		&= -T_{\al_A}^{-1}\cdot\statfn\cdot\Ql_A 
		+ T_{\al_A}^{-1}\cdot\statfn\cdot\Ql_A
		= 0,
	\end{aligned}
\end{equation}
and
\begin{equation}
	\begin{aligned}
		\opprod\circ(\id\ot S)\circ\coproduct(\Ql_A)
		&= \opprod\circ(\id\ot S)\bigl(
			\Ql_A\ot T_{\ar_A} + \statfn\cdot T_{\al_A}\ot\Ql_A
		\bigr) \\
		&= \opprod\bigl( \Ql_A\ot T_{\ar_A}^{-1} 
		- \statfn\cdot T_{\al_A}
			\ot T_{\al_A}^{-1}\cdot\statfn\cdot\Ql_A\cdot T_{\ar_A}^{-1}
		\bigr) \\
		&= \Ql_A\cdot T_{\ar_A}^{-1} - \Ql_A\cdot T_{\ar_A}^{-1}
		= 0,
	\end{aligned}
\end{equation}
while the last terms gives
\begin{equation}
	\eta\circ\counit(\Ql_A) = 0.
\end{equation}
Thus the identity \Ref{antipode} holds for the operator $\Ql_A$
with the definition \Ref{antipode-Q-P-T}.
Similar calculations show that $\Pl_\mu$ also obeys the identity.
As for $T_b$, we compute
\begin{equation}
	\opprod\circ(S\ot\id)\circ\coproduct(T_b)
	= \opprod\circ(S\ot\id)(T_b\ot T_b)
	= \opprod(T_b^{-1}\ot T_b)
	= \vid,
\end{equation}
and
\begin{equation}
	\opprod\circ(\id\ot S)\circ\coproduct(T_b)
	= \opprod\circ(\id\ot S)(T_b\ot T_b)
	= \opprod(T_b\ot T_b^{-1})
	= \vid,
\end{equation}
whereas
\begin{equation}
	\eta\circ\counit(T_b)
	= \vid,
\end{equation}
again showing the consistency.
Let us calculate the antipodes of products of operators
with the use of the identity \Ref{antipode}. Applying
the l.h.s.\ of the identity on $T_b\cdot T_c$,
\begin{equation}
	\begin{aligned}
		\opprod\circ(S\ot\id)\circ\coproduct(T_b\cdot T_c)
		&= \opprod\circ(S\ot\id)(T_b\cdot T_c\ot T_b\cdot T_c)
		= \opprod\Bigl(S(T_b\cdot T_c)\ot T_b\cdot T_c\Bigr) \\
		&= S(T_b\cdot T_c)\cdot (T_b\cdot T_c),
	\end{aligned}
\end{equation}
and the r.h.s.\ 
\begin{equation}
	\eta\circ\counit(T_b\cdot T_c) = \eta\bigl(\counit(T_b)\counit(T_c)\bigr)
	= \eta(1) = \vid,
\end{equation}
so that the identity reads
\begin{equation}
	S(T_b\cdot T_c) = (T_b\cdot T_c)^{-1} = T_c^{-1}\cdot T_b^{-1}
	= S(T_c)\cdot S(T_b),
\end{equation}
showing the anti-algebraic nature of the antipode.
Just in a similar manner can we show generally
\begin{equation}
	S(g_1\cdots g_n) = S(g_n)\cdots S(g_1),\qquad
	g_i = T_b\OR\statfn.
\label{antipode-gg}
\end{equation}
Applying next on $T_b\cdot\Ql_A$, the l.h.s.\ is
\begin{equation}
	\begin{aligned}
		\opprod\circ(S\ot\id)\circ\coproduct(T_b\cdot\Ql_A)
		&= \opprod\circ(S\ot\id)
		(T_b\cdot\Ql_A\ot T_b\cdot T_{\ar_A} 
		+ T_b\cdot\statfn\cdot T_{\al_A}\ot T_b\cdot \Ql_A) \\
		&= S(T_b\cdot\Ql_A)\cdot T_b\cdot T_{\ar_A}
			+ S\bigl(T_b\cdot\statfn\cdot T_{\al_A}\bigr)\cdot T_b\cdot\Ql_A,
	\end{aligned}
\end{equation}
while the r.h.s.\ is
\begin{equation}
	\eta\circ\counit(T_b\cdot \Ql_A)
	= \eta\bigl( \counit(T_b) \counit(\Ql_A)\bigr)
	= \eta(0) = 0,
\end{equation}
thus the identity gives
\begin{equation}
	S(T_b\cdot\Ql_A)
	= - S\bigl(T_b\cdot\statfn\cdot T_{\al_A}\bigr)
	\cdot T_b\cdot \Ql_A\cdot T_{\ar_a}^{-1}\cdot T_b^{-1}
	= - T_{\al_A}^{-1}\cdot\statfn\cdot\Ql_A\cdot T_{\ar_a}^{-1}
	\cdot T_b^{-1}
	= S(\Ql_A)\cdot S(T_b).
\end{equation}
Now we proceed to the calculation for $\Ql_A\cdot\Ql_B$: the l.h.s.\ reads
\begin{equation}
	\begin{aligned}
		& \opprod\circ(S\ot\id)\circ\coproduct(\Ql_A\cdot\Ql_B) \\
		&= \opprod\circ(S\ot\id)\Bigl(
			\Ql_A\cdot\Ql_B\ot T_{\ar_A}\cdot T_{\ar_B}
			+ \Ql_A\cdot\statfn\cdot T_{\al_B}\ot T_{\ar_A}\cdot\Ql_B \\
		&\phantom{= \opprod\circ(S\ot\vid)\Bigl(}
			+ \statfn\cdot T_{\al_A}\cdot\Ql_B\ot \Ql_A\cdot T_{\ar_B}
			+ \statfn\cdot T_{\al_A}\cdot\statfn\cdot T_{\al_B}
			\ot \Ql_A\cdot\Ql_B
		\Bigr) \\
		&=S(\Ql_A\cdot\Ql_B)\cdot T_{\ar_A}\cdot T_{\ar_B}
			+ S\bigl(\Ql_A\cdot\statfn\cdot T_{\al_B}\bigr)
			\cdot T_{\ar_A}\cdot\Ql_B \\
		&\	+ S\bigl(\statfn\cdot T_{\al_A}\cdot \Ql_B\bigr)
		\cdot \Ql_A\cdot T_{\ar_B}
			+ S\bigl(\statfn\cdot T_{\al_A}\cdot\statfn\cdot T_{\al_B}\bigr)
			\cdot \Ql_A\cdot\Ql_B,
	\end{aligned}
\end{equation}
and the r.h.s.\ 
\begin{equation}
		\eta\circ\counit(\Ql_A\cdot\Ql_B)
		= \eta\circ\bigl(\counit(\Ql_A)\counit(\Ql_B)\bigr) = \eta(0) = 0,
\end{equation}
so that the identity requires that
\begin{equation}
	\begin{gathered}
	\begin{aligned}
		S(\Ql_A\cdot\Ql_B)\cdot T_{\ar_A} T_{\ar_B}
		&= T_{\al_B}^{-1}\cdot\statfn\cdot T_{\al_A}^{-1}\cdot\statfn\cdot\Ql_A
		\cdot T_{\ar_A}^{-1} \cdot
			T_{\ar_A}\cdot \Ql_B \\
		&\ + T_{\al_B}^{-1}\cdot\statfn\cdot\Ql_B\cdot T_{\ar_B}^{-1}
		\cdot T_{\al_A}^{-1}\cdot\statfn\cdot
			\Ql_A\cdot T_{\ar_B} \\
		&\ - T_{\al_B}^{-1}\cdot\statfn\cdot
		T_{\al_A}^{-1}\cdot\statfn\cdot \Ql_A\cdot \Ql_B,
	\end{aligned} \\
	\text{i.e.}\qquad
	S(\Ql_A\cdot\Ql_B)
	= T_{\al_B}^{-1}\cdot\statfn\cdot\Ql_B\cdot T_{\ar_B}^{-1}\cdot
		T_{\al_A}^{-1}\cdot\statfn\cdot\Ql_A\cdot T_{\ar_A}^{-1}
	= S(\Ql_B)\cdot S(\Ql_A),
	\end{gathered}
\end{equation}
thus we find that the anti-algebraic nature of the antipode map
holds regardless of the fermionic nature of the supercharges.
The other identity
$\opprod\circ(\id\ot S)=\eta\circ\epsilon$ also gives
the same consequences.

With all these algebraic consistencies satisfied, we conclude
that the universal enveloping algebra, $\U(\A)$, i.e.\ our lattice 
superalgebra, can be regarded as a Hopf algebra.
We now move on to the construction of a field theory which has this
Hopf algebraic symmetry. We follow the general scheme formulated by
Oeckl \cite{Oeckl} as \emph{braided quantum field theory}
(BQFT).
To this purpose, we have to specify
the complete algebraic nature of the fields which is consistent to the
algebraic nature of the Hopf algebra. This requires one more
nontrivial ingredient called \emph{braiding}, or ``shifted
commutation'' in our language. It is in a way a generalization
of the statistics of fields.
In short, the covariance of the field theory under the Hopf
algebraic symmetry force us to introduce the braiding in a
consistent way. We will see this in what follows
for our specific case of the lattice superalgebra and
the corresponding field theory.

\subsection{Shift Structure as a Braiding}


Here we explain why we need the braiding or shift structure
in the space of fields, beginning with a simple illustration.
Suppose we are considering a normal supersymmetry with
a bosonic field $\phi$ and a fermionic field $\psi$.
Needless to say, bosonic fields commute with any other fields,
while fermionic fields anticommute only with other fermions.
Now take a supertransformation $Q\phi = \chi$ with a normal
supercharge $Q$ which is supposed to obey the Leibniz rule
$Q(\Vp_1\Vp_2) = Q\Vp_1 \Vp_2 + (-1)^\stat{\Vp_1}\Vp_1 Q\Vp_2$.
In the Hopf algebraic description we may say that
it has the coproduct
$\coproduct(Q) = Q\ot\vid+\statfn\vid\ot Q$ as before.
We ``know'' that the field $\chi$ is fermionic, as
a field of supertransformation of a boson $\phi$.
The point is that this fact is indeed inevitable;
we are forced to chose $\chi$ to be fermionic for the algebraic
consistency.
In fact, note that the quantity
$Q(\phi\psi)=\chi\psi + \phi(Q\psi)$ is equal to
$Q(\psi\phi)=(Q\psi)\phi - \psi\chi$, because
$\phi$ is defined as a boson, i.e.\ $\phi\psi=\psi\phi$.
Comparing these two relations, we find
$\chi\psi+\psi\chi = (Q\psi)\phi-\phi(Q\psi)$,
which is zero again due to that $\phi$ is bosonic.
This results in that $\chi\psi = -\psi\chi$, ``proving''
that $\chi$ is a fermion. The essence for this proof
is twofold: the one is the coproduct structure of the
transformation operator $Q$, especially the factor
$\statfn$, and the other is the covariance of
exchanging fields under the transformation $Q$, namely,
when we exchange the order of a product of fields
and then apply the transformation with $Q$,
the result is the same as the quantity obtained
by first applying the transformation with $Q$ on the product
and then exchanging the order of the transformed object.
The property of fields under exchanging the order in a product
is nothing but the \emph{statistics} of the fields. Here we have
just seen a natural and obvious fact
that the statistics of fields should be consistent
to the algebraic structure and covariance of transformations which
apply on the fields. It might be still worth stressing it,
however, because
it is the reason we need the braiding for our present application
of the Hopf algebraic symmetry on fields. It is also the reason
of that we think of the braiding as giving a
\emph{generalized statistics}.
We are going to investigate these issues in detail in the following.

Let us introduce the general notion of exchanging the order of
fields. We denote the exchanged object of $\Vp_1\ot\Vp_2$
as
\begin{equation}
	\Psi_{X_1, X_2}(\Vp_1\ot\Vp_2),\qquad
	\Vp_1\in X_1,\ \Vp_2\in X_2.
\end{equation}
The map $\Psi$ is called a braiding when it satisfies some
natural consistency conditions (see appendix~\ref{app-hopf}).
The trivial braiding is given as the normal transposition,
and, in the application to the link formalism, we assume
that the scalar fields on sites of the lattice
would have the trivial braiding nature;
\begin{equation}
	\Psi_{X_{\MR{s}}, X_{\MR{s}}}(\phi_1\ot\phi_2)
	= \phi_2\ot\phi_1,\qquad
	\phi_1,\ \phi_2\in X_\MR{s}:\ \text{scalar fields on sites}.
\label{triv-braiding}
\end{equation}
Repeating the argument above, we may apply
$\Ql_A$ on the product of scalar fields, or equivalently,
take the action of coproduct of $\Ql_A$ as
\begin{equation}
	\coproduct(\Ql_A)\act(\phi_1\ot\phi_2)
	= (\Ql_A\phi_1)\ot\phi_2(x+\ar_A)
	+ \phi_1(x+\al_A)\ot(\Ql_A\phi_2).
\end{equation}
Similarly on the exchanged product,
\begin{equation}
	\begin{aligned}
		\coproduct(\Ql_A)\act(\phi_2\ot\phi_1)
		&= \coproduct(\Ql_A)\act
		\Psi_{X_{\MR{s}}, X_{\MR{s}}}(\phi_1\ot\phi_2) \\
		&= (\Ql_A\phi_2)\ot\phi_1(x+\ar_A)
		+ \phi_2(x+\al_A)\ot(\Ql_A\phi_1).
	\end{aligned}
\label{lhs-braid}
\end{equation}
We now assume the covariance of the braiding
under symmetry transformations,
or, in other words, we assume that the braiding to be
an intertwiner of the transformations. In the present case,
this requires that
\begin{equation}
	\coproduct(\Ql_A)\act
		\Psi_{X_{\MR{s}}, X_{\MR{s}}}(\phi_1\ot\phi_2)
	= \Psi'\Bigl(\coproduct(\Ql_A)\act(\phi_1\ot\phi_2)\Bigr).
\end{equation}
The l.h.s.\ is given by \Ref{lhs-braid}, while
the r.h.s.\ is
\begin{equation}
	\begin{aligned}
		\Psi'\Bigl(\coproduct(\Ql_A)\act(\phi_1\ot\phi_2)\Bigr)
		&= \Psi_{X_{\MR{f}A},X_\MR{s}}\bigl(
				(\Ql_A\phi_1)\ot\phi_2(x+\ar_A)\bigr) \\
		&\phantom{=} + \Psi_{X_\MR{s},X_{\MR{f}A}}\bigl(
				\phi_1(x+\al_A)\ot(\Ql_A\phi_2)\bigr),
	\end{aligned}
\end{equation}
where we have denoted the space of fermionic fields of
the index $A$ as $X_{\MR{f}A}$ to which the transformed fields
$\Ql_A\phi_{1,2}$ are to belong.
Comparing these two equations, and noting that the fields
$\phi_1$ and $\phi_2$ could be completely independent,
we find the consequence, with a simple
identification, should be
\begin{equation}
	\begin{aligned}
		\Psi_{X_{\MR{f}A},X_\MR{s}}\bigl(
					(\Ql_A\phi_1)\ot\phi_2(x+\ar_A)\bigr)
		&=\phi_2(x+\al_A)\ot(\Ql_A\phi_1), \\
		\Psi_{X_\MR{s},X_{\MR{f}A}}\bigl(
				\phi_1(x+\al_A)\ot(\Ql_A\phi_2)\bigr)
		&=(\Ql_A\phi_2)\ot\phi_1(x+\ar_A).
	\end{aligned}
\end{equation}
These are not the trivial braiding as in \Ref{triv-braiding}.
Instead, these braiding mean that when we exchange
the order of the fermion $\Ql_A\phi$ with the other field,
it changes the argument of the other field by the amount
$\al_A - \ar_A$ under the exchange from the left to the right,
and by the opposite amount under the exchange
from the right to the left. Recalling that the scalar fields
obey the trivial braiding, we might interpret this fact
as that the transformed fields, fermions,
inherited the nontrivial braiding nature from the supercharge,
which, in a way, shows the nontrivial braiding already
in the structure of the coproduct. In fact, this kind of
nontrivial braiding is referred to as the shifted commutation
structure in the link formalism.

We have to emphasize here that the ``claimed inconsistency''
\cite{Dutch}
explained in section~\ref{rev}
no longer appears with incorporating this nontrivial braiding
in the non-gauged link formalism. Our approach which is purely based
on the Hopf algebraic description clarifies the necessity
of the braiding and shows how that problem criticized can be
resolved.

To confirm how the things work, let us compute another
example:
\begin{equation}
	\coproduct(\Ql_B)\act\bigl(\psi_{1A}(x)\ot\phi_2(x)\bigr)
	= (\Ql_B\psi_{1A})(x)\ot\phi_2(x+\ar_B)
	- \psi_{1A}(x+\al_B)\ot\psi_{2B}(x),
\end{equation}
where $\psi_{2B}:= \Ql_B\phi_2$, and thus
\begin{equation}
	\begin{aligned}
	\Psi'\Bigl(
		\coproduct(\Ql_B)\act\bigl(\psi_{1A}(x)\ot\phi_2(x)\bigr)
	\Bigr)
	&= \Psi_{X_{AB},X_\MR{s}}\bigl(
			(\Ql_B\psi_{1A})(x)\ot\phi_2(x+\ar_B)\bigr) \\
	&\phantom{=}
		- \Psi_{X_{\MR{f}A},X_{\MR{f}B}}\bigl(
			\psi_{1A}(x+\al_B)\ot\psi_{2B}(x)\bigr),
	\end{aligned}
\end{equation}
whereas
\begin{equation}
	\begin{aligned}
		\coproduct(\Ql_B)\act\bigl(\phi_2(x+\al_A-\ar_A)\ot\psi_{1A}(x)\bigr)
		&= \coproduct(\Ql_B)\act
		\Psi_{X_{\MR{f}A}, X_{\MR{s}}}\bigl(\psi_{1A}(x)\ot\phi_2(x)\bigr) \\
		&= \psi_{2B}(x+\al_A-\ar_A)\ot\psi_{1A}(x+\ar_B) \\
		&\phantom{=}
		+ \phi_2(x+\al_A-\ar_A+\al_B)\ot\bigl(\Ql_B\psi_{1A}(x)\bigr).
	\end{aligned}
\end{equation}
Here $X_{AB}$ is such that $\Ql_B\psi_{A}\in X_{AB}$.
Assuming again the covariance
\begin{equation}
	\Psi'\Bigl(
		\coproduct(\Ql_B)\act\bigl(\psi_{1A}(x)\ot\phi_2(x)\bigr)
	\Bigr)
	=
	\coproduct(\Ql_B)\act
		\Psi_{X_{\MR{f}A}, X_{\MR{s}}}\bigl(\psi_{1A}(x)\ot\phi_2(x)\bigr),
\end{equation}
we obtain the following braiding relations:
\begin{equation}
	\begin{aligned}
		\Psi_{X_{AB},X_\MR{s}}\bigl(
			(\Ql_B\psi_{1A})(x)\ot\phi_2(x+\ar_B)\bigr)
		&= \phi_2(x+\al_A-\ar_A+\al_B)\ot\bigl(\Ql_B\psi_{1A}(x)\bigr), \\
		\Psi_{X_{\MR{f}A},X_{\MR{f}B}}\bigl(
			\psi_{1A}(x+\al_B)\ot\psi_{2B}(x)\bigr)
		&= -\psi_{2B}(x+\al_A-\ar_A)\ot\psi_{1A}(x+\ar_B).
	\end{aligned}
\label{braiding-2nd}
\end{equation}
Notice, in passing, from the first equation of \Ref{braiding-2nd},
we have
\begin{equation}
	\begin{aligned}
		\Psi_{X_{AB},X_\MR{s}}\bigl(
			(\Ql_B\psi_{1A})(x)\ot\phi_2(x)\bigr)
 		&= \phi_2(x+\al_A-\ar_A+\al_B-\ar_B)\ot\bigl(\Ql_B\psi_{1A}\bigr)(x), \\
			\Psi_{X_{AB},X_\MR{s}}\bigl(
			(\Ql_A\psi_{1B})(x)\ot\phi_2(x)\bigr)
 		&= \phi_2(x+\al_B-\ar_B+\al_A-\ar_A)\ot\bigl(\Ql_A\psi_{1B}\bigr)(x),
	\end{aligned}
\end{equation}
so that, summing up these two,
\begin{equation}
	\begin{aligned}
		\Psi_{X_{AB},X_\MR{s}} &\Bigl(
			\bigl(\{\Ql_B,\Ql_A\}\phi_{1})(x)\ot\phi_2(x)\Bigr)
		= 2\tau_{AB}^\mu\Psi_{X_{AB},X_\MR{s}}\Bigl(
			(\Pl_\mu\phi_1)(x)\ot\phi_2(x)\Bigr) \\
 	&= \phi_2(x+\al_A-\ar_A+\al_B-\ar_B)\ot\bigl(
			\{\Ql_B,\Ql_A\}\phi_{1}\bigr)(x) \\
		&= 2\tau_{AB}^\mu
		\phi_2(x+\al_A-\ar_A+\al_B-\ar_B)\ot\bigl(
			\Pl_\mu\phi_{1}\bigr)(x).
	\end{aligned}
\label{diff-braiding}
\end{equation}
These examples show that the general braiding relation would be
similarly derived. As a rule of sum, we can write it as
\begin{equation}
	\begin{aligned}
	&\Psi\Bigl(
	\Vp_{A_0\cdots A_p}(x)\ot\Vp'_{B_0\cdots B_q}(y)
	\Bigr) \\
	&=
	(-1)^{pq}
	\Vp'_{B_0\cdots B_q}\left(y+\sum_{i=1}^p(\al_{A_i}-\ar_{A_i})\right)
	\ot
	\Vp_{A_0\cdots A_p}\left(x-\sum_{i=1}^q(\al_{B_i}-\ar_{B_i})\right),
	\end{aligned}
\label{gen-braid}
\end{equation}
where we have used the abbreviation
$\Vp_{A_0\cdots A_p}:= \Ql_{A_p}\cdots\Ql_{A_1}\Vp_{A_0}$,
which could just vanish, where $\Vp_{A_0}:=\phi$.
If we had introduced a scalar field which itself has
nontrivial braiding/shift structure, this relation
would have even been generalized.

The exchanging of a product of more than three fields
should be naturally introduced. In the case of the trivial
braiding,
\begin{equation}
	\begin{aligned}
	\Psi_{X_1\ot X_2, X_3}\bigl(
	(\phi_1\ot\phi_2)\ot\phi_3\bigr)
	&= \phi_3\ot(\phi_1\ot\phi_2)
	= \phi_3\ot\phi_1\ot\phi_2
	= \Psi_{X_1,X_3}\bigl((\phi_1\ot\phi_3)\ot\phi_2\bigr) \\
	&= \Psi_{X_1,X_3}\circ\Psi_{X_2,X_3}
	\bigl((\phi_1\ot\phi_2)\ot\phi_3)\bigr),
	\end{aligned}
\end{equation}
so that in the general case we extend it to
\begin{equation}
	\Psi_{X_1\ot X_2,X_3}
	= \Psi_{X_1, X_3}\circ\Psi_{X_2,X_3},\qquad
	\Psi_{X_1, X_2\ot X_3}
	= \Psi_{X_1, X_3}\circ\Psi_{X_1,X_2}.
\label{braiding-prod}
\end{equation}
For example,
\begin{equation}
\Psi_{X_1,X_2\ot X_3}\bigl(
	\phi_1(x)\ot\phi_2(x)\ot\psi_{3A}(x)\bigr)
	= \phi_2(x)\ot\psi_{3A}(x)\ot\phi_1(x-\al_A+\ar_A).
\label{phi-phi-psi}
\end{equation}
For the exchanging with the trivial or constant fields,
we should impose
\begin{equation}
	\Psi_{X^0_\MR{e}, X} = \Psi_{X, X^0_\MR{e}} = \id,
\label{braiding-id}
\end{equation}
where $X^0_\MR{e}$ denotes the space of trivial bosonic fields.
Using these rules, let us calculate one more example:
\begin{equation}
	\begin{aligned}
	(\id\ot\coproduct)\circ\coproduct(\Ql_A)\act
	(\phi_1\ot\phi_2\ot\psi_{3B})
	&= \psi_{1A}(x)\ot\phi_2(x+\ar_A)\psi_{3B}(x+\ar_A) \\
	&\phantom{=}+ \phi_1(x+\al_A)\ot\psi_{2A}(x)\ot\psi_{3B}(x+\ar_A) \\
	&\phantom{=}+ \phi_1(x+\al_A)\ot\phi_2(x+\al_A)\ot(\Ql_A\psi_{3B})(x),
	\end{aligned}
\end{equation}
where we have used the coassociativity \Ref{coassociativity-1}.
Applying $\Psi_{X_1,X_2\ot X_3}$, and comparing it with
\begin{equation}
	\begin{aligned}
	&(\id\ot\coproduct)\circ\coproduct(\Ql_A)\act
	\Bigl(\Psi_{X_1,X_2\ot X_3}\bigl(
	\phi_1(x)\ot\phi_2(x)\ot\psi_{3B}(x)\bigr)\Bigr) \\
	&=(\id\ot\coproduct)\circ\coproduct(\Ql_A)\act
	\Bigl(\phi_2(x)\ot\psi_{3B}(x)\ot\phi_1(x-\al_B+\ar_B)\Bigr) \\
	&= \psi_{2A}(x)\ot\psi_{3B}(x+\ar_A)\ot\phi_1(x+\ar_A-\al_B+\ar_B) \\
	&\phantom{=}
	+ \phi_2(x+\al_A)\ot(\Ql_A\psi_{3B})(x)\ot\phi_1(x+\ar_A-\al_B+\ar_B) \\
	&\phantom{=}
	- \phi_2(x+\al_A)\ot\psi_{3B}(x+\al_A)\ot\psi_{1A}(x-\al_B+\ar_B),
	\end{aligned}
\end{equation}
where we have used \Ref{phi-phi-psi} and \Ref{coassociativity-1} again,
we find that
\begin{equation}
	\begin{aligned}
	&\Psi_{X_1,X_2\ot X_3}	
	\Bigl(\psi_{1A}(x)\ot\bigl(\phi_2(x+\ar_A)\ot\psi_{3B}(x+\ar_A)\bigr)
	\Bigr) \\
	&= -\phi_2(x+\al_A)\ot\psi_{3B}(x+\al_A)\ot\psi_{1A}(x-\al_B+\ar_B), \\
	&\Psi_{X_1,X_2\ot X_3}
	\Bigl(\phi_1(x+\al_A)\ot\bigl(\psi_{2A}(x)\ot\psi_{3B}(x+\ar_A)\bigr)
	\Bigr) \\
	&= \psi_{2A}(x)\ot\psi_{3B}(x+\ar_A)\ot\phi_1(x+\ar_A-\al_B+\ar_B), \\
	&\Psi_{X_1,X_2\ot X_3}
	\Bigl(\phi_1(x+\al_A)\ot\bigl(\phi_2(x+\al_A)\ot(\Ql_A\psi_{3B})(x)\bigr)
	\Bigr) \\
	&= \phi_2(x+\al_A)\ot(\Ql_A\psi_{3B})(x)\ot\phi_1(x+\ar_A-\al_B+\ar_B).
	\end{aligned}
\end{equation}
These examples show that the braiding, i.e.\ 
the amount of shifts of the arguments of fields induced under exchanging,
is additive; for a field $\Vp_1$ with the shift
$a_1$ and another $\Vp_2$ with the shift $a_2$,
the product $\Vp_1\ot\Vp_2$ has the shift $a_1+a_2$.
This is a simple consequence of the natural braiding rule
\Ref{braiding-prod}.

These observations motivate us to introduce the notion
of shift structure of fields as a kind of an additive ``grading''
determined with how many supercharges are acting
on the scalar fields.
We may thus introduce, in addition to the normal
graded structure of fields, i.e.\ bosonic and fermionic
statistics, the graded structure which we call the shift
structure so that the space of elementary fields $X$
is decomposed in general as
\begin{equation}
	X = \bigoplus_{\text{grading}}\Xe\oplus\Xo.
\end{equation}
The space of whole fields, $\Xh$, is also decomposed
with respect to the shift/grading structure the same way;
\begin{equation}
	\Xh = \bigoplus_{n=0}^\infty\bigoplus_{\text{grading}}X^n.
\end{equation}
The field contents and their shift structure
are determined in each model, mainly with the use of
the Leibniz rule consistency conditions.
We have to emphasize that this grading structure is especially
crucial
to define the explicit form of the ``momentum'' operator
$\Pl_\mu$. As mentioned at the beginning of the previous
subsection, we might have started with
taking a difference operator as its representation:
$(\Pl_\mu\phi)(x)=a^{-1}(\phi(x+\al\muhat)-\phi(x+\ar\muhat))$.
This, however, doesn't satisfy the relation \Ref{diff-braiding},
since we have assumed that $\phi$ obeys a trivial braiding
and thus $a^{-1}(\phi(x+\al\muhat)-\phi(x+\ar\muhat))$ has the 
same trivial braiding. 
We thus need an expression like
$(\Pl_\mu\phi)(x)=a^{-1}(\phi'(x+\al\muhat)-\phi'(x+\ar\muhat))$
for which $\phi'$ has an additional grading to satisfy
the relation \Ref{diff-braiding}. To give a consistent
representation for them is important for the formulation and 
will be treated elsewhere. Here our claim is that the algebraic
description presented here can still formalize a field
theory with the Hopf algebraic symmetry even if we don't
have the explicit representation for these graded fields
and their ``momentum'' operators, as is seen in what follows.

Let us note also that our braiding satisfies that
\begin{equation}
	\Psi_{X_1,X_2}\circ\Psi_{X_2,X_1} = \id,
\label{triv-braid}
\end{equation}
or equivalently,
\begin{equation}
	\Psi_{X_2,X_1} = \Psi_{X_1,X_2}^{-1}.
\end{equation}
In a standard mathematical terminology this kind of
exchanging map $\Psi$ \emph{isn't} referred to as
a braiding, or one may distinguish it from
the \emph{strictly braided} case. Here we use the term
braiding in a broader sense, allowing a type of
simple nature \Ref{triv-braid}. We emphasize
that it is still nontrivial in the sense that $\Psi\neq\tau$,
where $\tau$ is the simple transposition:
$\tau(\Vp_1\ot\Vp_2)=\Vp_2\ot\Vp_1$. In fact,
our braiding is a transposition plus some shifts of
the arguments of fields up to the statistics factors.
This should be compared with the statistics
of usual bosons and fermions; for that case
the braid is nothing but the simple exchanging
up to the statistics.
We could therefore describe these facts as
that the fields which represent our Hopf algebraic lattice
superalgebra naturally obtain a braiding structure which
expresses slightly more generalized statistics than the usual one.%
\footnote{A well-known example of generalized statistics
is that of anyons, for which the exchanging map
is strictly braided in general. Our statistics is thus more like
the usual statistics than the anyonic one.}

According to the general discussion (see appendix~\ref{app-hopf}),
it seems that
the simple braiding structure \Ref{triv-braid} might be given
as an explicit formula \Ref{braid-R} when the corresponding
Hopf algebra is \emph{triangular}. We find that this is
indeed the case at least formally;
our symmetry algebra could be identified
as a triangular Hopf algebra with an additional grading structure,
and the braiding \Ref{gen-braid} be given with the corresponding
\emph{(quasi-)triangular structure} $\R$. 
To see this, let us first introduce a formal expression
for the shift operator $T_b$
\begin{equation}
	T_b = \exp(b^\mu\del_\mu).
\end{equation}
We write this as if the continuum derivative operator $\del_\mu$
were introduced on the lattice; however it must be understood
as a formal operator and only well-defined when exponentiated
to give the lattice proper operator $T_b$.
We may impose
\begin{equation}
	\coproduct(\del_\mu) = \del_\mu\ot\vid+\vid\ot\del_\mu,\qquad
	\counit(\del_\mu) = 0,\qquad
	S(\del_\mu) = -\del_\mu,
\end{equation}
which should be interpreted as formal equivalents of
the relations \Ref{counit-coproduct-T} and \Ref{antipode}
for $T_b$.
We then recall that the generator $\Ql_A$ has a kind of
grading as an amount of the shift $a_A:= \al_A-\ar_A$
induced under exchanging
$\Ql_A\Vp$ with other fields. We may express this fact
with introducing another operator $L^\mu$ such that
\begin{equation}
	a[L^\mu, \Ql_A] = (a_A)^\mu\Ql_A,\quad\IE\quad
	[L^\mu, \Ql_A] = l_A^\mu\Ql_A,
\label{grading-Q}
\end{equation}
where $l_A^\mu = a^{-1}(a_A)^\mu$.
Since $\Pl_\mu$ is given as $\Pl_\mu\sim\{\Ql_A,\Ql_B\}$, it also
has the grading as in
\begin{equation}
	a[L^\mu, \Pl_\nu] = a_P(\nuhat)^\mu\Pl_\nu = a_P\delta^\mu_\nu\Pl_\nu,
	\quad\IE\quad
	[L^\mu, \Pl_\nu] = l_P\delta^\mu_\nu\Pl_\nu,
\label{grading-P}
\end{equation}
where $a_P:= \al-\ar$ and $l_P:= a^{-1}a_P$.
We list the other relations
\begin{equation}
	[L^\mu, T_b] = [L^\mu, \statfn] = [L^\mu, L^\nu] = 0,
\label{algebra-Llr-2}
\end{equation}
where the first two are due to the fact that neither $T_b$ nor $\statfn$
induces shift and the latter one is automatic because
of the ``Abelian'' nature of \Ref{grading-Q}, \Ref{grading-P}
and the others. For completeness, we set
\begin{equation}
	\coproduct(L^\mu) = L^\mu\ot\vid + \vid\ot L^\mu,\quad
	\counit(L^\mu) = 0,\quad
	S(L^\mu) = -L^\mu.
\label{counit-coproduct-antipode-L}
\end{equation}
Now let
\begin{equation}
	\R := \exp\bigl(
		aL^\mu\ot\del_\mu - a\del_\mu\ot L^\mu + i\pi\fn\ot\fn
	\bigr).
\label{R}
\end{equation}
We can show that this formal operator $\R\in\U(\A)\ot\U(\A)$
is invertible and
satisfies the relations
\begin{equation}
	\begin{gathered}
		\tau\circ\coproduct(h)
		= \R\opprod\coproduct(h)\opprod\R^{-1},\\
		(\coproduct\ot\id)\R = \R_{13}\R_{23},\qquad
		(\id\ot\coproduct)\R = \R_{13}\R_{12}.
	\end{gathered}
\label{R-conditions}
\end{equation}
(See appendix~\ref{app-hopf} for the notation.)
Notice first that $\R^{-1}$ is given as
\begin{equation}
	\R^{-1} =	\exp\bigl(
		-aL^\mu\ot\del_\mu + a\del_\mu\ot L^\mu + i\pi\fn\ot\fn
	\bigr)
\end{equation}
(recall that $\fn$ only gives integer numbers), and so that
\begin{equation}
	\R_{21} =	\exp\bigl(
		a\del_\mu\ot L^\mu - aL^\mu\ot\del_\mu + i\pi\fn\ot\fn
	\bigr)
	= \R^{-1}.
\label{triangularity}
\end{equation}
As for the first relation in \Ref{R-conditions}, compute
\begin{equation}
	\R\opprod\coproduct(h)\opprod\R^{-1}
	= \sum_{n=0}^\infty\frac{1}{n!}
	\Bigl(\ad\bigl(
		L
	\bigr)\Bigr)^n\coproduct(h),
\end{equation}
where we have written
$L:= aL^\mu\ot\del_\mu - a\del_\mu\ot L^\mu + i\pi\fn\ot\fn$
just for simplicity, and used
$\ad$ to denote the Lie derivative.
For $h= \Ql_A$,
\begin{equation}
	\begin{aligned}
		&\ad(L)\coproduct(\Ql_A)
		=\bigl[
		aL^\mu\ot\del_\mu - a\del_\mu\ot L^\mu + i\pi\fn\ot\fn,\ 
		\Ql_A\ot T_{\ar_A} + \statfn\opprod T_{\al_A}\ot\Ql_A
		\bigr] \\
		&=
		\bigl[
			aL^\mu\ot\del_\mu + i\pi\fn\ot\fn,\ 
			\Ql_A\ot T_{\ar_A}
		\bigr]
		+
		\bigl[
			- a\del_\mu\ot L^\mu + i\pi\fn\ot\fn,\ 
			\statfn\opprod T_{\al_A}\ot\Ql_A
		\bigr] \\
		&=
			a[L^\mu, \Ql_A]\ot\del_\mu\opprod T_{\ar_A}
			+ i\pi[\fn,\Ql_A]\ot\fn\opprod T_{\ar_A} \\
		&\phantom{=}
			-a\del_\mu\opprod\statfn\opprod T_{\al_A}\ot[L^\mu,\Ql_A]
			+ i\pi\fn\opprod\statfn\opprod T_{\al_A}\ot[\fn,\Ql_A] \\
		&=
			\Ql_A\ot \bigl((a_A)^\mu\del_\mu + i\pi\fn\bigr)\opprod T_{\ar_A}
			+\bigl(-(a_A)^\mu\del_\mu + i\pi\fn\bigr)\opprod\statfn T_{\al_A}
				\ot\Ql_A,
	\end{aligned}
\end{equation}
so that
\begin{equation}
	\bigl(\ad(L)\bigr)^n\coproduct(\Ql_A)
	=
		\Ql_A\ot \bigl((a_A)^\mu\del_\mu + i\pi\fn\bigr)^n\opprod T_{\ar_A}
			+\bigl(-(a_A)^\mu\del_\mu + i\pi\fn\bigr)^n
				\opprod\statfn\opprod T_{\al_A}\ot\Ql_A.
\end{equation}
We therefore obtain
\begin{equation}
	\begin{aligned}
		&\R\opprod\coproduct(\Ql_A)\opprod\R^{-1} \\
		&= \Ql_A\ot\exp\bigl((a_A)^\mu\del_\mu + i\pi\fn\bigr)\opprod T_{\ar_A}
		+ \exp\bigl(-(a_A)^\mu\del_\mu + i\pi\fn\bigr)
			\opprod\statfn\opprod T_{\al_A}\ot\Ql_A \\
		&= \Ql_A\ot\statfn\opprod T_{\al_A} + T_{\ar_A}\ot\Ql_A
		= \tau\circ\coproduct(\Ql_A),
	\end{aligned}
\end{equation}
since
\begin{equation}
	\exp\bigl(\pm(a_A)^\mu\del_\mu+i\pi\fn\bigr)
	=\exp\bigl(\pm(\al_A)^\mu\del_\mu\bigr)\opprod
		\exp\bigl(\mp(\ar_A)^\mu\del_\mu\bigr)\opprod
		\exp(i\pi\fn)
	= T_{\al_A}^{\pm}\opprod T_{\ar_A}^{\mp}\opprod\statfn.
\end{equation}
A simpler calculation leads to similar result for $h=\Pl_\mu$ too.
For $h= T_b,\ \statfn,\ L^\mu$, it is rather clear that
\begin{equation}
	\R\opprod\coproduct(h)\opprod\R^{-1}
	= \coproduct(h)
	= \tau\circ\coproduct(h).
\end{equation}
Thus the first equation in \Ref{R-conditions} indeed holds
for the choice \Ref{R} of $\R$. The second relation follows as
\begin{equation}
	\begin{aligned}
		(\coproduct\ot\id)\R
		&=\exp\bigl(
			a\coproduct(L^\mu)\ot\del_\mu - a\coproduct(\del_\mu)\ot L^\mu 
			+ i\pi\coproduct(\fn)\ot\fn
		\bigr) \\
		&=\exp\Bigl(
			aL^\mu\ot\vid\ot\del_\mu - a\del_\mu\ot\vid\ot L^\mu 
			+ i\pi\fn\ot\vid\ot\fn \\
		&\phantom{=\exp\Bigl(\empbrckt)}
			+a\vid\ot L^\mu\ot\del_\mu - a\vid\ot\del_\mu\ot L^\mu 
			+ i\pi\vid\ot\fn\ot\fn
		\Bigr)	\\	
		&=\exp\bigl(
			aL^\mu\ot\vid\ot\del_\mu - a\del_\mu\ot\vid\ot L^\mu 
			+ i\pi\fn\ot\vid\ot\fn \bigr) \\
		&\phantom{=}
			\opprod\exp\bigl(
			a\vid\ot L^\mu\ot\del_\mu - a\vid\ot\del_\mu\ot L^\mu 
			+ i\pi\vid\ot\fn\ot\fn \bigr)
		= \R_{13}\opprod\R_{23}.
	\end{aligned}
\end{equation}
The third one is almost the same.

We have thus shown
that the formal operator $\R$ given as \Ref{R} is a quasitriangular
structure and, due to \Ref{triangularity},
our lattice superalgebra is identified as a triangular
Hopf algebra. The whole spaces of fields,
as representation spaces of a triangular Hopf algebra, would be
braided by $\R$ as in
\begin{equation}
	\Psi = \tau\circ\R\act,
\label{braid-R-0}
\end{equation}
which agrees with our formula \Ref{gen-braid} as now seen.
We need the representation of $L^\mu$ on the elementary fields.
First for normal scalar fields $\{\phi,\ \cdots\}$ let
\begin{equation}
	L^\mu\act\phi = 0\cdot\phi = 0.
\end{equation}
For the other fields in the irreducible supermultiplet to which
the above bosonic fields belong, the actions of $L^\mu$ are
automatically determined by the algebra \Ref{grading-Q}.
For instance, on $\psi_A:=\Ql_A\phi$, we find
\begin{equation}
	L^\mu\act\psi_A = L^\mu\act(\Ql_A\phi)
	= \bigl( [L^\mu,\Ql_A] + \Ql_A\opprod L^\mu \bigr)\act\phi
	= l_A^\mu \Ql_A\phi = l_A^\mu\psi_A.
\end{equation}
Then inductively, we find for
$\Vp_{A_1\cdots A_n}=\Ql_{A_n}\cdots\Ql_{A_1}\phi$ that
\begin{equation}
	L^\mu\act\Vp_{A_1\cdots A_n}
	= \bigl(l_{A_1}+\cdots+l_{A_n}\bigr)^\mu
	\Vp_{A_1\cdots A_n}.
\end{equation}
These relations express explicitly the grading structure of fields
explained above.
We thus compute
\begin{equation}
	\begin{aligned}
		&\R\act\bigl(
			\Vp_{A_1\cdots A_p}(x)\ot\Vp_{B_1\cdots B_q}(y)
		\bigr) \\
		&=
		\exp\Bigl(
			\vid\ot\bigl(
				(a_{A_1}+\cdots+a_{A_p})^\mu\del_\mu
			\bigr)
			-\bigl(
				(a_{B_1}+\cdots+a_{B_q})^\mu\del_\mu
			\bigr)\ot\vid
			+i\pi pq\vid\ot\vid
		\Bigr) \\
		&\phantom{=\exp}
		\act\,
		\bigl(
			\Vp_{A_1\cdots A_p}(x)\ot\Vp_{B_1\cdots B_q}(y)
		\bigr) \\
		&=
			(-1)^{pq}
			\bigl(\vid\ot T_{a_{A_1}+\cdots+a_{A_p}}\bigr)\opprod
			\bigl(T^{-1}_{a_{B_1}+\cdots+a_{B_q}}\ot\vid\bigr)\act
		\bigl(
			\Vp_{A_1\cdots A_p}(x)\ot\Vp_{B_1\cdots B_q}(y)
		\bigr) \\
		&=
			\Vp_{A_1\cdots A_p}\left(x-\sum_{i=1}^p a_{B_i}\right)
			\ot
			\Vp_{B_1\cdots B_q}\left(y+\sum_{i=1}^q a_{A_i}\right).
	\end{aligned}
\end{equation}
Since here $a_{A_i}=\al_{A_i}-\ar_{A_i}$ etc.,\ 
we have shown that the equation \Ref{braid-R-0} does reproduce
the general braiding rule \Ref{gen-braid}.

It is worth pointing out that our quasitriangular structure
$\R$ can be written as
\begin{equation}
		\R = \cocycle_{21}\opprod\R_0\opprod\cocycle^{-1},\qquad
		\R_0 := \exp\bigl(i\pi\fn\ot\fn\bigr),
\label{R-R0}
\end{equation}
with some invertible operator $\cocycle\in\U(\A)\ot\U(\A)$
which satisfies so-called the \emph{2-cocycle condition}
\begin{equation}
	(\cocycle\ot\vid)\opprod(\coproduct\ot\id)\cocycle
	= (\vid\ot\cocycle)\opprod(\id\ot\coproduct)\cocycle,
\end{equation}
and the \emph{counital condition}
\begin{equation}
	(\counit\ot\id)\cocycle = (\id\ot\counit)\cocycle = \vid.
\end{equation}
Such an operator is not necessarily unique. We take one specific
example to illustrate it:
\begin{equation}
	\begin{gathered}
		\cocycle 
			:= \exp\bigl(a\del_\mu\ot\Ll^\mu + a\Lr^\mu\ot\del_\mu\bigr),\\
		\cocycle_{21} 
			= \exp\bigl(a\Ll^\mu\ot\del_\mu + a\del_\mu\ot\Lr^\mu\bigr),\qquad
		\cocycle^{-1} 
			= \exp\bigl(-a\del_\mu\ot\Ll^\mu - a\Lr^\mu\ot\del_\mu\bigr),
	\end{gathered}
\end{equation}
where we have introduced two more operators $\Ll^\mu$ and $\Lr^\mu$
such that $L^\mu = \Ll^\mu - \Lr^\mu$, namely,
\begin{equation}
	a[L^{\MR{l,r}}{}^\mu, \Ql_A]
	= (a^{\MR{l,r}}_A)^\mu \Ql_A,\quad\text{etc.},
\label{algebra-Llr-1}
\end{equation}
with coproduct, counit and antipode formulae similar to those of
$L^\mu$.
It is easy to see that \Ref{R-R0} actually holds for this operator
$\cocycle$. The cocycle condition is fulfilled as
\begin{equation}
	\begin{aligned}
		\lhs
		&=
			\exp\bigl(a\del_\mu\ot\Ll^\mu\ot\vid
				+a\Lr^\mu\ot\del_\mu\ot\vid\bigr)
			\opprod
			\exp\bigl(a\coproduct(\del_\mu)\ot\Ll^\mu
				+a\coproduct(\Lr^\mu)\ot\del_\mu\bigr) \\
		&=
			\exp\Bigl(a\del_\mu\ot\Ll^\mu\ot\vid
			+a\Lr^\mu\ot\del_\mu\ot\vid \\
		&\phantom{=\exp\bigl(\empbrckt)}
			+ a\del_\mu\ot\vid\ot\Ll^\mu
			+ a\vid\ot\del_\mu\ot\Ll^\mu
			+ a\Lr^\mu\ot\vid\ot\del_\mu
			+ a\vid\ot\Lr^\mu\ot\del_\mu
			\Bigr) \\
		&=
			\exp\Bigl(\vid\ot a\del_\mu\ot\Ll^\mu
			+ \vid\ot a\Lr^\mu\ot\del_\mu \\
		&\phantom{=\exp\bigl(\empbrckt)}
			+ a\del_\mu\ot\Ll^\mu\ot\vid
			+ a\del_\mu\ot\vid\ot\Ll^\mu
			+ a\Lr^\mu\ot\del_\mu\ot\vid
			+ a\Lr^\mu\ot\vid\ot\del_\mu
			\Bigr) \\
		&=
			\exp\bigl(\vid\ot a\del_\mu\ot\Ll^\mu
				+\vid\ot a\Lr^\mu\ot\del_\mu\bigr)
			\opprod
			\exp\bigl(a\del_\mu\ot\coproduct(\Ll^\mu)
				+a\Lr^\mu\ot\coproduct(\del_\mu)\bigr) \\
			&= \rhs,
	\end{aligned}
\end{equation}
while the counitality is clear because
$\counit(\del_\mu)=\counit(L^{\MR{l,r}}{}^\mu)=0$.
We thus conclude from these results
that our lattice superalgebra $\U(\A)$ with the quasitriangular
structure $\R$ could be understood as so-called
the \emph{twist} 
by the \emph{cocycle element} $\cocycle$
of some other Hopf algebra
$\U(\A)_0$ with the simple quasitriangular structure $\R_0$.
The ``untwisted'' Hopf algebra $\U(\A)_0$ has the same algebra and counit
as those of $\U(\A)$ but its coproduct and antipode are such that
\begin{equation}
	\begin{aligned}
		\coproduct(h) 
		&= \cocycle\opprod\coproduct_0(h)\opprod\cocycle^{-1},\\
		S(h)
		&= U\opprod S_0(h)\opprod U^{-1},\qquad
		U := \opprod(\id\ot S)\cocycle,\quad
		U^{-1} = \opprod(S\ot\id)\cocycle^{-1}.
	\end{aligned}
\end{equation}
Thus for $h=T_b,\ \statfn,\ L^\mu$, we find
$\coproduct_0(h)=\coproduct(h)$, whereas for $h=\Ql_A,\ \Pl_\mu$,
we can show that
\begin{equation}
	\coproduct_0(\Ql_A) = \Ql_A\ot\vid + \statfn\ot\Ql_A,\qquad
	\coproduct_0(\Pl_\mu) = \Pl_\mu\ot\vid + \vid\ot\Pl_\mu.
\end{equation}
Since in the present case
\begin{equation}
	U = \exp\bigl(-a\Lp^\mu\opprod\del_\mu\bigr),\quad
	U^{-1} = \exp(a\Lp^\mu\opprod\del_\mu\bigr),\qquad
	\Lp^\mu := \Ll^\mu + \Lr^\mu,
\end{equation}
antipodes as well remain unchanged for
$h=T_b,\ \statfn,\ L^{\MR{l,r}}{}^\mu$, but changed again
for $h=\Ql_A,\ \Pl_\mu$:
\begin{equation}
	S_0(\Ql_A) = -\statfn\opprod\Ql_A,\qquad
	S_0(\Pl_\mu) = -\Pl_\mu,
\end{equation}
as seen with the use of
\begin{equation}
	U\opprod\Ql_A\opprod U^{-1}
	= \exp\bigl((\al_A+\ar_A)^\mu\del_\mu\bigr)\opprod\Ql_A
	= T_{\al_A}\opprod T_{\ar_A}\opprod\Ql_A
\end{equation}
and of similar for $\Pl_\mu$.

We have found that the (un)twisted Hopf algebra $(\U(\A)_0,\ \R_0)$
becomes much simpler and has the form of a normal universal
enveloping Lie superalgebra of normal supersymmetry.
This result might seem confusing because under the twisting
the algebraic structure of the original Hopf algebra remains the same
and operators themselves don't take any transformations; if
such simpler Hopf algebra exits, could we just begin with it
without taking the deformed one $(\U(\A),\ \R)$?
Actually we can equally formulate the whole story
with the simpler Hopf algebra $(\U(\A)_0,\ \R_0)$,
but notice that this twisting transformation is only
possible with the nontrivial ``charge'' or ``grading''
operators $L^{\MR{l,r}}{}^\mu$ at our disposal, and that the twisted
Hopf algebra keeps them as well. On our original Hopf algebra
$(\U(\A),\ \R)$, these have a natural interpretation as those
assigning how fields are geometrically put on the lattice
and how operators affect on such a geometrical structure.
On the twisted Hopf algebra $(\U(\A)_0,\ \R_0)$,
this kind of interpretation is less clear since $\coproduct_0, S_0$,
etc.,\ just have normal structure and, nevertheless,
these operators $L^{\MR{l,r}}{}^\mu$ must be included
for the whole algebra to be represented exactly.
This last observation would be quite crucial, particularly
when compared with the no-go theorem presented in
\cite{No-go:Kato-Sakamoto-So}, since in the twisted algebra
the ``momentum'' operator obeys the exact, not modified, Leibniz rule
for which no local and translationally covariant representation
is proved to exist.
We expect that the nontrivial grading of the ``momentum''
operator may fill the gap of difficulties.

One more aspect to be mentioned here is that
the multiplication rule for the fields representing
the algebra should become modified correspondingly under the twisting.
Let us denote by $(\Xh,\ \product)$ and $(\Xh_0,\ \product_0)$
the spaces of fields which represent, respectively, the deformed
algebra $(\U(\A),\ \R)$ and the twisted algebra $(\U(\A)_0,\ \R_0)$.
Here $\product$ and $\product_0$ are the multiplication maps on the spaces
$\Xh$ and $\Xh_0$, respectively.
On products for $\Xh$, $h\in\U(\A)$ acts covariantly as we
have seen in the previous subsection:
$h\act\product(\Vp\ot\Vp')
=\product\bigl(\coproduct(h)\act(\Vp\ot\Vp')\bigr)$.
According to the theory of twisting (see appendix~\ref{app-hopf}),
$h\in\U(\A)_0$ can act covariantly on products
of fields for $\Xh_0$ only with the product
\begin{equation}
	\product_0 := \product\circ\cocycle\act
\end{equation}
(note that the twisting from $(\U(\A),\ \R)$ to $(\U(\A)_0,\ \R_0)$
is given with $\cocycle^{-1}$), as in
\begin{equation}
	h\act\product_0(\Vp\ot\Vp')
= \product_0\bigl(\coproduct_0(h)\act(\Vp\ot\Vp')\bigr).
\end{equation}
Suppose that this product $\product_0$ is ``commutative''
in the sense that
\begin{equation}
	\product_0\circ\Psi_0 = \product_0,\quad\IE\quad
	\product_0\circ\tau\circ\R_0\act = \product_0,
\end{equation}
which means commutative up to the statistics factor induced by $\R_0$.
This assumption would be natural because the twisted
algebra $(\U(\A)_0,\ \R_0)$ has the simple Hopf algebraic structure
which is symmetric under exchanging orders of any objects.
It turns out that then the multiplication $\product$ is again
commutative up to the nontrivial statistics $\Psi$
(thus noncommutative in the standard sense):
\begin{equation}
	\begin{aligned}
		\product\circ\Psi
		&=\product_0\circ\cocycle^{-1}\act\circ\tau\circ\R\act
		=\product_0\circ\tau\circ
			\bigl(\cocycle^{-1}_{21}\opprod\R\opprod\cocycle\bigr)\act\circ
			\cocycle^{-1}\act \\
		&=\product_0\circ\Psi_0\circ\cocycle^{-1}\act
		=\product_0\circ\cocycle^{-1}\act
		=\product.
	\end{aligned}
\label{mild-nc}
\end{equation}
This consequence in a way shows that multiplication rule
should incorporate the statistics in the obvious 
manner so that it becomes commutative up to the statistics.
When the statistics is itself nontrivial, this notion of
the commutativity up to the statistics may be expressed as
just a noncommutativity in the standard sense. In our case,
we have
\begin{equation}
	\Vp_{A_1\cdots A_p}(x)\cdot\Vp_{B_1\cdots B_q}(y)
	= (-1)^{pq}
		\Vp_{B_1\cdots B_q}\left(
			y + \sum_{i=1}^p a_{A_i}
		\right)\cdot
		\Vp_{A_1\cdots A_p}\left(
			x - \sum_{i=1}^q a_{B_i}
		\right).
\label{gen-statistics}
\end{equation}
We regard it as the consequence of either the lattice-deformed
statistics, or the mild noncommutativity,
and may use the notation $\Vp\ncprod\Vp'$ to emphasize its
noncommutative nature.


We finally recall that the space of fields on the lattice $\Xh$,
defined in \Ref{space-fields}, forms an algebra.
It actually forms a Hopf algebra in a natural way
\cite{Oeckl, Sasai-Sasakura}:
\begin{equation}
	\begin{array}{ll}
	 \product(\Vp_1\ot\Vp_2) = \Vp_1\cdot\Vp_2 &
		(\text{product}), \\
		\eta(1) = \fid &
		(\text{unit}), \\
		\coproduct(\Vp) = \Vp\ot\fid + \fid\ot\Vp,\quad
		\coproduct(\fid) = \fid\ot\fid &
		(\text{coproduct}), \\
		\counit(\Vp) = 0,\quad
		\counit(\fid) = 1 &
		(\text{counit}), \\
		S(\Vp) = -\Vp,\quad
		S(\fid) = \fid &
		(\text{antipode}),
	\end{array}
\end{equation}
where $\Vp\in X$.
This Hopf algebraic structure shouldn't be confused
with that of the symmetry operators $\U(\A)$ acting on $\Xh$.
In addition to these Hopf algebraic structure, the space
$\Xh$ has the braiding/shift structure $\Psi$ which obeys
the consistency conditions \Ref{braiding-prod} and
\Ref{braiding-id}. With the use of the braiding,
the Hopf algebraic structure is extended to the whole
field space $\Xh$; coproduct, counit, and antipode
of a product of two elementary fields $\Vp_1,\ \Vp_2\in X$
are defined by
\begin{equation}
	\begin{aligned}
		\coproduct(\Vp_1\cdot\Vp_2)
		&:= (\product\ot\product)\circ(\id\ot\Psi\ot\id)
		\bigl(\coproduct(\Vp_1)\ot\coproduct(\Vp_2)\bigr), \\
		\counit(\Vp_1\cdot\Vp_2)
		&:= \counit(\Vp_1)\counit(\Vp_2), \\
		S(\Vp_1\cdot\Vp_2)
		&:= \product\circ\Psi\bigl(S(\Vp_1)\ot S(\Vp_2)\bigr),
	\end{aligned}
\end{equation}
and generalized inductively to any products in $\Xh$.
One of the most crucial nature for this braiding
structure is that it must be covariant under the symmetry operations.
In fact we recall that the braiding structure is inevitable
only for the covariant consistency under the Hopf algebraic
symmetry: $a\act\circ\Psi = \Psi\circ a\act$, $a\in\U(\A)$.
With all these properties, the space $\Xh$ is called
a braided Hopf algebra, or, more precisely,
Hopf algebra in a braided category.
We thus claim that the link formalism naturally
treats the space of fields as a braided Hopf algebra
with a Hopf algebraic symmetry,
for which the general BQFT formalism can apply.
We now see this application in the next subsection.

\subsection{Perturbative Definition of Supersymmetry
on the Lattice as a Braided Quantum Field Theory}


Following the general theory of BQFT
given in \cite{Oeckl},
we are now constructing a lattice theory which has
the Hopf algebraic symmetry introduced in the previous
subsections. Before going to concrete examples, let us
here briefly review the general framework. The crucial
ingredient to define a quantum field theory is
the path integral. For defining a perturbation theory
it is enough to introduce it as a formal Gaussian
integral, such that the total functional derivative
under it is supposed to be zero. We therefore need
the functional derivative, which is defined as below.

We now introduce the functional derivative with respect to
$\Vp\in X$ as in
\begin{equation}
	\frac{\delta}{\delta\Vp(x)}\Vp(y) = \delta^D(x-y).
\end{equation}
Following more abstract definition in \cite{Oeckl},
we write this as
\begin{equation}
	\ev\left(\fd{\Vp(x)}\ot\Vp(y)\right)
	:= \frac{\delta}{\delta\Vp(x)}\Vp(y),
\end{equation}
introducing the evaluation map $\ev$.
It is a kind of natural contraction
of $X$ and $X^\ast$, where $X^\ast$ is the dual space to $X$
composed of $\delta/\delta\Vp$.
Similarly we might introduce the opposite one, a kind of 
completeness relation, as
\begin{equation}
	\coev(\lambda)
	:= \lambda\sum_x \Vp(x)\ot\fd{\Vp(x)}.
\end{equation}
These maps are characterized with the identities:
\begin{equation}
	(\ev\ot\id)(\id\ot\coev) = \id_{X^\ast},\qquad
	(\id\ot\ev)(\coev\ot\id) = \id_{X}.
\end{equation}
The functional derivative can be naturally extended to the one
which acts on the whole space of fields $\Xh$ as in the following
way.
On a product of two elementary fields $\Vp_1,\ \Vp_2\in X$,
the functional derivative
acts with the use of a braided Leibniz rule as in
\begin{equation}
	\fd{\Vp(x)}\bigl(
	\Vp_1(x_1)\cdot\Vp_2(x_2)
	\bigr)
	= \fd{\Vp(x)}\Vp_1(x_1)\cdot\Vp_2(x_2)
	+\cdot \left[\Psi^{-1}\left(
		\fd{\Vp(x)}\ot\Vp_1(x_1)
	\right)\left(\fid\ot\Vp_2(x_2)\right)\right].
\end{equation}
On products of more than three fields are extended inductively.
Needless to say, the derivative trivially commutes with
a constant field (see \Ref{braiding-id}), and
gives zero when it acts on a constant. More rigorous
definition of the functional derivative is given
in \cite{Oeckl, Sasai-Sasakura}.

Now we can introduce a Gaussian integration with the following
property:
\begin{equation}
	\int \fd{\Vp}\Bigl(\MC{O}[\Vp]e^{-S_0}\Bigr)
	= 0,\qquad
	\MC{O}[\Vp]\in\Xh,\quad
	\fd{\Vp}\in X^\ast,
\label{Gaussian}
\end{equation}
where $\exp(-S_0)\in\Xh$ is the corresponding Gaussian factor.
In the application to the field theory, $S_0$
is interpreted as the free part of the action.
Notice that this integration is formally understood
as the one which satisfies the property \Ref{Gaussian}
without referring to its real values. This way of
abstract definition is already enough to define
a perturbation theory and to compute correlation functions
with arbitrary order, since for such computations
only the ratio of the integral to another integral,
partition function, is needed (this is nothing
different from the path integral in a usual
field theory), and that ratio can be computed
only with these algebraic properties.

We now introduce a kind of propagator. Letting
\begin{equation}
	\fd{\Vp(x)}e^{-S_0}
	= -\gamma\left(\fd{\Vp(x)}\right)e^{-S_0},
\label{gamma}
\end{equation}
we define an object $\gamma: X^\ast\RA X$. More specifically,
it is given as
\begin{equation}
	\gamma\left(\fd{\Vp(x)}\right)
	= \fd{\Vp(x)}S_0,
\end{equation}
which roughly corresponds to the inverse propagator,
so that the propagator is in a way given as
$\gamma^{-1}$. This \naive argument can be justified
shortly.

The free $n$-point correlation function is now defined by
\begin{equation}
	Z^{(0)}_n(\alpha_n)
	:= \frac{\DS\int\alpha_n e^{-S_0}}{\DS\int e^{-S_0}},\qquad
	\alpha_n\in X^n.
\end{equation}
The superscript $(0)$ stands for the free theory.
In this definition, the denominator, denoted here
tentatively as $\Zz$, might be interpreted as
the free partition function, but in the general case
we don't have any definition to directly compute it
as mentioned above.
Still this definition is enough to calculate
the correlation functions of any order.
To see this argument, notice first that
\begin{equation}
	\alpha_n\Vp e^{-S_0}
	= \alpha_n \gamma\bigl(\gamma^{-1}(\Vp)\bigr)e^{-S_0}
	= -\alpha_n \gamma^{-1}(\Vp)\bigl(e^{-S_0}\bigr),
\end{equation}
where we have used the definition \Ref{gamma} and
the fact that $\gamma^{-1}(\Vp)\in X^\ast$ and so
is a functional derivative. We then find, using
the braided Leibniz rule, that
\begin{equation}
	-\alpha_n\gamma^{-1}(\Vp)\bigl(e^{-S_0}\bigr)
	= -\Bigl(
		\gamma^{-1}(\Vp^{\alpha_n})\bigl(
		\alpha_n^{\Vp}e^{-S_0}\bigr)
		-\gamma^{-1}(\Vp^{\alpha_n})\bigl(
		\alpha_n^{\Vp}\bigr) e^{-S_0}
	\Bigr)
\end{equation}
where we have denoted the ``shifted'' field
as $\Vp^{\alpha_n}$ and $\alpha_n^\Vp$, with the superscripts
implying the amount of shifts%
\footnote{This notational simplicity
can only apply to our present case for the specific braiding/shift
structure. The general expression with general braiding
$\Psi$ is given in \cite{Oeckl}.}. We thus find
\begin{equation}
	\begin{aligned}
	\int \alpha_n\Vp e^{-S_0}
	&= -\int
	\Bigl(
		\gamma^{-1}(\Vp^{\alpha_n})\bigl(
		\alpha_n^{-\Vp}e^{-S_0}\bigr)
		-\gamma^{-1}(\Vp^{\alpha_n})\bigl(
		\alpha_n^{\Vp}\bigr) e^{-S_0}
	\Bigr) \\
	&=\int\gamma^{-1}(\Vp^{\alpha_n})\bigl(
		\alpha_n^{\Vp}\bigr) e^{-S_0}.
	\end{aligned}
\end{equation}
In the second equality, the first term vanishes because
its a total derivative under the path integral.
We therefore obtain a basic formula
\begin{equation}
	\Zz(\alpha_n\Vp)
	= \Zz\bigl(\gamma^{-1}(\Vp^{\alpha_n})(\alpha_n^\Vp)\bigr).
\label{Wick-basic}
\end{equation}

For example, putting $\alpha = \vid\ (n=0)$ in
the formula above \Ref{Wick-basic}, it is clear that
\begin{equation}
	\Zz_1(\Vp) = 0.
\end{equation}
The simplest nontrivial example is given for $n=1$
by taking $\alpha_1 = \Vp_1(x_1)\in X$ and
$\Vp = \Vp_2(x_2)\in X$ in \Ref{Wick-basic}, so that
\begin{equation}
	\begin{aligned}
	\Zz_2\bigl(\Vp_1(x_1)\Vp_2(x_2)\bigr)
	&=\Zz_2\Bigl(\gamma^{-1}\bigl(\Vp_2(x_2+a_{\Vp_1})\bigr)
	\bigl(\Vp_1(x+a_{\Vp_2})\bigr)\Bigr) \\
	&=\gamma^{-1}\bigl(\Vp_2(x_2+a_{\Vp_1})\bigr)
	\bigl(\Vp_1(x+a_{\Vp_2})\bigr).
	\end{aligned}
\end{equation}
The other formulae can be computed inductively using
\Ref{Wick-basic}.
The general results are summarized as follows:
\begin{align}
	Z^{(0)}_2
	&= \ev\circ(\gamma^{-1}\ot\id)\circ\Psi, \\
	\Zz_{2n}
	&= (\Zz_2)^n\circ[2n-1]'_\Psi!!, \\
	\Zz_{2n+1}
	&= 0,
\end{align}
where
\begin{equation}
	\begin{aligned}
		{}[2n-1]'_\Psi!!
		&:= ([1]'_\Psi\ot\id^{2n-1})\circ([3]'_\Psi\ot\id^{2n-3})\circ
		\cdots\circ([2n-1]'_\Psi\ot\id), \\
		[n]'_\Psi
		&:= \id^n + \id^{n-2}\ot\Psi^{-1} + \cdots
		+ \Psi^{-1}_{1,n-1}.
	\end{aligned}
\end{equation}
These are the Wick's theorem in the BQFT formalism.

When an interaction is turned on, we can treat the theory
perturbatively. Let the action be $S=S_0 + \lambda \Sint$.
The $n$-point correlation function now reads
\begin{equation}
	\begin{aligned}
	Z_n(\alpha_n)
	&:= \frac{\DS\int\alpha_n e^{-S}}{\DS\int e^{-S}} \\
	&= \frac{\DS\int\alpha_n(1-\lambda\Sint+\cdots)e^{-S_0}}{%
	\DS\int(1-\lambda\Sint+\cdots)e^{-S_0}},
	\qquad
	\alpha_n\in X^n.
	\end{aligned}
\end{equation}
Dividing both the numerator and denominator by
the ``partition function'' $\Zz$, we find
\begin{equation}
	Z_n =
	\frac{\DS \Zz_n -\lambda\Zz_{n+k}\circ(\id^n\ot\Sint)
	+\frac{1}{2}\lambda^2\Zz_{n+2k}\circ(\id^n\ot\Sint\ot\Sint)
	+\cdots}{\DS%
	1-\lambda\Zz_k\circ\Sint
	+\frac{1}{2}\lambda^2\Zz_{2k}\circ(\Sint\ot\Sint)
	+\cdots},
\end{equation}
where $k$ is the order of the interaction $\Sint$,
i.e.\ $\Sint\in X^k$, and we have put a map
$\Sint:\MBB{C}\RA X^k$ with the abuse of notation.

Let us give an example to see the formalism above more explicitly.
We here consider $\MC{N}=(2,2)$ Wess--Zumino model in two dimensions
in the Dirac-\Kahler twisted basis.
Superalgebra is given as before
\begin{equation}
	\{\Ql, \Ql_\mu\} = i\del_{+\mu},\qquad
	\{\Qtl, \Ql_\mu\} = -i\epsilon_{\mu\nu}\del_{-\nu}.
\end{equation}
Bosonic fields include scalars
$\phi,\ \sigma$ and auxiliary fields
$\phit,\ \sigmat$, whereas fermionic fields are
$\psi,\ \psit, \psi_\mu$.
Supertransformations are given in Appendix~\ref{app-WZ}.
The action is given as
\begin{equation}
	\begin{aligned}
		S &= \sum_x\Bigl[
		(\dell_{+\mu}\sigma)(x-a\muhat)\fprod(\dell_{-\mu}\phi)(x)
		- \sigmat(x+a_1+a_2)\fprod\phit(x)\\
		&\phantom{=\sum_x\Bigl[}
		- i\psi(x-a)\fprod\dell_{-\mu}\psi_\mu(x)
		- i\epsilon_{\mu\nu}\psit(x-\at)\fprod\dell_{+\mu}\psi_\nu(x) \\
		&\phantom{=\sum_x\Bigl[}
		-\del_\phi W(x+a_1+a_2)\fprod\phit(x)
		-\del_\sigma V(x+a+\at)\fprod\sigmat(x) \\
		&\phantom{=\sum_x\Bigl[}
		+\del_\sigma^2 V(x+a+\at)\fprod\psi(x+\at)\fprod\psit(x)
		-i\epsilon^{\mu\nu}\del_\phi^2 W(x+a_1+a_2)
			\fprod\psi_\mu(x+a_\nu)\fprod\psi_\nu(x)
		\Bigr],
	\end{aligned}
\end{equation}
where $W$ and $V$ are potentials in the twisted basis.
The invariance of the action can be unambiguously
seen using the modified Leibniz rule
taking care of the specific ``staggered'' configurations
of arguments of the fields as well as of the mildly generalized
statistics \Ref{gen-statistics}.

\subsection{Ward--Takahashi Identities}

Here we follow \cite{Sasai-Sasakura}.
The invariance of the correlation functions can be
written as
\begin{equation}
	Z_n(a\act\chi) = \counit(a)Z_n(\chi),\qquad
	a\in\U(\A),\quad
	\chi\in X^n,
\label{WTI-gen}
\end{equation}
which is the Ward--Takahashi identity corresponding to
the Hopf algebraic symmetry $\U(\A)$.
Just as in the usual field theory, the invariance
of the correlation functions follows
from the invariance of the action. One obvious
difference from the usual case is that, with the nontrivial
braiding, the symmetry operators must act on the fields
in a manner consistent to the braiding structure.
In fact it is shown that the identity \Ref{WTI-gen}
follows when the following four conditions are satisfied
\cite{Sasai-Sasakura}:
\begin{enumerate}
\item Invariance of the free action:
\begin{equation}
	a\act\gamma^{-1}(\Vp) = \gamma^{-1}(a\act\Vp).
\end{equation}

\item Invariance of the interaction:
\begin{equation}
	a\act\Sint = \counit(a)\Sint.
\end{equation}

\item Covariance of the braiding:
\begin{equation}
	\Psi\bigl(a\act(X_1\ot X_2)\bigr)
	= a\act\Psi(X_1\ot X_2).
\end{equation}

\item Invariance of the delta function:
\begin{equation}
	\ev\bigl(a\act(X^\ast\ot X)\bigr)
	= \counit(a)\ev(X^\ast\ot X).
\end{equation}

\end{enumerate}

In our current application, the general formula
of Ward--Takahashi identity \Ref{WTI-gen} naturally gives
the correct identities on the lattice.
It is important that the general formula \Ref{WTI-gen}
can be proved unambiguously only using the algebraic
relations.

\subsection{Nonperturbative Definition?}


In this section, we first extracted the essential requirements
for the symmetry operators in the link formalism, concluding
that this symmetry is Hopf algebraic. Then we utilized
the general framework of BQFT formulated in \cite{Oeckl},
showing that supersymmetric theory on a lattice in the link 
formalism can be treated with a formal definition of path integral.
This path integral approach, however, only gives a
perturbative formulation in general, due to the lack of
explicit definition of the path integral.
As a field theory on a lattice, this situation
wouldn't be satisfactory at all, especially for the application
to numerical simulations. It is known in some cases
one can define a ``braided integral'' explicitly
\cite{Majid}. We might be able to apply such an approach to the current
problem to define a rigorous path integral on the lattice,
which, if possible, should give the nonperturbative
definition in this formulation based on the Hopf
algebraic symmetry. 
As we have shown in subsection 3.2, it is also crucial to 
accommodate the explicit representation of grading nature 
for the lattice momentum operator.

%% file: concl.tex
\section{Conclusion and Discussion}

We have shown how the link formalism
is treated as a field theory on a lattice with
deformed or modified algebraic symmetry. The deformation
of the algebra is indeed identified as the one
naturally treated in the framework of the Hopf algebra.
We showed this argument explicitly, defining
the corresponding Hopf algebraic structures of the supersymmetry algebra
for the link formalism. The modified Leibniz rule,
which is the crucial notion in the original link formalism,
was incorporated as the coproduct structure of the Hopf
algebra, whose consistency is assured with the other
relevant structures of the algebra.
The Hopf algebra introduced this way in fact turned out to
be a (quasi)triangular Hopf algebra, which has a nontrivial universal
$R$-matrix. When represented on the space of fields, this
quasitriangular structure inevitably induce a nontrivial statistics,
or a noncommutativity,
which is the key ingredient for the consistent representation.
With these algebraic descriptions, we could identify
the link formalism as a representation theory of a quasitriangular
Hopf algebra.
On the other hand, it is known that there is a general
scheme to construct a quantum field theory which has
a Hopf algebraic symmetry, called braided quantum field theory.
We applied this general
formulation to the link formalism. The construction is purely
algebraic. In particular, it defines a path integral
using only algebraic properties. One can show that
it still gives a well-defined perturbative
description of the theory, providing full methods
for calculating correlation functions in any order.
It also gives a concise formulation to derive
the possible Ward--Takahashi identities corresponding
to the Hopf algebraic symmetry.
We therefore realized the link formalism as a quantum field theory
which has the quasitriangular Hopf algebraic symmetry at least
in the perturbative sense.


From the consistency of the Hopf algebraic structure, it is 
required that the lattice momentum operator which is proportional to 
the difference operator should carry a grading compatible with the 
shifting nature of the difference operator. 
In this paper we have not given a concrete representation of 
this grading structure which may be needed to give an explicit 
nonperturbative definition of this formulation. We leave this 
issue for the future investigation. 

The algebraic inconsistency pointed out in \cite{Dutch}, which is 
connected with the ordering ambiguity of component fields when 
applying supersymmetry transformation, is solved by the introduction 
of braiding structure according to the notion of coproduct for 
the lattice super charges and the momentum operator in Hopf algebra. 


It is then important to ask the question how the continuum limit of 
this formulation is realized. If one can formulate the braided quantum 
field theory which respects the Hopf algebraic structure as a concrete 
representation for modified path integral, the twisted lattice 
supersymmetry will be kept in the continuum limit since the lattice 
twisted supersymmetry is exactly kept. 
As we have shown the lattice supersymmetry is kept in the perturbative 
level of braided quantum field theory. 
It is still nontrivial question how the symmetry is recovered 
even in the nonperturbative level. 
In any case we expect that fine tuning is not needed to keep the 
supersymmetry in the continuum limit if the formulation of deformed 
supersymmetry algebra is concretely constructed.  

In the formulation of orbifold construction of lattice field theories 
only a subset of lattice super charges in particular the nilpotent 
scalar super charge which corresponds to the shiftless charge in the 
link construction is exactly preserved on the lattice 
\cite{deconstruction, Unsal, twist-lattsusy, Sugino,
Takimi, Giedt, Damgaard-Matsuura, Unsal2, review-lat-SUSY}. 
The lattice super algebra in this case is identified as the same as 
the continuum twisted supersymmetry algebra. 
It was stressed that the super charges carrying a shift break the 
lattice super symmetry in the sense of the continuum twisted 
superalgebra~\cite{Damgaard-Matsuura, Unsal2}. 
Our claim in this paper is that these supercharges 
carrying the shift may break the continuum twisted supersymmetry but 
preserve exactly the Hopf algebraic supersymmetry. Thus in the link 
approach all the lattice super charges are claimed to preserve 
exactly in the framework of Hopf algebraic supersymmetry. 
The supersymmetry algebra is deformed from the continuum twisted 
supersymmetry to Hopf algebraic supersymmetry. 



We have not considered the gauge extension of deformed supersymmetry 
in this paper. It was pointed out that there is similar ordering 
ambiguity for the lattice super Yang-Mills formulation of link 
approach~\cite{Dutch}. We consider that this problem can be solved 
similar as non-gauge case by identifying the lattice supersymmetry 
with gauge symmetry of link approach as Hopf algebraic symmetry. 
There is, however, yet another problem in the gauge extension; 
the loss of the gauge invariance due to the link nature of the 
lattice super charges. A possible solution was proposed by 
introducing covariantly constant super parameters 
$\eta_A$~\cite{DKKN:3}: 
$$
\{\nabla_B,\eta_A\}=0
$$
where $\nabla_B$ is super covariant derivative. This is highly 
nontrivial relation in the sense that the fermionic parameter 
$\eta_A$ carrying a shift should carry an internal space-time 
dependence caused by the super covariant derivative $\nabla_B$ 
to keep the covariant constancy. Here we may consider that 
fermionic link variables are defined on the links of internal 
space-time. In other words the space-time distortion of internal 
space-time may compensate the required dependence of the fermionic 
parameter. There is a possibility that gravity may play a role 
in these questions. 


It has been pointed out that the breakdown of the Leibniz rule 
for the lattice difference operator is inevitable under 
reasonable assumptions for algebraic property on the 
lattice~\cite{No-go:Kato-Sakamoto-So}. 
Recent renormalization group analyses confirms this statement 
from different point of view~\cite{G-W Bruckmann}. 
In order to realize supersymmetry 
algebra which includes the momentum operator on the lattice it 
is most natural to introduce the difference operator in the 
lattice supersymmetry algebra. While the exact supersymmetry of 
continuum supersymmetry algebra was realized only for the 
nilpotent super charge which is the scalar part of the twisted 
supersymmetry algebra but does not include the crucial momentum 
dependence. We claim that the deformation of the Lie algebraic 
continuum supersymmetry to Hopf algebraic supersymmetry on the 
lattice is inevitable to accommodate the difference operator in 
the algebra. 

It is obviously very important to find concrete representation of 
the Hopf algebraic supersymmetry algebra on the lattice to obtain 
"modified" path integral definition of the QFT with this 
particular braiding structure. As we have already shown this 
type of mild noncommutativity with shifting nature may be well 
accommodated by a matrix formulation of lattice 
noncommutativity\cite{ADFKS,Bars-Minic}. This part of concrete 
proposal with the 
necessary formulation of graded momentum lattice operator will 
be given elsewhere. It would be also interesting to compare the 
formulation of the link approach with other noncommutative 
approach~\cite{Ambjorn:nclattice} and non-lattice 
formulations~\cite{non-lattice,Tsuchiya}.

\subsection*{Acknowledgments}

We are grateful for useful discussions with S. Arianos, T. Asakawa, 
P. H. Damgaard, A. Feo, H. B. Nielsen, Y. Sasai and N. Sasakura. 
This work is supported in part by Japanese Ministry of Education, 
Science, Sports and Culture under the grant No. 18540245 and INFN 
research fund. 
 

%% file: app-hopf.tex
\section{Brief Summary of Hopf Algebra}
\label{app-hopf}

Here we briefly list the axioms of Hopf algebra and some related
notions which are used in this article.
For rigorous and complete descriptions,
see, for example, \cite{Majid:book, Chari:book, Chaichian:book}.

\subsection{Hopf Algebra}

A \emph{Hopf algebra} over a field $k\ (=\MBB{C}\ \text{or}\ \MBB{R})$
is a vector space $H$ over $k$ which has the following
properties
\ref{axiom-algebra}, \ref{axiom-coalgebra}, \ref{axiom-bialgebra}
and \ref{axiom-antipode}.

\begin{enumerate}

\item \label{axiom-algebra}
$H$ is a \emph{unital associative algebra}, so that
\begin{itemize}

	\item it has a $k$-linear multiplication (or product) map%
	\footnote{In what follows we take, unless otherwise specified,
	$h,\ h_1,\ h_2,\ \cdots$, to be arbitrary elements of $H$.} 
	\begin{equation}
		\opprod: H\otimes H \RA H,\qquad
		\opprod(h_1\ot h_2) = h_1\opprod h_2,
	\end{equation}
	which is associative
	\begin{equation}
		\opprod\circ(\opprod\ot\id)=\opprod\circ(\id\ot\opprod),\IE
		(h_1\opprod h_2)\opprod h_3 = h_1\opprod (h_2\opprod h_3);
	\end{equation}

	\item it has unit element $\vid$ which satisfies
	$\vid\cdot h = h\cdot\vid = h$, whose existence can be
	formally expressed as the existence of a $k$-linear map
	\begin{equation}
		\eta:k\RA H,\qquad
		\eta(\lambda) = \lambda\vid,\ \lambda\in k.
	\end{equation}

\end{itemize}

\item \label{axiom-coalgebra}
$H$ is a \emph{coalgebra}. Namely,
\begin{itemize}

	\item it has a $k$-linear map called \emph{coproduct}:
	\begin{equation}
		\coproduct: H\RA H\ot H,\qquad
		\coproduct(h) = \sum_i h_{i(1)}\ot h_{i(2)},
		\quad h_{i(1)},\ h_{i(2)}\in H,
	\end{equation}
	which satisfies the \emph{coassociativity}%
	\footnote{We use below much simpler abbreviation
	$\coproduct(h)= h_{(1)}\ot h_{(2)}$ known as the Sweedler's
	notation.}
	\begin{equation}
		(\coproduct \ot\id)\circ\coproduct
		= (\id\ot\coproduct)\circ\coproduct,\IE
		h_{(1)(1)}\ot h_{(1)(2)}\ot h_{(2)}
		=
		h_{(1)}\ot h_{(2)(1)}\ot h_{(2)(2)};
	\end{equation}

	\item it has also another $k$-linear map called \emph{counit}
	\begin{equation}
		\counit: H\RA k,
	\end{equation}
	which obeys the relation
	\begin{equation}
		(\counit\ot\id)\circ\coproduct 
		= (\id\ot\counit)\circ\coproduct 
		= \id,\IE
		\counit\bigl(h_{(1)}\bigr) h_{(2)}
		= \counit\bigl(h_{(2)}\bigr) h_{(1)}
		= h.
	\end{equation}

\end{itemize}

\item \label{axiom-bialgebra}
These structures of algebra and coalgebra 
are compatible with each other. Namely,
\begin{itemize}

	\item the coproduct and the counit
	are both algebra maps:
	\begin{equation}
		\coproduct(h_1\cdot h_2) = \coproduct(h_1)\cdot\coproduct(h_2),
		\qquad
		\counit(h_1\cdot h_2) = \counit(h_1)\counit(h_2).
	\end{equation}

\end{itemize}

\item \label{axiom-antipode}
$H$ has one more map called \emph{antipode}:
\begin{itemize}

	\item it has a $k$-linear map
	\begin{equation}
		\antipode: H \RA H,
	\end{equation}
	which obeys the identity
	\begin{equation}
		\opprod(\antipode\ot\id)\circ\coproduct 
		= \opprod(\id\ot\antipode)\circ\coproduct
		= \eta\circ\counit,\IE
		\antipode\bigl(h_{(1)}\bigr)\opprod h_{(2)}
		=
		h_{(1)}\opprod\antipode\bigl(h_{(2)}\bigr)
		=
		\counit(h)\vid.
	\end{equation}

\end{itemize}
	
\end{enumerate}
If the $k$-linear space $H$ satisfies these properties
\ref{axiom-algebra}, \ref{axiom-coalgebra}, \ref{axiom-bialgebra}, but
not
\ref{axiom-antipode}, it is called a \emph{bialgebra}.

\subsection{Quasitriangular Structure}

A Hopf algebra $H$ is said to be \emph{quasitriangular}
if there exists an invertible element
$\R\in H\otimes H$ which satisfies
\begin{equation}
	\begin{gathered}
		\tau\circ\coproduct h
		= \R\cdot(\coproduct h)\cdot\R^{-1},\\ 
		(\coproduct\otimes\id)\R
		= \R_{13}\R_{23},\qquad
		(\id\otimes\coproduct)\R
		= \R_{13}\R_{12},
	\end{gathered}
\label{Rmatrix}
\end{equation}
where
\begin{equation}
	\begin{gathered}
		\R = \sum\R^{(1)}\otimes\R^{(2)},\\
		\R_{12} 
		= \sum\R^{(1)}\ot\R^{(2)}\ot\vid,\qquad
		\R_{13} 
		= \sum\R^{(1)}\ot\vid\ot\R^{(2)},\qquad
		\R_{23} 
		= \sum\vid\ot\R^{(1)}\ot\R^{(2)},
	\end{gathered}
\end{equation}
and
$\tau: H\otimes H \RA H\otimes H$ is the transposition map
\begin{equation}
	\tau(h_1\otimes h_2) = h_2\otimes h_1,\qquad h_1,\ h_2\in H.
	\label{transposition}
\end{equation}
The element $\R$, if exists, is
called the \emph{quasitriangular structure}
or \emph{universal R-matrix}.

If a quasitriangular structure $\R$ of a quasitriangular
Hopf algebra $H$ obeys further the following condition,
the Hopf algebra is said to be \emph{triangular}:
\begin{equation}
	\R_{21}\R = \vid\ot\vid,\IE
	\R_{21} = \R^{-1},\quad\text{where}\quad
	\R_{21} = \sum \R^{(2)}\ot\R^{(1)}.
\end{equation}

\subsection{Action on Algebras}

A (left) \emph{action} of a Hopf algebra $H$ on
an associative algebra $X$
is a representation $\rho:H\RA\Lin(X)$, where $\Lin(X)$ is
the algebra of linear maps on $X$, which satisfies the covariance
in the following sense
\begin{equation}
	\begin{cases}
		h\act (\Vp\cdot\Vp') = m\bigl(\coproduct(h)\act(\Vp\ot\Vp')\bigr), \\
		h\act\fid = \counit(h)\fid,
	\end{cases}
	\qquad
	\Vp,\ \Vp'\in X.
	\label{covariance}
\end{equation}
We have here introduced the notation $h\act\Vp:=\rho(h)(\Vp)$,
the product $m$ of $X$ with the abbreviation
$\Vp\cdot\Vp':=m(\Vp\ot\Vp')$,
and the unit $\fid\in X$.

\subsection{Braiding}

Let us consider a formal collection%
\footnote{The notion of braiding would be most suitably
defined in terms of category theory. Here instead
we just give a simple and intuitive description.}
of representation spaces ($\bs{1},\ X,\ Y,\ Z,\ \cdots$)
of a Hopf algebra $H$
together with the collection of tensor products of
the representation spaces ($\bs{1}\ot X\cong X\ot\bs{1}\cong X,\ 
X\ot Y,\ (X\ot Y)\ot Z \cong X\ot(Y\ot Z),\ \cdots$)
for which $H$ acts with the coproduct structure
($(\coproduct h)\act(\Vp\ot\chi),
\ h\in H,\ \Vp\in X,\ \chi\in Y$).
If there exists an invertible intertwiner (isomorphism)
\begin{equation}
	\Psi_{X,Y}:X\ot Y\RA Y\ot X,\qquad
	\Psi_{X,Y}\bigl(
		\coproduct(h)\act(\Vp\ot\chi)
	\bigr)
	=
	\coproduct(h)\act\Psi_{X,Y}(\Vp\ot\chi)
\label{intertwiner}
\end{equation}
with the properties
\begin{equation}
	\Psi_{X\ot Y, Z} = \Psi_{X,Z}\circ\Psi_{Y,Z},\qquad
	\Psi_{X, Y\ot Z} = \Psi_{X,Z}\circ\Psi_{X,Y},
\label{hexagon}
\end{equation}
it unambiguously relates the two representations on $X\ot Y$ and $Y\ot X$.
It should be compatible
with any maps which intertwine the representation spaces
as in
\begin{equation}
	\Psi_{Z,W}\circ(g_{XZ}\ot g_{YW})=(g_{YW}\ot g_{XZ})\circ\Psi_{X,Y},\qquad
	g_{XZ}: X\RA Z,\quad
	g_{YW}: Y\RA W.
\label{fanctoriality}
\end{equation}
We call this isomorphism $\Psi$ a \emph{braid}.
Strictly speaking, a braid should be such that
\begin{equation}
	\Psi\circ\Psi \neq \id,\quad\OR\quad
	\Psi_{X,Y}\neq\Psi_{Y,X}^{-1},
\label{nontriv-braid}
\end{equation}
which means there are two distinct ways in relating
$X\ot Y$ to $Y\ot X$.
It gives a nontrivial rule of exchanging factors of a tensor product,
and generalizes the statistics of the representation spaces.
If, on the other hand, it satisfies $\Psi\circ\Psi=\id$,
the isomorphism is more like a simple transposition 
and said to be symmetric.

When the Hopf algebra $H$ is quasitriangular,
we can express the braiding more explicitly using
the quasitriangular structure $\R$ of $H$
and the transposition map $\tau$ \Ref{transposition}
as in
\begin{equation}
	\Psi_{X,X'}(\Vp\ot\Vp')
	= \tau\circ\R\act(\Vp\ot\Vp'),\qquad
	\Vp\in X,\ \Vp'\in X'.
\label{braid-R}
\end{equation}
This indeed becomes an invertible intertwiner \Ref{intertwiner}
and satisfies the conditions \Ref{hexagon}
and \Ref{fanctoriality}.
We find that
\begin{equation*}
	\Psi_{X,X'}\circ\Psi_{X',X}
	= \tau\circ\R\act(\tau\circ\R\act)
	= \tau\bigl((\R\cdot\R_{21})\act\circ\tau\bigr).
\end{equation*}
Thus the condition \Ref{nontriv-braid}, that is for $\Psi$ to be
strictly braided, is equivalent to
\begin{equation}
	\R\cdot\R_{21}\neq\vid\ot\vid,
\end{equation}
namely that the universal $R$-matrix is really
quasitriangular. Equivalently, a symmetric isomorphism $\Psi$
corresponds to the triangular structure $\R\cdot\R_{21}=\vid\ot\vid$.

\subsection{Twist}

Let $H$ be a Hopf algebra.
An invertible element $\cocycle\in H\ot H$ is called a \emph{(2-)cocycle}
when it satisfies that
\begin{equation}
	(\cocycle\ot\vid)(\coproduct\ot\id)\cocycle
	=
	(\vid\ot\cocycle)(\id\ot\coproduct)\cocycle
	\qquad(\text{(2-)cocycle condition}).
\end{equation}
A cocycle $\cocycle$ is said to be \emph{counital} if%
\footnote{Here actually only one of the two conditions is suffice.}
\begin{equation}
	(\counit\ot\id)\cocycle = \vid\quad\AND\quad
	(\id\ot\counit)\cocycle = \vid\qquad
	(\text{counital condition}).
\end{equation}

For a quasitriangular Hopf algebra $(H,\ \R)$ 
and a counital 2-cocycle $\cocycle$,
there exists a new Hopf algebra $(H_\cocycle,\ \R_\cocycle)$
which has
\begin{itemize}
	\item the same algebra and counit as those for $(H,\ \R)$,

	\item coproduct: $\coproduct_\cocycle h = \cocycle(\coproduct h)\cocycle^{-1}$,

	\item antipode: $S_\cocycle h = U(S h)U^{-1}$, where
	$U = \cdot(\id\ot S)\cocycle$,
	$U^{-1} = \cdot(S\ot\id)\cocycle^{-1}$,

	\item quasitriangular structure:
	$\R_\cocycle = \cocycle_{21}\R\cocycle^{-1}$, where
	$\cocycle_{21} = \tau(\cocycle)$ with $\tau$ given
	in \Ref{transposition}.

\end{itemize}
The process obtaining the Hopf algebra $(H_\cocycle,\ \R_\cocycle)$
from the original one $(H,\ \R)$ is called a \emph{twist}
with the element $\cocycle$ called a twist element.
If $(H,\ \R)$ is triangular, so is $(H_\cocycle,\ \R_\cocycle)$.

When a Hopf algebra $H$ acts on an associative algebra $X$
covariantly as in \Ref{covariance},
the twisted Hopf algebra $H_\cocycle$ with a twist element $\cocycle$
acts covariantly on a new algebra $X_\cocycle$ with a new product
\begin{equation}
	\Vp\ncprod\Vp' := m\circ\cocycle^{-1}\act(\Vp\ot\Vp')
\end{equation}
and with the same unit.
The new product $\ncprod$ is associative and
in general noncommutative even if the original product $\cdot$
is commutative.

%% file: app-wz.tex

\section{$\MC{N}=(2,2)$ Wess--Zumino Model in Two Dimensions}
\label{app-WZ}

We list here the explicit supertransformation formulae
for $\MC{N}=(2,2)$ Wess--Zumino model in two dimensions.
The superalgebra is
	\begin{equation*}
			\begin{gathered}
		\{\Ql, \Ql_\mu\} = \Pl_{+\mu},\quad
			\{\Qtl, \Ql_\mu\} = -\epsilon_{\mu\nu}\Pl_{-\nu}, \\
			(\Pl_{\pm\mu} := i\del_{\pm\mu}),
			\end{gathered}
	\end{equation*}
with the other commutators just vanishing.
The field contents are
$\{\phi,\ \sigma,\ \psi,\ \psi_\mu,\ \psit,\ \phit,\ \sigmat\}$,
for which the supertransformations are as follows:
	\begin{alignat*}{3}
	\Ql\phi &= 0, &\quad \Ql_\mu\phi &= \psi_\mu, &\quad \Qtl\phi &= 0, \\
	\Ql\psi_\nu &= i\del_{+\nu}\phi, &\quad
	\Ql_\mu\psi_\nu &= -\epsilon_{\mu\nu}\phit, &\quad
	\Qtl\psi_\nu &= -i\epsilon_{\nu\mu}\del_{-\mu}\phi, \\
	\Ql\phit &= -i\epsilon_{\mu\nu}\del_{+\mu}\psi_\nu, &\quad
	\Ql_\mu\phit &= 0, &\quad
	\Qtl\phit &= i\del_{-\mu}\psi_\mu, \\
	\Ql\sigma &= -\psi, &\quad \Ql_\mu\sigma &= 0, &\quad
	\Qtl\sigma &= -\psit, \\
	\Ql\psi &= 0, &\quad \Ql_\mu\psi &= -i\del_{+\mu}\sigma, &\quad
	\Qtl\psi &= -\sigmat, \\
	\Ql\psit &= \sigmat, &\quad
	\Ql_\mu\psit &= i\epsilon_{\mu\nu}\del_{-\nu}\sigma, &\quad
	\Qtl\psit &= 0, \\
	\Ql\sigmat &= 0, &\quad
	\Ql_\mu\sigmat &= \begin{aligned}[t]
	&i\epsilon_{\mu\nu}\del_{-\nu}\psi \\&+i\del_{+\mu}\psit,
	\end{aligned} &\quad
	\Qtl\phit &= 0.
	\end{alignat*}

%% file: bib.tex
%

%% file: main.bbl
\begin{thebibliography}{99}



\bibitem{No-Go}
H.~B.~Nielsen and M.~Ninomiya,
``Absence of Neutrinos on a Lattice I. Proof by 
Homotopy Theory"
Nucl.\ Phys. {\bf B185} (1981) 20; erratum:
{\bf B195} (1981) 541; ``Absence of Neutrinos on a 
Lattice II. Intuitive Topology Proof", 
Nucl.\ Phys. {\bf B193} (1981) 173. 

\bibitem{Anomaly}
L.~H.~Karsten and J.~Smit,
``Lattice Fermions: Species Doubling, Chiral Invariance 
and the Triangle Anomaly", 
Nucl.\ Phys. {\bf B183} (1981) 2649.

\bibitem{Gins-Wil}
P.~H.~Ginsparg and K.~G.~Wilson,
``A Remnant of Chiral Symmetry on the Lattice",
Phys.\ Rev. {\bf D25} (1982) 2649.

\bibitem{Domai-Wall}
D.~B.~Kaplan,
``A Method for Simulating Chiral Fermions on the Lattice",
Phys.\ Lett. {\bf B288} (1992) 342.

\bibitem{chiral-fermion}
P.~Hasenfratz, 
``Prospects for perfect actions", 
Nucl.\ Phys.{\bf(Proc. Suppl.) 63A-C} (1998) 53. \\
H.~Neuberger,
``More about exactly massless quarks on the lattice",
Phys.\ Lett. {\bf B427} (1998) 353. \\
M.~L\"uscher,
``Exact chiral symmetry on the lattice and the Ginsparg-Wilson 
relation, 
Phys.\ Lett. {\bf B428} (1998)342. 


\bibitem{DKKN:1}
A.~D'Adda, I.~Kanamori, N.~Kawamoto and K.~Nagata,
``Twisted superspace on a lattice,''
Nucl.\ Phys.\  B {\bf 707} (2005) 100
[arXiv:hep-lat/0406029];
``Twisted N=2 exact SUSY on the lattice for BF and Wess-Zumino,''
Nucl.\ Phys.\ Proc.\ Suppl.\  {\bf 140} (2005) 754
[arXiv:hep-lat/0409092];
``N=D=2 twisted supersymmetry on a lattice,''
Nucl.\ Phys.\ Proc.\ Suppl.\  {\bf 140} (2005) 757.


\bibitem{DKKN:2}
A.~D'Adda, I.~Kanamori, N.~Kawamoto and K.~Nagata,
``Exact extended supersymmetry on a lattice: Twisted N = 2 super Yang-Mills
in two dimensions,''
Phys.\ Lett.\  B {\bf 633} (2006) 645
[arXiv:hep-lat/0507029].
\bibitem{DKKN:3}
A.~D'Adda, I.~Kanamori, N.~Kawamoto and K.~Nagata,
``Exact Extended Supersymmetry on a Lattice: Twisted N=4 Super Yang-Mills in
Three Dimensions,''
Nucl.\ Phys.\  B {\bf 798}, 168 (2008)
[arXiv:0707.3533 [hep-lat]];
``Lattice Formulation of the N=4 D=3 Twisted Super Yang-Mills,''
PoS {\bf LAT2007} (2007) 271
[arXiv:0709.0722 [hep-lat]].
\bibitem{Catt:1}
S.~Catterall, 
``A Geometrical approach to N=2 super Yang-Mills theory on the two 
dimensional lattice.", 
JHEP0411:006 (2004); 
``Dirac-K\"ahler fermions and exact lattice supersymmetry", 
PoS{\bf LAT2005} (2006) 006 
[arXiv:0509136[hep-lat]].


\bibitem{Kawamoto:DK}
N.~Kawamoto and T.~Tsukioka,
``N = 2 supersymmetric model with Dirac-K\"ahler fermions from generalized
gauge theory in two dimensions,''
Phys.\ Rev.\  D {\bfseries 61} (2000) 105009
[arXiv:hep-th/9905222].
\\
J.~Kato, N.~Kawamoto and Y.~Uchida,
``Twisted superspace for N = D = 2 super BF and Yang-Mills with
Dirac-K\"ahler fermion mechanism,''
Int.\ J.\ Mod.\ Phys.\  A {\bfseries 19} (2004) 2149
[arXiv:hep-th/0310242].
\\
J.~Kato, N.~Kawamoto and A.~Miyake,
``N = 4 twisted superspace from Dirac-K\"ahler twist and off-shell SUSY
invariant actions in four dimensions,''
Nucl.\ Phys.\  B {\bfseries 721} (2005) 229
[arXiv:hep-th/0502119].
\\
J.~Kato and A.~Miyake,
``Topological hypermultiplet on N = 2 twisted superspace in four
dimensions,''
Mod.\ Phys.\ Lett.\  A {\bfseries 21} (2006) 2569
[arXiv:hep-th/0512269];
%
``Vafa-Witten Theory on N=2 and N=4 Twisted Superspace in Four Dimensions,''
arXiv:0808.2538 [hep-th].
\\
J.~Saito,
``Superspace Formulation of $N=4$ Super Yang-Mills
Theory with a Central Charge,''
Soryushiron Kenkyu (Kyoto) 111 (2005) 117,
[arXive:hep-th/0512226].

\bibitem{Marcus}
N.~Marcus,
``The Other topological twisting of N=4 Yang-Mills",
Nucl.\ Phys.\ {\bf B452} (1995) 331,
[arXiv:hep-th/9506002].



\bibitem{Nagata:CS}
K.~Nagata,
``On the Continuum and Lattice Formulations of N=4 D=3 Twisted Super
Yang-Mills,''
JHEP {\bf 0801} (2008) 041
[arXiv:0710.5689 [hep-th]].
\\
K.~Nagata and Y.~S.~Wu,
``Twisted SUSY Invariant Formulation of Chern-Simons Gauge Theory on a
Lattice,''
Phys.\ Rev.\  D {\bf 78} (2008) 065002
[arXiv:0803.4339 [hep-lat]].


\bibitem{Nagata:nclattice}
K.~Nagata,
``Exact Lattice Supersymmetry at Large N,''
JHEP {\bf 0810} (2008) 036
[arXiv:0805.4235 [hep-lat]].
\bibitem{old-Dirac-Kahler}
S.~Elitzur, E.~Rabinovici and A.~Schwimmer, 
``Supersymmetric Models On The Lattice''
Phys. Lett. {\bf B 199} (1982) 165.
\\
N.~Sakai and M.~Sakamoto,
``Lattice Supersymmetry And The Nicolai Mapping''
Nucl.\ Phys. {\bf B229} (1983) 173.
\\
V.~A.~Kostelecky, J.~M.~Rabin, 
``Supersymmetry on a superlattice", 
J.\ Math.\ Phys.\ {\bf 25} (1984) 2744. 
\\
D.~M.~Scott, 
``Lattices, supersymmetry, and K\"ahler fermions", 
J.\ Phys.\ {\bf A17} (1984) 1123.  
\\
H.~Aratyn, A.~H.~Zimerman, 
``Lattice supersymmetry for N=4 Yang-Mills model", 
J.\ Phys.\ {\bf A18} (1985)L487.

\bibitem{Becher-Joos}
P.~Becher and H.~Joos,
``The Dirac-K\"ahler equation and fermions on the lattice",
Zeit.\ Phys.\ {\bf C15} (1982) 343.
\\
T.~Banks and P.~Windy, 
``Supersymmetric lattice theories", 
Nucl.\ Phys.\ {\bf B198} (1982) 226,
\\
J.~M.~Rabin,
``Homology theory of lattice fermion doubling", 
Nucl.\ Phys.\ {\bf B201} (1982) 315.

\bibitem{Kanamori-Kawamoto}
I.~Kanamori and N.~Kawamoto,
``Dirac-K\"ahler fermion from Clifford product with noncommutative
differential form on a lattice,''
Int.\ J.\ Mod.\ Phys.\  A {\bf 19} (2004) 695
[arXiv:hep-th/0305094];
%
``Dirac-K\"ahler fermion with noncommutative differential forms on a
lattice,''
Nucl.\ Phys.\ Proc.\ Suppl.\  {\bf 129} (2004) 877
[arXiv:hep-lat/0309120].

%
%
%
%
\bibitem{Pauli}
W.~Pauli,
``The Connection Between Spin and Statistics",
Phys.\ Rev.\ {\bf 58} (1940) 716.
\bibitem{Dondi-Nicolai}
P.~H.~Dondi and H.~Nicolai,
``Lattice Supersymmetry,''
Nuovo Cim.\  A {\bf 41} (1977) 1.
\bibitem{Montvay}
I.~Montvay,
``Supersymmetric Yang-Mills theory on the lattice",
Int.\ J.\ Mod.\ Phys.\ {\bf A17} (20002) 2377.
\bibitem{Feo}
A. Feo,
``Supersymmetry on the lattice,''
Nucl. Phys. Proc. Suppl. {\bf 119} (2003) 198
[{\tt hep-lat/0112052}];
``Supersymmetry on the lattice: Where do predictions and results stand?''
[{\tt hep-lat/0311037}], and references therein;
``Predictions and recent results in SUSY on the lattice,''
  Mod.\ Phys.\ Lett.\  A {\bf 19} (2004) 2387
  [arXiv:hep-lat/0410012].




\bibitem{other-lattsusy1}
K.~Fujikawa and M.~Ishibashi,
``Lattice chiral symmetry and the Wess-Zumino model,''
Nucl.\ Phys. {\bf B622} (2002) 115 [{\tt hep-th/0109156}],
Phys.\ Lett. {\bf B528} (2002) 295 [{\tt hep-lat/0112050}].
\\
Y.~Kikukawa and Y.~Nakayama,
``Nicolai mapping vs. exact chiral symmetry on the lattice,''
Phys.\ Rev. {\bf D66} (2002) 094508
[{\tt hep-lat/0207013}].
K.~Fujikawa,
``N = 2 Wess-Zumino model on the d = 2 Euclidean lattice,''
Phys.\ Rev. {\bf D66} (2002) 074510
[{\tt hep-lat/0208015}].
\\
M.~Bonini and A.~Feo,
``Wess-Zumino model with exact supersymmetry on the lattice,''
JHEP {\bf 0409} (2004) 011
[{\tt hep-lat/0402034}].
\\
J.~W.~Elliott and G.~D.~Moore,
  ``Three dimensional N = 2 supersymmetry on the lattice,''
		JHEP {\bf 0511}, 010 (2005)
		[{\tt hep-lat/0509032}].
\bibitem{other-lattsusy2}
K.~Itoh, M.~Kato, H.~Sawanaka, H.~So and N.~Ukita,
``Novel approach to super Yang-Mills theory on lattice: Exact fermionic 
symmetry and 'Ichimatsu' pattern,''
JHEP {\bf 0302} (2003) 033
[{\tt hep-lat/0210049}],
``Towards the super Yang-Mills theory on the lattice,''
Prog.\ Theor.\ Phys.\  {\bf 108} (2002) 363
[{\tt hep-lat/0112052}].
\bibitem{Japanese-recent}
H.~Suzuki and Y.~Taniguchi,
``Two-dimensional N=(2,2) super Yang-Mills theory on the lattice via
dimensional reduction,''
JHEP {\bf 0510}, 082 (2005)
[{\tt hep-lat/0507019}].
\\
H.~Suzuki,
``Two-dimensional $\mathcal{N}=(2,2)$ super Yang-Mills theory on computer,''
arXiv:0706.1392 [hep-lat].
\\
Y.~Kikukawa and H.~Suzuki,
``A local formulation of lattice Wess-Zumino model with exact U(1)R
symmetry,''
JHEP {\bf 0502} (2005) 012
[{\tt hep-lat/0412042}].
\\
M.~Harada and S.~Pinsky,
``N = 1 super Yang-Mills on a (3+1) dimensional transverse lattice with  one
exact supersymmetry,''
Phys.\ Rev.\ D {\bf 71} (2005) 065013
[{\tt hep-lat/0411024}].

\bibitem{Fujikawa:Leibniz}
K.~Fujikawa,
``Supersymmetry on the lattice and the Leibniz rule,''
Nucl.\ Phys.\  B {\bf 636} (2002) 80
[arXiv:hep-th/0205095].
\bibitem{D-W-lattsusy}
J.~Nishimura,
``Four-dimensional N = 1 supersymmetric Yang-Mills theory on the lattice  
without fine-tuning'',
Phys.\ Lett. {\bf B406} (1997) 215
[{\tt hep-lat/9701013}].
\\
N.~Maru and J.~Nishimura,
``Lattice formulation of supersymmetric Yang-Mills theories without 
fine-tuning'',
Int.\ J.\ Mod.\ Phys. {\bf A13} (1998) 2841
[{\tt hep-th/9705152}].
\\
H.~Neuberger,
``Vector like gauge theories with almost massless fermions on the  lattice''
Phys.\ Rev. {\bf D57} (1998) 5417
[{\tt hep-lat/9710089}].
\\
D.~B.~Kaplan and M.~Schmaltz,
``Supersymmetric Yang-Mills theories from domain wall fermions''
Chin.\ J.\ Phys.\  {\bf 38} (2000) 543
[{\tt hep-lat/0002030}].
\\
G.~T.~Fleming, J.~B.~Kogut and P.~M.~Vranas,
``Super Yang-Mills on the lattice with domain wall fermions''
Phys.\ Rev. {\bf D64} (2001) 034510
[{\tt hep-lat/0008009}].
\bibitem{Igarashi-So}
Y.~Igarashi, H.~So and N.~Ukita, 
``Ginsparg-Wilson relation and lattice chiral symmetry in 
fermionic interacting theories", 
Phys.\ Lett.\ {\bf B535} (2002) 363
[{\tt hep-lat/0203019}].
\\
G.~Bergner, T.~Kaestner, S.~Uhlmann, and A.~Wipf, 
``Low-dimensional supersymmetric lattice models", 
Annals Phys.\ {\bf 323} (2008) 946 
[{\tt arXiv:0705.2212}].
\\
T.~Kastner, G.~Bergner, S.~Uhlmann, A.~Wipf, and C.~Wozar, 
``Two-Dimensional Wess-Zumino Models at Intermediate Couplings", 
Phys.\ Rev.\ {\bf D78} (2008) 095001 
[{\tt arXiv:0807.1905}].
\\
F.~Synatschke, G.~Bergner, H.~Gies, and A.~Wipf, 
``Flow Equation for Supersymmetric Quantum Mechanics", 
[{\tt arXiv:0809.4396v2}].
\bibitem{G-W Bruckmann}
G.~Bergner, F.~Bruckmann, and J.~M.~Pawlowski, 
``Generalising the Ginsparg-Wilson relation: 
Lattice supersymmetry from blocking transformations", 
[{\tt arXiv:0807.1110}].

\bibitem{deconstruction}
D.~B.~Kaplan, E.~Katz and M.~Unsal,
``Supersymmetry on a spatial lattice,''
JHEP {\bf 0305} (2003) 037
[arXiv:hep-lat/0206019].
\\
D.~B.~Kaplan,
``Lattice supersymmetry,''
Nucl.\ Phys.\ Proc.\ Suppl.\  {\bf 119} (2003) 900
[arXiv:hep-lat/0208046].
\\
A.~G.~Cohen, D.~B.~Kaplan, E.~Katz and M.~Unsal,
``Supersymmetry on a Euclidean spacetime lattice. I: A target theory with
four supercharges,''
JHEP {\bf 0308} (2003) 024
[arXiv:hep-lat/0302017];
%
``Supersymmetry on a Euclidean spacetime lattice. II: Target theories  with
eight supercharges,''
JHEP {\bf 0312} (2003) 031
[arXiv:hep-lat/0307012].
\\
D.~B.~Kaplan,
``Recent developments in lattice supersymmetry,''
Nucl.\ Phys.\ Proc.\ Suppl.\  {\bf 129} (2004) 109
[arXiv:hep-lat/0309099].
\\
D.~B.~Kaplan and M.~Unsal,
``A Euclidean lattice construction of supersymmetric Yang-Mills theories
with sixteen supercharges,''
JHEP {\bf 0509} (2005) 042
[arXiv:hep-lat/0503039].
\\
M.~G.~Endres and D.~B.~Kaplan,
``Lattice formulation of (2,2) supersymmetric gauge theories with matter
fields,''
JHEP {\bf 0610} (2006) 076
[arXiv:hep-lat/0604012].


\bibitem{Unsal}
M.~Unsal,
``Compact gauge fields for supersymmetric lattices,''
JHEP {\bf 0511}, 013 (2005)
[arXiv:hep-lat/0504016];
%
``Supersymmetric deformations of type IIB matrix model as matrix
regularization of N = 4 SYM,''
JHEP {\bf 0604}, 002 (2006)
[arXiv:hep-th/0510004];
%
``Twisted supersymmetric gauge theories and orbifold lattices,''
JHEP {\bf 0610}, 089 (2006)
[arXiv:hep-th/0603046].
%

\bibitem{twist-lattsusy}
S.~Catterall and S.~Karamov,
``Exact lattice supersymmetry: the two-dimensional N = 2 Wess-Zumino  
model,''
Phys.\ Rev. {\bf D65} (2002) 094501
[{\tt hep-lat/0108024}];
``A lattice study of the two-dimensional Wess Zumino model,''
Phys.\ Rev. {\bf D68} (2003) 014503
[{\tt hep-lat/0305002}].
\\
S.~Catterall,
``Lattice supersymmetry and topological field theory,''
JHEP {\bf 0305} (2003) 038
[{\tt hep-lat/0301028}].
\\
S.~Catterall and S.~Ghadab,
``Lattice sigma models with exact supersymmetry,''
JHEP {\bf 0405} (2004) 044
[{\tt hep-lat/0311042}],
``Twisted supersymmetric sigma model on the lattice,''
JHEP {\bf 0610}, 063 (2006)
[{\tt hep-lat/0607010}].
\\
S.~Catterall,
``A geometrical approach to N = 2 super Yang-Mills theory on the two
dimensional lattice,''
JHEP {\bf 0411} (2004) 006
[{\tt hep-lat/0410052}];
%
``LATTICE FORMULATION OF N=4 SUPER YANG-MILLS THEORY,''
JHEP {\bf 0506}, 027 (2005)
[{\tt hep-lat/0503036}];
``Simulations of ${\cal N}=2$ super Yang-Mills theory in two dimensions,''
JHEP {\bf 0603} (2006) 032
[{\tt hep-lat/0602004}];
JHEP {\bf 0704}, 015 (2007)
[{\tt hep-lat/0612008}].
\\
S.~Catterall and T.~Wiseman,
``Towards lattice simulation of the gauge theory duals to black holes and hot
strings,''
arXiv:0706.3518 [hep-lat].

\bibitem{Sugino}
F.~Sugino,
``A lattice formulation of super Yang-Mills theories with 
exact supersymmetry,''
JHEP {\bf 0401} (2004) 015 [{\tt hep-lat/0311021}];
``Super Yang-Mills theories on the two-dimensional lattice with 
exact supersymmetry,''
JHEP {\bf 0403} (2004) 067 [{\tt hep-lat/0401017}];
``Various super Yang-Mills theories with exact supersymmetry on the 
lattice,''
JHEP {\bf 0501} (2005) 016 [{\tt hep-lat/0410035}];
``Two-Dimensional Compact N=(2,2) Lattice Super Yang-Mills Theory 
with Exact Supersymmetry,''
Phys.Lett. {\bf B635} (2006) 218
[{\tt hep-lat/0601024}].

\bibitem{Takimi}
T.~Onogi and T.~Takimi,
Phys.\ Rev.\  D {\bf 72} (2005) 074504
[arXiv:hep-lat/0506014].
\\
K.~Ohta and T.~Takimi,
``Lattice formulation of two dimensional topological field theory,''
Prog.\ Theor.\ Phys.\  {\bf 117}, 317 (2007)
[{\tt hep-lat/0611011}].
\\
T.~Takimi,
``Relationship between various supersymmetric lattice models,''
arXiv:0705.3831 [hep-lat].
\bibitem{Giedt}
J.~Giedt,
``Advances and applications of lattice supersymmetry",
PoS{\bf LAT2006} 008,2006
[{\tt hep-lat/0701006}];
``R-symmetry in the q-exact (2,2) 2d lattice Wess-Zumino model",
Nucl.\ Phys.\ {\bf B726} (20005) 210
[{\tt hep-lat/0507016}];
``Deconstruction and other approaches to supersymmetric lattice field
theories".
Int.\ J.\ Mod.\ Phys.\ {\bf A21} (2006) 3039
[{\tt hep-lat/0602007}].


\bibitem{Damgaard-Matsuura}
P.~H.~Damgaard and S.~Matsuura,
``Classification of Supersymmetric Lattice Gauge Theories by Orbifolding,''
JHEP {\bf 0707} (2007) 051
[arXiv:0704.2696 [hep-lat]];
%
``Relations among Supersymmetric Lattice Gauge Theories via Orbifolding,''
JHEP {\bf 0708} (2007) 087
[arXiv:0706.3007 [hep-lat]];
%
``Lattice Supersymmetry: Equivalence between the Link Approach and
Orbifolding,''
JHEP {\bf 0709} (2007) 097
[arXiv:0708.4129 [hep-lat]];
%
``Geometry of Orbifolded Supersymmetric Lattice Gauge Theories,''
Phys.\ Lett.\  B {\bf 661} (2008) 52
[arXiv:0801.2936 [hep-th]].

\bibitem{Unsal2}
M.~Unsal,
``Deformed matrix models, supersymmetric lattice twists and N=1/4
supersymmetry,''
arXiv:0809.3216 [hep-lat].

\bibitem{review-lat-SUSY}
S.~Catterall, D.~Kaplan, and M.~Unsal, 
``Exact lattice supersymmetry",
[{\tt arXiv:0903.4881v1}].

\bibitem{accidental symmetry}
D.~B.~Kaplan,
``Dynamical Generation Of Supersymmetry''
Phys.\ Lett. {\bf B136} (1984) 162.




\bibitem{Kanamori}
  H.~Fukaya, I.~Kanamori, H.~Suzuki, M.~Hayakawa and T.~Takimi,
  ``Note on massless bosonic states in two-dimensional field theories,''
  Prog.\ Theor.\ Phys.\  {\bf 116}, 1117 (2007)
  [arXiv:hep-th/0609049].
\\
  H.~Fukaya, I.~Kanamori, H.~Suzuki and T.~Takimi,
  ``Numerical results of two-dimensional N=(2,2) super Yang-Mills theory,''
  PoS {\bf LAT2007}, 264 (2007)
  [arXiv:0709.4076 [hep-lat]].
\\
  I.~Kanamori, H.~Suzuki and F.~Sugino,
  ``Euclidean lattice simulation for the dynamical supersymmetry breaking,''
  Phys.\ Rev.\  D {\bf 77}, 091502 (2008)
  [arXiv:0711.2099 [hep-lat]];
%
  ``Observing dynamical supersymmetry breaking with euclidean lattice
  simulations,''
  Prog.\ Theor.\ Phys.\  {\bf 119}, 797 (2008)
  [arXiv:0711.2132 [hep-lat]].
\\
  I.~Kanamori,
  ``Observing dynamical SUSY breaking with lattice simulation,''
  AIP Conf.\ Proc.\  {\bf 1078}, 423 (2009)
  [arXiv:0809.0646 [hep-lat]];
%
  ``RHMC simulation of two-dimensional N=(2,2) super Yang-Mills with exact
  supersymmetry,''
  arXiv:0809.0655 [hep-lat].
\\
  I.~Kanamori and H.~Suzuki,
  ``Restoration of supersymmetry on the lattice: Two-dimensional
  $\mathcal{N}=(2,2)$ supersymmetric Yang-Mills theory,''
  Nucl.\ Phys.\  B {\bf 811}, 420 (2009)
  [arXiv:0809.2856 [hep-lat]];
  ``Some physics of the two-dimensional $\mathcal{N}=(2,2)$ supersymmetric
  Yang-Mills theory: Lattice Monte Carlo study,''
  Phys.\ Lett.\  B {\bf 672}, 307 (2009)
  [arXiv:0811.2851 [hep-lat]].
\\
  I.~Kanamori,
  ``Vacuum energy of two-dimensional N=(2,2) super Yang-Mills theory,''
  arXiv:0902.2876 [hep-lat].

\bibitem{Witten:tft}
E.~Witten,
``Topological Quantum Field Theory,''
Commun.\ Math.\ Phys.\  {\bf 117} (1988) 353;
%
``Topological Sigma Models,''
Commun.\ Math.\ Phys.\  {\bf 118} (1988) 411.

\bibitem{ADFKS}
S.~Arianos, A.~D'Adda, N.~Kawamoto and J.~Saito,
``Lattice supersymmetry in 1D with two supercharges,''
PoS {\bf LATTICE2007} (2006) 259
[arXiv:0710.0487 [hep-lat]].
\\
S.~Arianos, A.~D'Adda, A.~Feo, N.~Kawamoto and J.~Saito,
``Matrix formulation of superspace on 1D lattice with two supercharges,''
arXiv:0806.0686 [hep-lat].

\bibitem{Oeckl}
R.~Oeckl,
``Braided quantum field theory,''
Commun.\ Math.\ Phys.\  {\bf 217} (2001) 451
[arXiv:hep-th/9906225];
%
``Introduction To Braided Quantum Field Theory,''
Int.\ J.\ Mod.\ Phys.\  B {\bf 14} (2000) 2461.

\bibitem{Dutch}
F.~Bruckmann and M.~de Kok,
``Noncommutativity approach to supersymmetry on the lattice: SUSY quantum
mechanics and an inconsistency,''
Phys.\ Rev.\  D {\bf 73} (2006) 074511
[arXiv:hep-lat/0603003].
\\
F.~Bruckmann, S.~Catterall and M.~de Kok,
``A critique of the link approach to exact lattice supersymmetry,''
Phys.\ Rev.\  D {\bf 75} (2007) 045016
[arXiv:hep-lat/0611001].
\bibitem{No-go:Kato-Sakamoto-So}
M.~Kato, M.~Sakamoto and H.~So,
``Taming the Leibniz Rule on the Lattice,''
JHEP {\bf 0805}, 057 (2008)
[arXiv:0803.3121 [hep-lat]];
%
``Leibniz rule and exact supersymmetry on lattice: A case of supersymmetrical
quantum mechanics,''
PoS {\bf LAT2005} (2006) 274
[arXiv:hep-lat/0509149];
%
``No-Go Theorem of Leibniz Rule and Supersymmetry on the Lattice,''
arXiv:0810.2360 [hep-lat].

\bibitem{SLAC}
S.~D.~Drell, M.~Weinstein and S.~Yankielowicz,
``Strong Coupling Field Theories. 2. Fermions And Gauge Fields On A
Lattice,''
Phys.\ Rev.\  D {\bf 14} (1976) 1627.

\bibitem{Osterwalder-Schrader}
K.~Osterwalder and R.~Schrader,
``AXIOMS FOR EUCLIDEAN GREEN'S FUNCTIONS,''
Commun.\ Math.\ Phys.\  {\bf 31} (1973) 83;
%
\bibitem{Osterwalder:1974tc}
K.~Osterwalder and R.~Schrader,
``Axioms For Euclidean Green's Functions. 2,''
Commun.\ Math.\ Phys.\  {\bf 42} (1975) 281.

\bibitem{reflection}
K.~Osterwalder and E.~Seiler,
``Gauge Field Theories On The Lattice,''
Annals Phys.\  {\bf 110} (1978) 440.
\bibitem{Kawamot-Smit}
N.~Kawamoto and J.~Smit, 
``Effective Lagrangian and dynamical symmetry
 breaking in strongly coupled lattice QCD ", 
Nucl. Phys. \textbf{B192} (1981), 100.
\bibitem{G}
F.~Gliozzi, 
``Spinor algebra of the one component lattice fermions", 
Nucl. Phys. \textbf{B204} (1982), 419.
\bibitem{KMN}
H.~Kluberg-Stern, A.~Morel, O.~Napoly, and B.~Petersson, 
``Flavors of Lagrangian Susskind fermions", 
Nucl. Phys. \textbf{B220} (1983), 447.
\bibitem{Kogut-Susskind}
J.~B.~Kogut and L.~Susskind,
Phys.\ Rev.\ {\bf D11} (1975) 395,
L.~ Susskind,
Phys.\ Rev.\ {\bf D16} (1977) 3031. 

\bibitem{twisted-sym}
M.~Chaichian, P.~P.~Kulish, K.~Nishijima and A.~Tureanu,
``On a Lorentz-invariant interpretation of noncommutative space-time and  its
implications on noncommutative QFT,''
Phys.\ Lett.\  B {\bf 604} (2004) 98
[arXiv:hep-th/0408069].
\\
J.~Wess,
``Deformed coordinate spaces: Derivatives,''
arXiv:hep-th/0408080.
\\
F.~Koch and E.~Tsouchnika,
``Construction of theta-Poincare algebras and their invariants on
M(theta),''
Nucl.\ Phys.\  B {\bf 717} (2005) 387
[arXiv:hep-th/0409012].
\\
R.~Oeckl,
``Untwisting noncommutative R**d and the equivalence of quantum field
theories,''
Nucl.\ Phys.\  B {\bf 581} (2000) 559
[arXiv:hep-th/0003018].
\\
M.~Chaichian, P.~Presnajder and A.~Tureanu,
``New concept of relativistic invariance in NC space-time: Twisted  Poincare
symmetry and its implications,''
Phys.\ Rev.\ Lett.\  {\bf 94} (2005) 151602
[arXiv:hep-th/0409096].
\\
G.~Fiore and J.~Wess,
``On 'full' twisted Poincare' symmetry and QFT on Moyal-Weyl spaces,''
Phys.\ Rev.\  D {\bf 75} (2007) 105022
[arXiv:hep-th/0701078].
\\
A.~P.~Balachandran, A.~R.~Queiroz, A.~M.~Marques and P.~Teotonio-Sobrinho,
``Quantum Fields with Noncommutative Target Spaces,''
Phys.\ Rev.\  D {\bf 77} (2008) 105032
[arXiv:0706.0021 [hep-th]].

\bibitem{Asakawa}
T.~Asakawa, M.~Mori and S.~Watamura,
``Hopf Algebra Symmetry and String,''
arXiv:0805.2203 [hep-th];
%
``Twist Quantization of String and B Field Background,''
arXiv:0811.1638 [hep-th].




\bibitem{Sasai-Sasakura}
Y.~Sasai and N.~Sasakura,
``Braided quantum field theories and their symmetries,''
Prog.\ Theor.\ Phys.\  {\bf 118} (2007) 785
[arXiv:0704.0822 [hep-th]].

\bibitem{Riccardi-Szabo}
M.~Riccardi and R.~J.~Szabo,
``Duality and Braiding in Twisted Quantum Field Theory,''
JHEP {\bf 0801} (2008) 016
[arXiv:0711.1525 [hep-th]].
\bibitem{Majid}
S.~Majid,
``Examples of braided groups and braided matrices,''
J.\ Math.\ Phys.\  {\bf 32} (1991) 3246;
%
``Quantum and braided linear algebra,''
J.\ Math.\ Phys.\  {\bf 34} (1993) 1176
[arXiv:hep-th/9208006];
%
``Braided momentum in the Q Poincare group,''
J.\ Math.\ Phys.\  {\bf 34} (1993) 2045
[arXiv:hep-th/9210141];
%
``Free braided differential calculus, braided binomial theorem and the
braided exponential map,''
J.\ Math.\ Phys.\  {\bf 34} (1993) 4843
[arXiv:hep-th/9302076];
%
``Beyond supersymmetry and quantum symmetry: An Introduction to braided
groups and braided matrices,''
arXiv:hep-th/9212151;
\\
A.~Kempf and S.~Majid,
``Algebraic q integration and Fourier theory on quantum and braided spaces,''
J.\ Math.\ Phys.\  {\bf 35} (1994) 6802
[arXiv:hep-th/9402037].


\bibitem{Bars-Minic}
I.~Bars and D.~Minic,
``Non-commutative geometry on a discrete periodic lattice and gauge
theory,''
Phys.\ Rev.\  D {\bf 62}, 105018 (2000)
[arXiv:hep-th/9910091].

\bibitem{Ambjorn:nclattice}
J.~Ambjorn, Y.~M.~Makeenko, J.~Nishimura and R.~J.~Szabo,
``Finite N matrix models of noncommutative gauge theory,''
JHEP {\bf 9911} (1999) 029
[arXiv:hep-th/9911041];
%
``Nonperturbative dynamics of noncommutative gauge theory,''
Phys.\ Lett.\  B {\bf 480} (2000) 399
[arXiv:hep-th/0002158];
%
``Lattice gauge fields and discrete noncommutative Yang-Mills theory,''
JHEP {\bf 0005} (2000) 023
[arXiv:hep-th/0004147].




\bibitem{non-lattice}
M.~Hanada, J.~Nishimura and S.~Takeuchi
``Non-lattice simulation for supersymmetric gauge theories in one dimension,''
arXiv:0706.1647 [hep-lat].
\\
J.~W.~Elliott, J.~Giedt and G.~D.~Moore,
Phys.\ Rev.\  D {\bf 78} (2008) 081701
[arXiv:0806.0013 [hep-lat]].

\bibitem{Tsuchiya}
G.~Ishiki, Y.~Takayama and A.~Tsuchiya,
``N = 4 SYM on R x S**3 and theories with 16 supercharges,''
JHEP {\bf 0610} (2006) 007
[arXiv:hep-th/0605163].
\\
G.~Ishiki, S.~Shimasaki, Y.~Takayama and A.~Tsuchiya,
``Embedding of theories with SU(2|4) symmetry into the plane wave matrix
model,''
JHEP {\bf 0611} (2006) 089
[arXiv:hep-th/0610038].
\\
T.~Ishii, G.~Ishiki, K.~Ohta, S.~Shimasaki and A.~Tsuchiya,
``On relationships among Chern-Simons theory, BF theory and matrix model,''
Prog.\ Theor.\ Phys.\  {\bf 119} (2008) 863
[arXiv:0711.4235 [hep-th]].
\\
T.~Ishii, G.~Ishiki, S.~Shimasaki and A.~Tsuchiya,
``Fiber Bundles and Matrix Models,''
Phys.\ Rev.\  D {\bf 77} (2008) 126015
[arXiv:0802.2782 [hep-th]];
%
``N=4 Super Yang-Mills from the Plane Wave Matrix Model,''
Phys.\ Rev.\  D {\bf 78} (2008) 106001
[arXiv:0807.2352 [hep-th]].
\\
G.~Ishiki, S.~W.~Kim, J.~Nishimura and A.~Tsuchiya,
``Deconfinement phase transition in N=4 super Yang-Mills theory on RxS\verb+^+3 from
supersymmetric matrix quantum mechanics,''
arXiv:0810.2884 [hep-th].
\\
G.~Ishiki, K.~Ohta, S.~Shimasaki and A.~Tsuchiya,
``Two-Dimensional Gauge Theory and Matrix Model,''
arXiv:0811.3569 [hep-th].







\bibitem{Chaichian:book}
M.~Chaichian and A.~P.~Demichev,
``Introduction To Quantum Groups,''
{\it  Singapore, Singapore: World Scientific (1996) 343~p}.

\bibitem{Majid:book}
S.~Majid,
``Foundations of quantum group theory,''
{\it  Cambridge, UK: Univ.\ Pr.\ (1995) 607~p}.

\bibitem{Chari:book}
V.~Chari and A.~Pressley,
{\it  Cambridge, UK: Univ. Pr. (1994) 651 p}.



\end{thebibliography}
